\documentclass[longauth]{aa}  
\usepackage{natbib}
\usepackage{graphicx}
\usepackage{gensymb}
\usepackage[toc,page]{appendix}
\usepackage{pbox}
\usepackage{makecell}
\usepackage[flushleft]{threeparttable}
\usepackage{hyperref}
\usepackage{multirow}
\usepackage{multicol}
\usepackage{subcaption}
\hypersetup{colorlinks=true,breaklinks=true,citecolor=blue,urlcolor=blue}

\usepackage{xcolor}
\usepackage{lscape}
\usepackage{hhline}
\usepackage[export]{adjustbox} 
\usepackage{comment}
\usepackage{txfonts}
\bibpunct{(}{)}{;}{a}{}{,}

		\newcommand{\xmm}{XMM-\textit{Newton}}
		 
		\newcommand{\chandra}{\textit{Chandra}}

		\newcommand{\z}{\textit{z}}
		\newcommand{\kashz}{KASH\textit{z}}
		\newcommand{\cosmos}{COSMOS}
		\newcommand{\cdfs}{CDFS}
		\newcommand{\xuds}{X-UDS}

		\newcommand{\oiii}{[O{ III}]}

		\newcommand{\counozero}{CO(1-0)}
		\newcommand{\codueuno}{CO(2-1)}
		\newcommand{\cotredue}{CO(3-2)}
		\newcommand{\coquattrotre}{CO(4-3)}
		\newcommand{\lpco}{$L'_{\rm CO}$}
		\newcommand{\lpcotre}{$L'_{\rm \cotredue}$}
		\newcommand {\Msun}{\,{{\rm  M}_{\odot}}}
		\newcommand {\Lsun}{\,{{\rm  L}_{\odot}}}
		\newcommand {\kms} {\,{\rm km\,s}^{-1}}

		\newcommand{\Mstar}{M_{\ast}}
		\newcommand{\mstar}{$M_{\ast}$}

		\newcommand{\Lfir}{L_{\rm FIR}}

		\newcommand{\lumcgs}{erg~s$^{-1}$}
		\newcommand{\Lumcgs}{\rm erg~s^{-1}}

		\newcommand{\lbol}{{$L_{\rm bol}$}}
		\newcommand{\Lbol}{{L_{\rm bol}}}

		\newcommand{\alphaco}{\alpha_{\rm CO}}	
		
		\defcitealias{circosta2018}{C18}
		\defcitealias{circosta2021}{C21}
		\defcitealias{perna2018}{P18}

\begin{document} 

\title{\kashz+SUPER:  Evidence for cold molecular gas depletion in AGN hosts at cosmic noon
	}

\titlerunning{\kashz+SUPER: Cold molecular gas depletion at cosmic noon}

   \author{E. Bertola\inst{1}\fnmsep\inst{2}\fnmsep\inst{3}\thanks{elena.bertola@inaf.it},
          C. Circosta\inst{4}\fnmsep\inst{5}, 
          M. Ginolfi\inst{6}\fnmsep\inst{1}\fnmsep\inst{7}, 
          V. Mainieri\inst{7}, 
          C. Vignali\inst{2}\fnmsep\inst{3}, 
          G. Calistro Rivera\inst{7}\fnmsep\inst{8}, 
          S. R. Ward\inst{9}\fnmsep\inst{10}\fnmsep\inst{11}\fnmsep\inst{7}, 
          I. E. Lopez\inst{2}\fnmsep\inst{3},
          A. Pensabene\inst{12},
          D. M. Alexander\inst{13},
          M. Bischetti\inst{14}\fnmsep\inst{15},
          M. Brusa\inst{2},
          M. Cappi\inst{3},
          A. Comastri\inst{3},
          A. Contursi\inst{16},
          C. Cicone\inst{17},
          G. Cresci\inst{1},
          M. Dadina\inst{3},
          Q. D’Amato\inst{1},
          A. Feltre\inst{1},
          C. M. Harrison\inst{9},
          D. Kakkad\inst{18}\fnmsep\inst{19}\fnmsep\inst{20}, 
          I. Lamperti\inst{21},
          G. Lanzuisi\inst{3},
          F. Mannucci\inst{1},
          A. Marconi\inst{6}\fnmsep\inst{1},
          M. Perna\inst{21},
          E. Piconcelli\inst{22},
          A. Puglisi\inst{23}\thanks{Anniversary Fellow},
          F. Ricci\inst{24}\fnmsep\inst{22}, 
          J. Scholtz\inst{25}\fnmsep\inst{26},
          G. Tozzi\inst{16},
          G. Vietri\inst{27},
          G. Zamorani\inst{3},
          L. Zappacosta\inst{22},
          }

   \institute{
   INAF--OAA, Osservatorio Astrofisco di Arcetri, largo E. Fermi 5, 50127, Firenze, Italy
   \and Dipartimento di Fisica e Astronomia ``Augusto Righi'', Universit\`a degli Studi di Bologna, via P. Gobetti 93/2, 40129 Bologna, Italy
    \and INAF--OAS, Osservatorio di Astrofisica e Scienza dello Spazio di Bologna, via P. Gobetti 93/3, 40129 Bologna, Italy 
    \and European Space Agency (ESA), European Space Astronomy Centre (ESAC), Camino bajo del Castillo s/n, Villanueva de la Canada, E-28692 Madrid, Spain
    \and Department of Physics \& Astronomy, University College London, Gower Street, London WC1E 6BT, UK
    \and   Dipartimento di Fisica e Astronomia, Universit\`a di Firenze, Via G. Sansone 1, I-50019, Sesto F.no (Firenze), Italy
    \and European Southern Observatory, Karl–Schwarzschild–Straße 2, D-85748 Garching bei Munchen, Germany
    \and German Aerospace Center (DLR), Institute of Communications and Navigation, Wessling, Germany
    \and School of Mathematics, Statistics and Physics, Newcastle University, Newcastle upon Tyne, NE1 7RU, UK
    \and Excellence Cluster ORIGINS, Boltzmannstraße 2, 85748 Garching bei München, Germany%
    \and Ludwig Maximilian Universität, Professor-Huber-Platz 2, 80539 München, Germany
    \and Dipartimento di Fisica “G. Occhialini”, Università degli Studi di Milano-Bicocca, Piazza della Scienza 3, I-20126, Milano, Italy
    \and Centre for Extragalactic Astronomy, Department of Physics, Durham University, South Road, Durham, DH1 3LE, UK
    \and Dipartimento di Fisica, Universit\`a di Trieste, Sezione di Astronomia, Via G.B. Tiepolo 11, I-34131 Trieste, Italy
    \and INAF—Osservatorio Astronomico di Trieste, Via G.B. Tiepolo 11, I–34143 Trieste, Italy
    \and Max-Planck-Institut für Extraterrestrische Physik (MPE), Giessenbachstraße 1, D-85748 Garching, Germany
    \and Institute of Theoretical Astrophysics, University of Oslo, Postboks 1029, Blindern, 0315 Oslo
    \and Space Telescope Science Institute, 3700 San Martin Drive, Baltimore, MD
21218, USA
    \and European Southern Observatory, Alonso de Cordova 3107, Vitacura,
Casilla, 19001 Santiago de Chile, Chile
    \and Department of Physics, University of Oxford, Denys Wilkinson Building,
Keble Road, Oxford OX1 3RH, UK
    \and Centro de Astrobiolog\'ia (CAB), CSIC--INTA, Cra. de Ajalvir Km.~4, 28850 -- Torrej\'on de Ardoz, Madrid, Spain  
    \and INAF – Osservatorio Astronomico di Roma, Via Frascati 33, 00040 Monte Porzio Catone, Roma, Italy      
    \and School of Physics and Astronomy, University of Southampton, Highfield SO17 1BJ, UK
    \and Dipartimento di Matematica e Fisica, Università Roma Tre, via della Vasca Navale 84, I-00146, Roma, Italy
    \and Kavli Institute for Cosmology, University of Cambridge, Madingley Road, Cambridge, CB3 OHA, UK
    \and Cavendish Laboratory - Astrophysics Group, University of Cambridge, 19 JJ Thomson Avenue, Cambridge, CB3 OHE, UK
    \and INAF – Istituto di Astrofisica Spaziale e Fisica Cosmica Milano, Via
A. Corti 12, 20133 Milano, Italy
 }

\authorrunning{Bertola, Circosta, Ginolfi et al.}

   \date{Received ; accepted }
 
 \abstract
  {
  The energy released by active galactic nuclei (AGN) has the potential to heat or remove the gas of the ISM, thus likely impacting the cold molecular gas reservoir of host galaxies at first, with star formation following as a consequence on longer timescales. Previous works on high-\z\ galaxies, which compared the gas content of those without identified AGN, have yielded conflicting results, possibly due to selection biases and other systematics.
  To provide a reliable benchmark for galaxy evolution models at cosmic noon ($z=1-3$), two surveys were conceived: SUPER and \kashz, both targeting unbiased, X-ray-selected AGN at $z>1$ that span a wide bolometric luminosity range. 
  In this paper, we assess the effects of AGN feedback on the molecular gas content of host galaxies in a statistically-robust, uniformly-selected, coherently-analyzed sample of AGN at $z=1-2.6$, drawn from the \kashz\ and SUPER surveys. 
  By using targeted and archival ALMA data in combination with dedicated SED modeling, we retrieve CO and FIR luminosity as well as $\Mstar$ of SUPER and \kashz\ host galaxies. 
  We selected non-active galaxies from PHIBBS, ASPECS and multiple ALMA/NOEMA surveys of sub-millimeter galaxies in the COSMOS, UDS and ECDF fields. By matching the samples in redshift, stellar mass and FIR luminosity, we compared the properties of AGN and non-active galaxies within a Bayesian framework.
  We find that AGN hosts at given FIR luminosity are on average CO depleted compared to non-active galaxies, thus confirming what previously found in the SUPER survey. Moreover, the molecular gas fraction distributions of AGN and non-active galaxies are statistically different, with the distribution of AGN being skewed to lower values. Our results indicate that AGN can indeed reduce the total cold molecular gas reservoir of their host galaxies. Lastly, by comparing our results with predictions from three cosmological simulations (TNG, Eagle and Simba) filtered to match the properties of observed AGN, AGN hosts and non-active galaxies, we confirm already known and highlight new discrepancies between observations and simulations. 
  }


  \maketitle
%
\section{Introduction}
\label{intro}
The sphere of gravitational influence of supermassive black holes (SMBHs) is spatially limited to the innermost regions of galaxies ($\sim10$pc for a $10^8\Msun$ black hole; \citealp{alexanderhickox2012}). Yet, SMBH accretion can release large amounts of energy that, if efficiently coupled to the surrounding material, allows active galactic nuclei (AGN) to shape galaxy growth by heating, exciting and/or removing the gas of the interstellar medium (ISM). Such an interplay between AGN activity and the life of galaxies is referred to as AGN feedback \citep[e.g.,][]{silk_rees1998,fabian2012}. 
Astronomers found multiple observational evidence for a connection between observational and physical properties of the central AGN and its host galaxy \citep[e.g.,][]{strateva2001,mcconell2011,rodighiero2015}, as also predicted by theoretical models \citep[e.g.,][]{somerville2008,lapi2014,costa2018b}, leading to the so-called ``AGN/galaxy co-evolution'' scenario. 
However, we still lack a full understanding of AGN feedback despite its importance \citep[e.g.,][]{kormendy2013,harrison2017,harrison2018,ward2022}, for instance whether galaxy evolution is influenced by AGN activity as a whole or whether AGN-driven winds and jets are key properties in driving AGN feedback effects or whether we are using sensitive observational tracers for AGN feedback.  
AGN feedback is believed to be highly dominant when both cosmic SMBH growth and star formation rate (SFR) were at their peaks \citep[the ``cosmic noon'', $z \simeq 1-3$; e.g.,][]{madau2014,aird2015}. 
Hence, sources at the cosmic noon are the most promising and most informative targets to investigate AGN feedback, both in terms of ``causes'' (e.g., AGN-driven winds and jets) and ``effects'' (e.g., reduced gas fraction and disturbed gas kinematics in AGN hosts). 

A way to test the impact of AGN feedback on the host galaxy is to observe its molecular gas content and conditions. 
In particular, the Schmidt-Kennicutt law \citep{schmidt1959,kennicutt1989} is a fundamental relation that links the molecular gas content and the level of star formation (SF) in galaxies, that is often used in its integrated form, i.e., global SFR of galaxies vs. their total molecular mass \citep[e.g.,][]{carilli_2013,sargent2014,perna2018,bischetti2021,circosta2021}. 
If AGN influence the evolution of galaxies as a whole \citep[e.g.,][]{fabian2012,harrison2017}, they will reasonably impact the molecular gas reservoir first, and then, consequently, SF on longer timescales. 
Besides addressing AGN feedback as ``caught in the act'' (i.e., AGN-driven outflows or radio jets), comparative studies of the molecular gas properties in AGN host galaxies and matched non-active galaxies (i.e., galaxies not hosting an AGN) can shed light on the impact of AGN on galaxy evolution, regardless of whether outflows are concurrently detected in a given source.

A powerful proxy to measure the molecular content of galaxies is represented by the carbon monoxide 
(CO), the second most abundant molecule after H$_{2}$, and hence the primary tracer of molecular gas \citep[e.g.,][]{carilli_2013,hodge_dacunha2020_reviewALMA}. 
The molecular gas mass can be derived from the luminosity of the ground state transition CO(1-0) ($M_{\rm H_2}=\alpha_{\rm CO}L'_{\rm\counozero}$). 
Yet, the value of $\alphaco$ has to be assumed and it is very uncertain as it depends on different quantities, for instance the distance from the main sequence (MS, \citealp{elbaz2007,noeske2007}) and the metallicity \citep[e.g.,][]{accurso2017}. 
The value calibrated on the Milky Way ($\alphaco=4 \Msun\rm\,(K\ km\ s^{-1}\ pc^2)^{-1}$, \citealp{bolatto2013}) can be a good approximation for regular star-forming galaxies \citep{carilli_2013}, while a lower value 
($\alphaco=0.8-1 \Msun\rm\,(K\ km\ s^{-1}\ pc^2)^{-1}$, 
\citealp{downes_solomon_1998,solomonvandenbout2005,calistrorivera2018,amvrosiadis2023} or $\alphaco=1.8-2.5 \Msun\rm\,(K\ km\ s^{-1}\ pc^2)^{-1}$, \citealp{cicone2018_ngc6240,Herrero-Illana2019,MontoyaArroyave2023}), could be more representative for high-\z\ sub-\textit{mm} galaxies, which usually have an extremely dense ISM and often intense SF activity \citep[e.g.,][]{birkin2021}. 
Observations of massive MS galaxies at cosmic noon resulted in an $\alphaco=3.6 \Msun\rm\ (K\ km\ s^{-1}\ pc^2)^{-1}$ \citep[e.g.,][]{daddi2010a,genzel2015}. 

Measuring the ground transition of CO is not always possible, especially at high redshift. However, observations of CO transitions from higher $J$ levels can be converted to the  \counozero\ flux by assuming a CO Spectral Line Energy Distribution (CO-SLED), which expresses the relative strength of the CO rotational lines as a function of the quantum number $J$ \citep[e.g.,][]{pozzi2017,mingozzi2018,kirkpatrick2019,boogaard2020_aspecs,valentino2021,pensabene2021,esposito2022,molyneux2024}. Even though AGN mostly contribute to high-$J$ levels, CO ladders of AGN hosts and non-active galaxies are seen to vary also at $J<5$, despite the large associated uncertainties, with AGN showing larger ratios that increase for increasing AGN luminosity   \citep{carilli_2013,kirkpatrick2019,vallini2019,boogaard2020_aspecs}. 

Results regarding the effects of AGN on the molecular gas phase are controversial and seem to hint at a dichotomy between low-\z\ and high-\z\ scenarios. 
For what concerns the integrated properties of galaxies, AGN hosts in the local Universe are usually similar to non-active galaxies \citep[e.g.,][but see also \citealp{mountrichas2024}]{rosario2018,koss2021,salvestrini2022} or even gas richer, in the sense that more powerful AGN are hosted in gas richer host galaxies \citep[e.g.,][]{vito_2014,husemann2017}. Yet, significant effects of negative AGN feedback are seen in local AGN hosting molecular outflows \citep{fiore2017}. Additionally, some results hint at enhanced SF efficiencies in both local and high-\z\ AGN hosts possibly linked to large molecular gas reservoirs \citep[e.g.,][]{shangguan2020b_molgas,jarvis2020,bischetti2021}.Spatially-resolved studies of local galaxies unveiled gas depletion in the nuclear regions of AGN hosts, and not on galactic scales \citep{sabatini2018,rosario2019,fluetsch2019,feruglio2020,ellison2021,zanchettin2021}, with indications that the central SF activity (within few 100~pc) is indeed reduced due to the presence of AGN, while the galaxy-wide SFR is unaffected \citep[e.g.,][but see also \citealp{molina2023}, who find enhanced SF activity in the proximity of the AGN]{sanchez2018,garciaburillo2021_gatos,lammers2023}.
At high redshift, the observed impact of AGN on the molecular ISM of their hosts is debated. Several studies find that AGN hosts are significantly CO-depleted 
when compared to control samples of non-active galaxies, either in terms of gas fraction\footnote{The definition of gas fraction varies within different studies. Throughout this work, we assume the definition $f_{\rm gas}=M_{\rm gas}/\Mstar$, also dubbed $\mu_{\rm mol}$ in the literature. } or of depletion timescales \citep[e.g.,][but see also \citealp{herrera-camus2019,spingola2020}]{brusa2016,brusa2018,kakkad2017,perna2018,bischetti2021}. Yet, other studies find limited (\citealp[e.g.,][]{circosta2021}, hereafter, \citetalias{circosta2021}) or no significant difference between the gas fraction of non-active galaxies and AGN hosts \citep[e.g.,][]{valentino2021} and some other works find no clear link between gas content and AGN power \citep[e.g.,][]{kirkpatrick2019}. An additional issue in comparing the properties of AGN hosts and non-active galaxies resides in the intrinsic variability of AGN and their flickering activity \citep{harrison2017}, which can also affect the distinction of galaxies that host and that do not host an AGN based on when the system was targeted, for instance, in the X-rays. 

Predictions from cosmological simulations seem to be likewise controversial. In particular, \citet{ward2022} recently analyzed the output of three key cosmological simulations (Illustris-TNG, hereafter TNG, \citealp{springel2018,pillepich2018,naiman2018,nelson2018,marinacci2018}; EAGLE, \citealp{crain2015,schaye2015}; Simba, \citealp{dave2019}) applying the same methods employed to analyze observations of galaxies and AGN hosts. 
Considering samples of local and $z=2$ targets, the authors conclude that none of the selected simulations predicts strong negative correlations between AGN power and molecular gas fraction or specific SFR (sSFR~$=$~SFR/$\Mstar$). Conversely, powerful AGN seem to preferentially reside in gas-rich, highly-star-forming galaxies, whereas gas-depleted and quenched fractions are higher in the control samples of regular galaxies than for the AGN hosts. \citet{ward2022}, however, argue that there is a quantifiable difference among the predictions of the three selected simulations and that the bolometric luminosity range covered by simulations and observational efforts hardly overlap, especially at cosmic noon. This is mainly due to the difficulty in reproducing sizable samples of high-luminosity AGN ($\Lbol\gtrsim10^{43.5}$ \lumcgs\ at $z=0$ and $\Lbol\gtrsim10^{45}$ \lumcgs\ at $z=2$) in cosmological simulations, since they are rare and too short-lived to be captured in the simulated volumes, and to the observational cost of observing sizable samples of less luminous AGN (e.g., $\Lbol\lesssim10^{44}$ \lumcgs\ at $z>1$). 

The dichotomy between results on local and high-\z\ targets, and between different samples of high-\z\ sources, could arise from selection effects. High-\z\ studies usually rely on few, high-luminosity ($\log(L_{\rm Bol}/\rm erg~s^{-1})>46$) and heterogeneous targets, both because high-power AGN are the best candidates to produce efficient feedback \citep[e.g.,][]{fiore2017} and because of the challenging observations required for distant objects. 

Two surveys of X-ray selected AGN were recently conceived to provide multi-wavelength studies of AGN samples blindly selected with respect to the presence of ionised outflows or jets at cosmic noon: \kashz\ \citep[KMOS AGN Survey at High redshift, PI: D. Alexander;][Scholtz et al., in prep]{harrison2016} at $z=0.6-2.6$ and SUPER \citep[SINFONI Survey for Unveiling the Physics and Effect of Radiative feedback, PI: V. Mainieri;][hereafter, \citetalias{circosta2018}]{circosta2018} at $z=2-2.5$. Both surveys aim at studying the interplay between AGN and galaxy evolution in terms of AGN-driven ionized winds, traced through [OIII] emission lines, and properties of the host galaxy (see Sect. \ref{kashz:sec:samp_sel}). 
SUPER relied on adaptive-optics-assisted, spatially-resolved studies of ionized gas, which require longer exposures, leading to a reduced sample size and sparser sampling of AGN with $\log(\Lbol/$\lumcgs$)<45$ compared to the \kashz\ survey, with its larger sample size and lower bolometric luminosity cut. 
In this context, the different observational capabilities and selection criteria make \kashz\ and SUPER complementary surveys, thus using them jointly allows us to capitalize on their strengths. 

As part of the SUPER survey, \citetalias{circosta2021} provided the first systematic analysis of the molecular gas content of AGN host galaxies at $z\simeq2$ through a dedicated follow-up in \cotredue\ with ALMA (Atacama Large Millimeter Array). With robust measurements of physically relevant quantities (SFR, $M_{\rm gas}$) traced through observational proxies ($L_{\rm FIR}$, \lpcotre), the authors find hints for negative AGN feedback effects in their sample. Yet, the sample size of SUPER ALMA AGN did not allow to discriminate whether such hints of gas depletion were indicative of less prominent AGN feedback effects in moderate luminosity AGN or due to a large fraction of upper limits combined with a limited sample size. 
As Figure \ref{kashz:fig:lx_kashz} shows, combining SUPER targets with \kashz\ AGN allows for a robust sampling of moderate-luminosity AGN ($44<\log(\Lbol/$\lumcgs$)<46$), still covering a wide range in AGN bolometric luminosity by means of the different sample selection criteria (see Fig. \ref{kashz:fig:lx_kashz}, but also Fig. 1 of \citetalias{circosta2018}). 

The aim of this work is to investigate the molecular gas properties of a statistically-robust, consistently-selected, coherently-analyzed sample of AGN at cosmic noon ($z=1-2.6$) available in the ALMA archive, improving the sampling of the $\log(\Lbol/$\lumcgs$)<44-46$ bolometric range. We build upon the methods and analysis of \citetalias{circosta2021}, and build our sample combining SUPER ALMA AGN with targets from the \kashz\ survey (see Sect. \ref{kashz:sec:samp_sel}). The control sample of non-active galaxies is presented in Sect. \ref{kashz:sec:control}. 
The ALMA data selection and reduction of \kashz\ targets is presented in Sect. \ref{kashz:sec:kashz_data} and Appendix \ref{app:chap:kashz_alma}, while the SED fitting and photometry collection of both SUPER and \kashz\ AGN are described in Appendix  \ref{app:sec:sedfit}. 
We then apply a quantitative analysis in the Bayesian framework developed by \citetalias{circosta2021} to compare our enlarged AGN sample with the control sample of non-active galaxies in Sect. \ref{kashz:sec:results}. In Sect. \ref{sec:ward22}, we compare our observed sample with the output of cosmological simulations. Results are discussed and summarized in Sects. \ref{sec:discussion} and \ref{sec:summary}. We assume a flat ${\Lambda}$CDM cosmology \citep{planckcoll2020}, with $H_{\rm 0}=67.7\ {\rm km\ s^{-1}\ Mpc^{-1}}$ and $\Omega_{\rm m,0}=0.31$ throughout the paper. 

\section{Sample selection}
\label{kashz:sec:samp_sel}
To build a robust sample of moderate-to-high-luminosity AGN ($\log(\Lbol$/\lumcgs) $\simeq44-47$), we considered all the SUPER ALMA targets presented in \citetalias{circosta2021} and complement them with archival ALMA observations targeting the CO emission with $J<5$ (\counozero, \codueuno, \cotredue, \coquattrotre) of 
\kashz\ AGN (see Sects. \ref{kashz:sec:survey} and \ref{kashz:sec:kashz_data_sel}, Appendix \ref{app:chap:kashz_alma} and Table \ref{kashz:tab:almatargets}). 
The other needed parameters for our analysis (mainly $\Mstar$ and $\Lfir$) are estimated from SED fitting performed with CIGALE \citep{boquien2019_cigale,yang2020_xcigale} using the most up-to-date multi-wavelength, broadband photometry released by the deep fields' collaborations (see Appendix \ref{app:sec:sedfit}). 
Figure \ref{kashz:fig:lx_kashz} shows the $L_{\rm X}$ vs. \z\ distribution of the full \kashz\ and SUPER survey samples. We also show the $\Lbol$ vs. \z\ distribution of the AGN used in this work, i.e., ALMA targets from \kashz\ and SUPER, as derived from SED fitting (see Appendix \ref{app:chap:kashz_alma}).  

For the analysis presented in Sect. \ref{kashz:sec:results}, we discard 6 AGN from the SUPER and \kashz\ ALMA samples because they are missing at least one of the parameters of interest needed for our analysis ($\Mstar$ or $\Lfir$; we mark these targets with an asterisk in Table \ref{kashz:tab:super+kashz} and in the bottom panel of Fig.  \ref{kashz:fig:lx_kashz}). These six targets are extremely bright Broad Line AGN with sparse or missing FIR photometry coverage. A reasonable decoupling of AGN and galaxy emission in the SED fitting is thus not possible, hampering also the determination of meaningful upper limits for stellar mass and FIR luminosity of the host galaxy \citepalias{circosta2018,circosta2021}. 

\begin{figure}[!h]
	\includegraphics[width=0.48\textwidth]{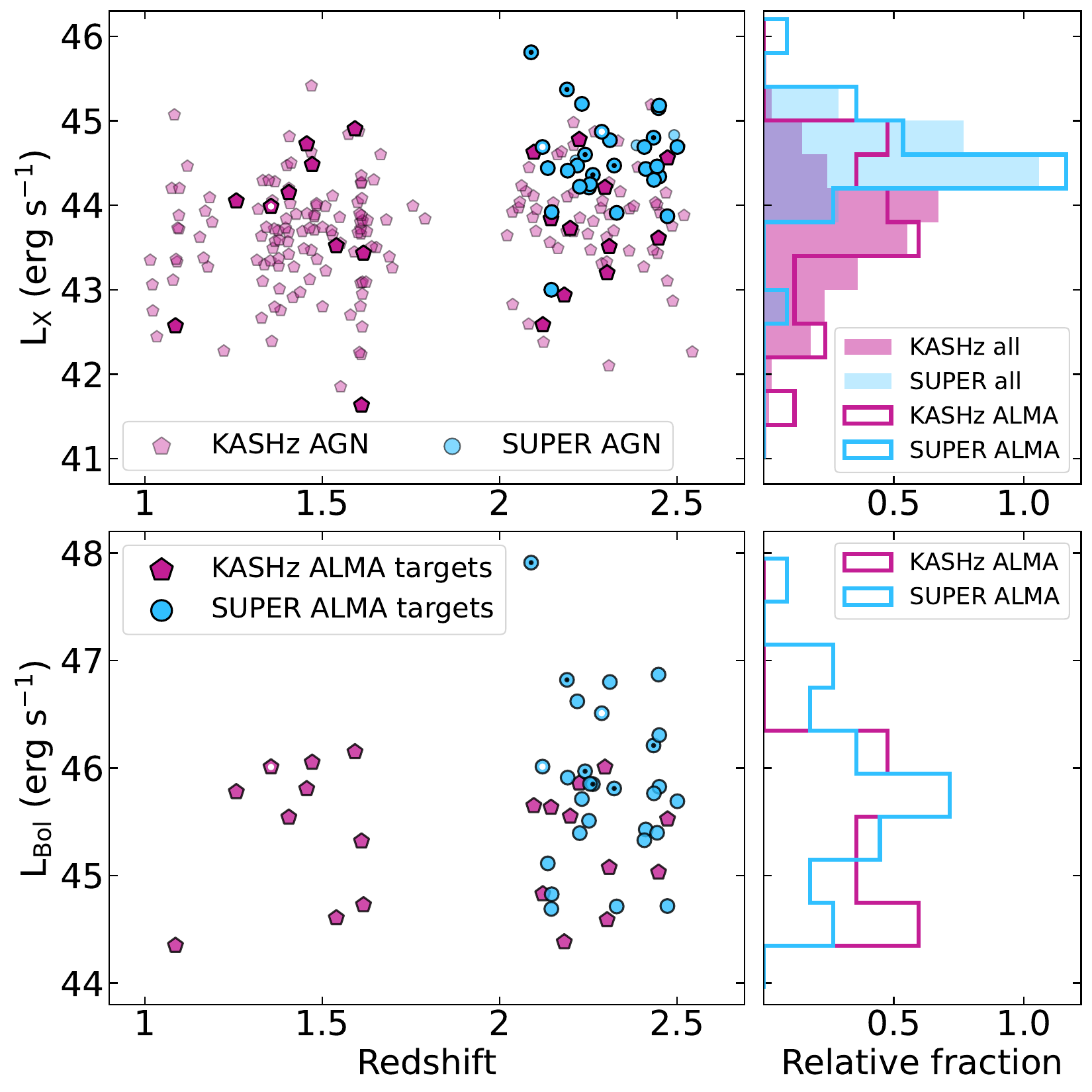}
	\caption{\textit{Top}: Intrinsic 2-10 keV rest-frame X-ray luminosity $L_{\rm X}$ vs. redshift \z\ distribution of the full \kashz\ (purple pentagons) and SUPER (blue circles) surveys. SUPER ALMA targets are taken from \citetalias{circosta2021}. The side panel shows the density distribution of SUPER (blue) and KASHz (purple) targets. \textit{Bottom}: Bolometric luminosity $L_{\rm Bol}$ vs. redshift \z\ distribution of the AGN targeted by ALMA and included in this work. For \kashz\ (purple pentagons) and SUPER (blue circles) targets included in this panel, the bolometric luminosity was derived from SED fitting in this work (see Sect. \ref{app:sec:sedfit}) and  in  \citetalias{circosta2018}, \citetalias{circosta2021}. The side panel shows the density distribution of the samples. Black dots mark the six targets discarded because missing at least one parameter of interest for our analysis, white dots mark the three targets that do not allow to compute $f_{\rm gas}$. }
	\label{kashz:fig:lx_kashz}
\end{figure}

\subsection{\kashz\ survey}
\label{kashz:sec:survey}
The \kashz\ survey (\citealp[Scholtz et al., in prep.]{harrison2016}) is a VLT/KMOS GTO survey of X-ray selected AGN (PI: D. Alexander), built with the aim of investigating the impact of AGN feedback on the ionized gas phase as traced by the H$\beta$, [OIII]$\lambda5007$, H$\alpha$, [NII]$\lambda6583$ and [SII]$\lambda\lambda6717,6730$ optical emission lines. The final catalog of the survey will be presented in Scholtz et al. (in prep), alongside results from the spectral analysis of the ionized gas emission. We summarize here some important properties of the survey sample. 

The \kashz\ survey totals $\simeq$230 AGN, spanning the redshift range $z=0.6-2.6$ and the $L_{\rm X}$ range $\log (L_{\rm X}/\rm erg~s^{-1})=41-45.4$, which corresponds to a bolometric luminosity range of $\log (L_{\rm Bol}/\rm erg~s^{-1})\simeq42-47$, based on the bolometric correction derived by \citealp{duras_2020}. 
Near-IR (NIR) spectroscopic data is available for all \kashz\ targets, predominantly from our KMOS program but also including archival SINFONI data. 
\kashz\ targets were drawn from five different X-ray deep fields: \textit{i)} \chandra\ Cosmic evolution survey fields \citep[C-COSMOS and COSMOS-Legacy;][name convention: cid\_ID and lid\_ID]{civano2016,marchesi2016}; \textit{ii)} \chandra\ Deep Field South \citep[CDFS;][name convention: cdfs\_ID]{luo2017}; \textit{iii)} \textit{Subaru}/\xmm\ Deep Survey \citep[SXDS;][name convention: sxds\_ID]{ueda2008,akiyama2015}; \textit{iv)} \chandra\ Legacy Survey of the UKIDSS Ultra Deep Survey field \citep[\xuds;][name convention: xuds\_ID]{kocevski2018}; \textit{v)} SSA22 protocluster field \citep[][name convention: ssa22\_ID]{lehmer2009}. 
Based on our spectroscopic redshift identification using NIR spectra, we confirm or update the redshift listed in the parent X-ray survey catalogs and the respective rest-frame X-ray properties, as described in Appendix \ref{app:xraylum}. 
\begin{table}[!h]
    \centering
    \caption{Summary of the AGN sample}
    \begin{tabular}{lcccc}
	\hline\hline
	ID                 &   RA[J2000] &  DEC[J2000] &      $z_{\rm spec}$ & sample \\ \hline
	J1333              & 13:33:35.79 & +16:49:04.0 &  2.089 &      2 \\
	X\_N\_102\_35      &  02:29:05.94 & -04:02:43.0 &   2.19 &      2 \\
	X\_N\_104\_25      & 02:30:24.47 &  -04:09:13.4 &  2.241 &      2 \\
	X\_N\_128\_48      &  02:06:13.54 & -04:05:43.2 &  2.323 &      2 \\
	X\_N\_44\_64       &  02:27:01.46 & -04:05:06.7 &  2.252 &      2 \\
	X\_N\_53\_3        &  02:20:29.84 & -02:56:23.4 &  2.434 &      2 \\
	X\_N\_6\_27        & 02:23:06.32 & -03:39:11.1 &  2.263 &      2 \\
	X\_N\_81\_44       &  02:17:30.95 & -04:18:23.7 &  2.311 &      2 \\
	cdfs\_258          &  03:32:14.43 & -27:51:10.7 &   1.54 &      1 \\
	cdfs\_313          &  03:32:17.44 & -27:50:03.1 &  1.611$^c$ &  1 \\
	cdfs\_419$^a$      &  03:32:23.44 & -27:42:55.0 & 2.1416 &      3 \\
	cdfs\_427$^a$      &  03:32:24.20 & -27:42:57.5 & 2.3021 &      3 \\
	cdfs\_458          &  03:32:25.68 & -27:43:05.6 &  2.297 &      1 \\
	cdfs\_522$^a$      &  03:32:28.50 & -27:46:58.0 & 2.3085 &      3 \\
	cdfs\_587          &  03:32:31.46 & -27:46:23.1 & 2.2246 &      1 \\
	cdfs\_614$^a$      &  03:32:33.02 & -27:42:00.3 & 2.4525 &      3 \\
	cdfs\_794          &  03:32:43.18 & -27:55:14.7 & 2.1219 &      1 \\
	cid\_38            &  10:01:02.83 & +02:03:16.6 &  2.192 &      2 \\
	cid\_72            & 10:00:21.97 & +02:23:56.7 & 2.4734 &      1 \\
	cid\_86            & 10:00:28.70 & +02:17:45.3 & 2.0965 &      1 \\
	cid\_108           & 10:00:14.08 & +02:28:38.7 & 1.2582 &      1 \\
	cid\_166           & 09:58:58.68 &  +02:01:39.2 &  2.448 &      2 \\
	cid\_178           & 09:58:20.45 & +02:03:04.1 &  1.356 &      1 \\
	cid\_247           & 10:00:11.23 & +01:52:00.3 &  2.412 &      2 \\
	cid\_337           & 09:59:30.39 & +02:06:56.1 &  2.226 &      2 \\
	cid\_346$^b$       & 09:59:43.41 & +02:07:07.4 &  2.219 &      3 \\
	cid\_357$^b$       &  09:59:58.02 & +02:07:55.1 &  2.136 &      3 \\
	cid\_451$^b$       & 10:00:00.61 & +02:15:31.1 &   2.45 &      3 \\
	cid\_467           & 10:00:24.48 & +02:06:19.8 &  2.288 &      2 \\
	cid\_499           & 09:59:40.74 & +02:19:38.9 & 1.4566 &      1 \\
	cid\_852           & 10:00:44.21 & +02:02:06.8 &  2.232 &      2 \\
	cid\_864           & 09:59:31.58 & +02:19:05.5 & 1.6166 &      1 \\
	cid\_970$^b$       & 10:00:56.52 & +02:21:42.4 &  2.501 &      3 \\
	cid\_971$^b$       & 10:00:59.45 & +02:19:57.4 &  2.473 &      3 \\
	cid\_1205$^b$      & 10:00:02.57 & +02:19:58.7 &  2.255 &      3 \\
	cid\_1215$^b$      & 10:00:15.49 & +02:19:44.6 &   2.45 &      3 \\
	cid\_1253          & 10:01:30.57 & +02:18:42.6 &  2.147 &      2 \\
	cid\_1286          & 10:00:34.08 & +02:15:54.3 & 2.1992 &      1 \\
	cid\_1605          & 09:59:19.82 & +02:42:38.7 &  2.121 &      2 \\
	cid\_2682          & 10:00:08.81 &  +02:06:37.7 &  2.435 &      2 \\
	lid\_206           & 10:01:15.56 & +02:37:43.4 &   2.33 &      2 \\
	lid\_1289          & 09:59:14.65 &  +01:36:35.0 &  2.408 &      2 \\
	lid\_1565          & 10:02:11.28 & +01:37:06.5 & 1.5926 &      1 \\
	lid\_1639          & 10:02:58.41 & +02:10:13.9 &  1.472$^d$ &      1 \\
	lid\_1852          & 09:58:26.57 & +02:42:30.2 &  2.444 &      2 \\
	lid\_3456          & 09:58:38.40 &  +01:58:26.8 &  2.146 &      2 \\
	xuds\_358          & 02:17:17.43 & -05:13:48.1 & 2.1824 &      1 \\
	xuds\_477          & 02:18:02.50 & -05:00:32.9 & 1.0867 &      1 \\
	xuds\_481          & 02:18:37.77 & -04:58:50.3 & 1.4062 &      1 \\ \hline
\end{tabular}
\tablefoot{The sample flag indicates the parent sample of the targets: 1 for \kashz\ only, 2 for SUPER only, 3 for targets shared by the two surveys. $^a$ Target shared with the SUPER survey, ALMA data analysis is presented in this work. $^b$ Target shared with the SUPER survey, ALMA data analysis is presented in \citetalias{circosta2021}. $^c$ Redshift from \citet{luo2017}. $^d$ Redshift from \citet{marchesi2016}. }
    \label{tab:sample}
\end{table}

By dropping the AGN at $z<1$, the \kashz\ survey comprises $\simeq200$ sources of which only $\sim10\%$ were targeted in CO at $J<5$ and are thus included in this work (see Sect. \ref{kashz:sec:kashz_data_sel}). We present in Sect. \ref{kashz:sec:kashz_data} and Appendix \ref{app:chap:kashz_alma} the reduction and analysis of ALMA data of \kashz\ AGN. Through dedicated SED fitting (see Appendix \ref{app:sec:sedfit}), we measure for \kashz\ ALMA targets stellar masses in the $\log(\Mstar/\Msun)\simeq10.3-11.8$ range, FIR luminosities in the $\log(\Lfir/$\lumcgs$)\simeq44.3-46.2$ range (see Table \ref{kashz:tab:super+kashz}) and bolometric luminosity in the $\log(\Lbol/$\lumcgs$)\simeq44.3-46.6$. 

\subsection{SUPER survey}
\label{kashz:sec:supersurvey}
The SUPER survey\footnote{\url{http://www.super-survey.org/}} total sample includes 48 X-ray selected AGN at $z=2-2.5$ with $\log(L_{\rm X}/\rm erg~s^{-1})\gtrsim42$, without prior knowledge on the presence of AGN-driven outflows (\citetalias{circosta2018}, \citetalias{circosta2021}). 
This SINFONI Large Program allowed to \textit{i)} provide a spatially-resolved, systematic study of the occurrence and properties of ionized AGN-driven winds \citep{kakkad2020,tozzi2024arXiv}, \textit{ii)} investigate the effects of AGN activity on the available molecular gas reservoir using
ALMA Band 3 observations of the CO(3-2) transition (\citetalias{circosta2021}), and \textit{iii)} investigate the impact of ionized outflows on SFR \citep{kakkad2023} in cosmic noon host galaxies. 
Targets were drawn from both deep and large-area X-ray surveys: COSMOS-Legacy, CDFS, XMM-Newton XXL survey \citep[XMM-XXL,][name convention X\_N\_ID]{pierre2016}, Stripe 82 X-ray survey \citep[Stripe82X][name convention S82X\_ID]{lamassa2016,ananna2017}. The sample includes also AGN selected from the WISE/SDSS selected Hyper-luminous quasars sample \citep[WISSH,][]{bischetti2017} based on their X-ray luminosity and redshift, as obtained by the WISSH collaboration from proprietary X-ray observations by \chandra\ and \xmm\ \citep{martocchia2017}. 
Given the similar selection strategy and the common goal, the SUPER survey can be considered the high-resolution version of \kashz\ at $z\simeq2-2.5$. In fact, there are 14 AGN that are shared between the two samples\footnote{Shared targets between SUPER and \kashz: cdfs\_36, cdfs\_419, cdfs\_427, cdfs\_522, cdfs\_614, cid\_1205, cid\_1215, cid\_346, cid\_357, cid\_451, cid\_970, cid\_971, cid\_1057, cid\_1143}. We refer to \citetalias{circosta2021} for what concerns the analysis and results of the ALMA sample of SUPER AGN (27 targets), and to \citetalias{circosta2018} and \citetalias{circosta2021} for what concerns the SED fitting of ALMA targets drawn from the XMM-XXL field. The updates in dedicated SED fitting of SUPER AGN selected from CDFS and COSMOS are presented in Appendix \ref{app:sec:sedfit}. We measure for SUPER ALMA targets stellar masses in the $\log(\Mstar/\Msun)\simeq10.2-11.7$ range, FIR luminosity in the $\log(\Lfir/$\lumcgs$)\simeq44.3-46.4$ range and bolometric luminosity in the  $\log(\Lbol/$\lumcgs$)\simeq44.7-46.9$ (see Table \ref{kashz:tab:super+kashz}).

\begin{figure}[!h]
	    \centering
	    \includegraphics[width=0.5\textwidth]{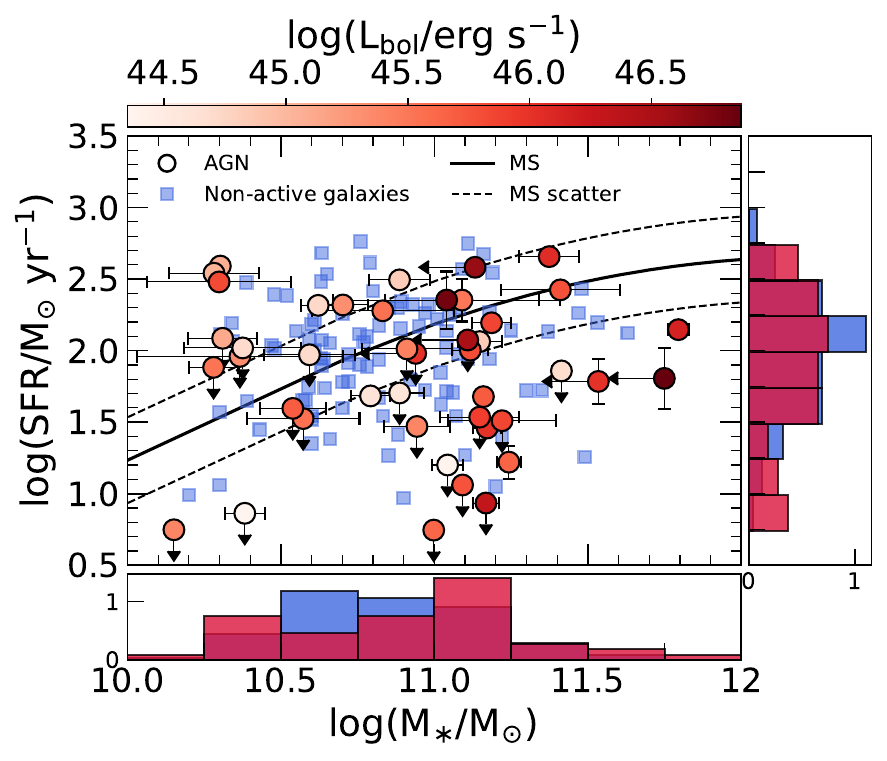} 
	    \caption{Comparison of star formation rate and stellar mass of the AGN host galaxies (red circles) and the non-active galaxies of the control sample (blue squares). The red circles are color-coded based on their AGN bolometric luminosity, as retrieved from SED fitting from this work or from the literature. The solid line marks the MS at $z=2$ from \citet{schreiber2015} and dashed lines show its scatter (equal to 0.3 dex). The distribution of SFR (right) and $\Mstar$ (bottom) for AGN (red) and non-active galaxies (blue) in the two side panels are intended for illustration purposes only, since the upper limits are considered at face value. A robust comparison of the distribution of SFR and \mstar\ applying the hierarchical method described in Sect. \ref{kashz:sec:distrofgas} is presented in Fig. \ref{kashz:fig:sfr_mstar_distros}. }
	    \label{kashz:fig:MS}
    \end{figure} 
\section{Control sample of non-active galaxies}
\label{kashz:sec:control}
We build the comparison sample using literature measurements of non-active galaxies. We select star-forming, non-active galaxies from the Plateau de Bure high-\z\ Blue Sequence Survey catalog \citep[PHIBSS,][]{tacconi2018}, which aims at assessing the gas properties of galaxies across cosmic time, employing a sample of 1444 targets placed at $z=0-4.4$. Each target is complemented with estimates of the molecular gas mass, as traced by CO emission (including CO non-detections), and FIR luminosity, as measured either from SED fitting or from the dust continuum luminosity. We include also the ALMA/NOEMA survey of sub-$mm$ galaxies in the COSMOS, UDS, and extended CDFS (ECDFS) fields by \citet{birkin2021} and the non-active galaxies of ASPECS \citep[and references therein]{boogaard2020_aspecs}, to better sample the higher-end and the lower-end, respectively, of the stellar mass range spanned by our AGN host galaxies. The PHIBSS project includes also the pilot ALMA program of ASPECS that was presented in \citet{decarli2016}; for those galaxies, we use the latest estimates provided by the ASPECS survey, as reported in \citet{boogaard2020_aspecs}.

	We include in the control sample all the non-active galaxies observed in \cotredue\ and \codueuno, as obtained by ALMA and NOEMA, and retrieve the \counozero\ using $r_{21}=0.6$ and $r_{31}=0.5$, i.e., excitation ratios commonly found for star-forming galaxies in the same redshift range of our AGN \citep{daddi2010b,tacconi2013,kakkad2017}. CO fluxes were retrieved from the literature (\citealp{boogaard2020_aspecs,birkin2021}) or provided by the PHIBSS collaboration (L. Tacconi, private communication) since \citet{tacconi2018} only report the final molecular gas masses.
	We also checked the nature of the galaxies in the control sample and excluded those that are flagged as AGN. We do not include such targets in our AGN sample because their stellar masses and SFRs were not corrected accounting for the AGN contribution and could thus be overestimated, and because they are identified as AGN with methods different than ours. 

    All the galaxies in the comparison sample have available estimates of stellar mass and FIR luminosity. However, the SFR of a small subsample of PHIBSS galaxies was estimated from H$\alpha$ fluxes. Based on the reasonable agreement between the SFRs derived from H$\alpha$ and from the FIR for MS galaxies at cosmic noon \citep[within $\simeq$0.4~dex;][]{rodighiero2014,puglisi2016,shivaei2016}, we convert such SFRs to FIR luminosity applying the \citet{kennicutt1998} relation corrected for a \citet{chabrier2003} IMF (i.e., reduced by 0.23 dex). 
	
    We build the control sample of non-active galaxies as follows. 
    First, we divide our AGN sample into two redshift bins ($z=1-1.8$ and $z=1.8-2.55$), which we use as redshift constraints for the match in stellar mass and SFR to take into account the evolution of the MS with redshift \citep[e.g.,][]{genzel2015,tacconi2018}. For each redshift bin, we only consider non-active galaxies within the same stellar mass range covered by our AGN host galaxies within the uncertainties ($\log\Mstar/\Msun\simeq10-12$), and we further divide the samples in bins of stellar mass of 0.5~dex width. For each stellar mass bin, we then select all those non-active galaxies with a SFR consistent with (within $+0.2$dex to consider the uncertainties) or lower than that of AGN hosts in a given mass (and redshift) bin, so to also account for the SFR upper limits in the AGN sample. We show in Fig. \ref{kashz:fig:MS} the comparison of our AGN (in red) and control sample (in blue) in terms of SFR vs. $\Mstar$. Figure \ref{kashz:fig:MS_zbins} in Appendix \ref{app:mstar_sfr_distros} shows the same comparison in the low-\z\ and high-\z\ bins. 
    Figures \ref{kashz:fig:MS} and \ref{kashz:fig:MS_zbins} also show the distribution of $\Mstar$ and SFR both for AGN hosts and non-active galaxies for illustration purposes, since the upper limits are considered, for these plots only, at face value. In fact, a robust comparison of the distribution of SFR and \mstar{} applying the hierarchical method described in Sect. \ref{kashz:sec:distrofgas} and the Kolmogorov-Smirnov (KS) test is presented in Appendix \ref{app:mstar_sfr_distros}. We find that AGN hosts and non-active galaxies selected as presented in this Section are consistent with being drawn from the same distribution for what concerns the stellar mass in all redshift bins (low-\z, high-\z, total) and for what concerns the SFR in the low-\z\ bin and total redshift range. However, they are significantly different for what concerns SFR distribution of the high-\z\ bin (p-value$\simeq$1\%) . In fact, the galaxy database used to build the control sample is missing non-active galaxies with CO observations that fall below the MS at $z>1.8$ ($\Delta\rm MS<-0.5 dex$; see Fig. \ref{kashz:fig:MS_zbins}, bottom) due to the rareness of low-SFR star-forming galaxies not hosting AGN and the time-consuming observations required to target their molecular phase. Since not even dust-continuum observations cover non-active galaxies in such a region of the MS \citep[e.g.,][]{tacconi2020}, we run our comparative analysis twice: we compare the full AGN and non-active galaxy samples built as described in this section and in Sect. \ref{kashz:sec:samp_sel}, and then we compare them again excluding those AGN lacking a match in the control sample in the high-\z\ bin (i.e., $z>1.8$, $\log(\Mstar/\Msun)>11$, $\rm \log (SFR/\Msun yr^{-1})<2$; see Fig. \ref{kashz:fig:MS_zbins}) and those galaxies in the low-\z\ bin that do not have a match in the AGN sample (i.e., $z<1.8$, $\log(\Mstar/\Msun)$<$10.5$, $\rm \log (SFR/\Msun yr^{-1})<1.5$; see Fig. \ref{kashz:fig:MS_zbins}) as a sanity check of our results.

\section{ALMA observations of \kashz\ targets}
\label{kashz:sec:kashz_data}
\subsection{Data selection}
\label{kashz:sec:kashz_data_sel}
We mined the ALMA archive for observations of \kashz\ AGN at $1<z<2.6$ (202 sources). We narrowed our query to ALMA Band 3, 4 or 5, since these are the ones that cover low-J CO transitions ($J_{\rm up}=2,3,4$), best suitable to derive total molecular gas masses. 
ALMA observations in Band 3 and 4 are available for $\simeq$10\% of the \kashz\ sample (to which we will refer to also as the ALMA \kashz\ sample), for a total of 22 ALMA fields from 13 ALMA projects. No suitable ALMA data is available in Band 5. We report in Table \ref{app:tab:almaobs} all observations analyzed in this work. We refer to the results and analysis in \citetalias{circosta2021} for those \kashz\ targets shared with SUPER that are already presented in \citetalias{circosta2021} (marked in Table \ref{kashz:tab:almatargets}). Some \kashz\ AGN were observed multiple times in more than one CO transition (cdfs\_427, cdfs\_587, cdfs\_614) or in the same transition at different angular resolutions (cdfs\_794, cdfs\_587). For these targets, we consider the ALMA observation targeting the CO transition at the lowest-J and/or with the largest beam size, to maximize the sensitivity per beam and simultaneously minimize the resolution in order to best recover  the total CO flux. The only exception is cdfs\_794: since this is a merger system observed twice in \cotredue\ in low- and high-angular-resolution set up, we select the observation at the highest resolution \citep{calistrorivera2018} to ensure no contamination from the nearby companion. Moreover, we exclude cdfs\_718, the brightest target in the ALMA Spectroscopic Survey in the \textit{Hubble} Ultra Deep Field \citep[ASPECS ID 1mm.1;][]{decarli2019_aspecs,gonzalezlopez2019_aspecs,aravena2019_aspecs,boogaard2020_aspecs}, from the \kashz\ ALMA sample, due to the presence of a dust lane as seen in HST data \citep{boogaard2019} that contaminates the FIR emission, thus preventing from a good determination of its FIR parameters and decoupling of AGN and host components in the SED fitting.

\begin{figure*}[!h]
	\centering
	\includegraphics[width=0.35\linewidth]{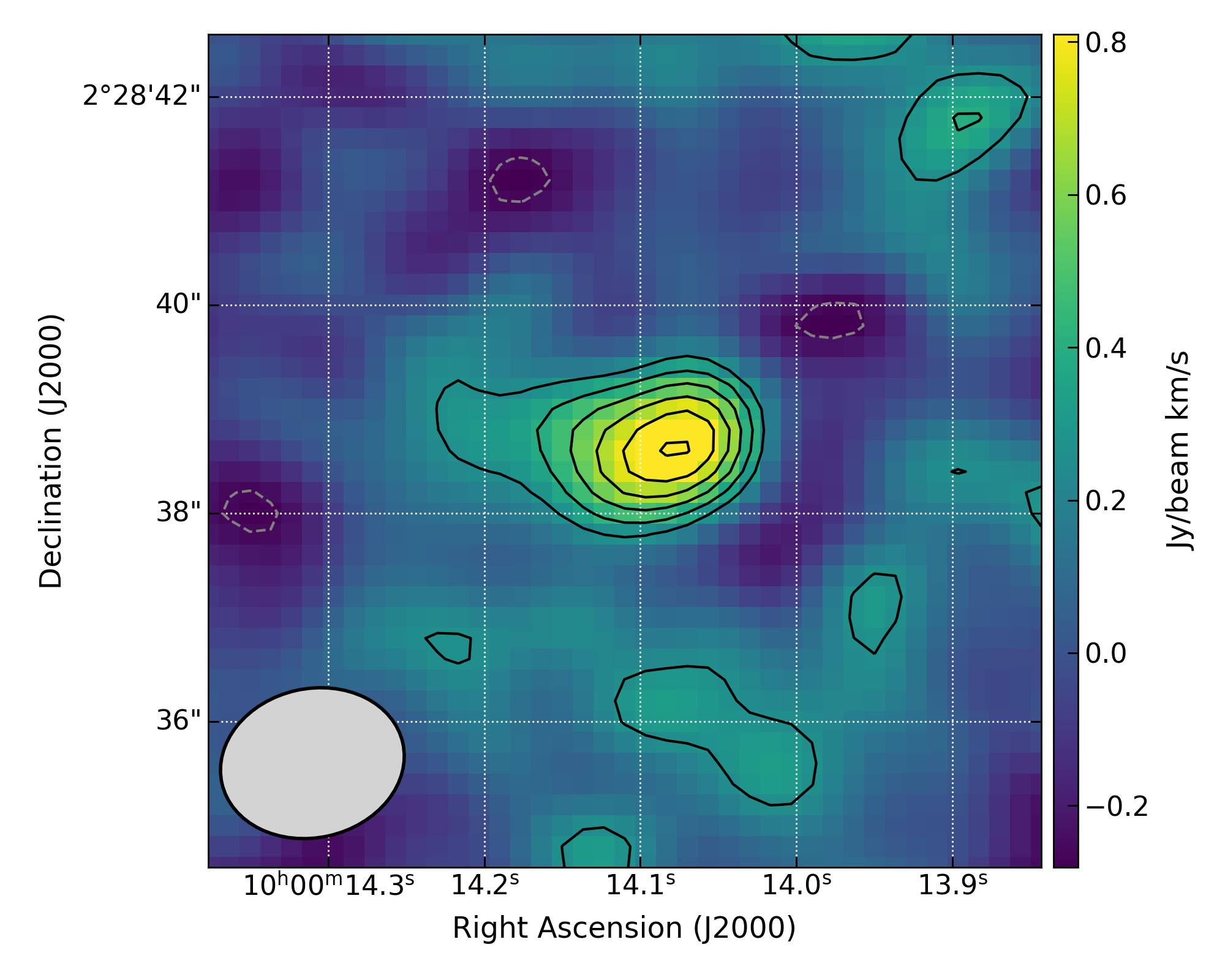}
	\hfil
	\includegraphics[width=0.5\linewidth]{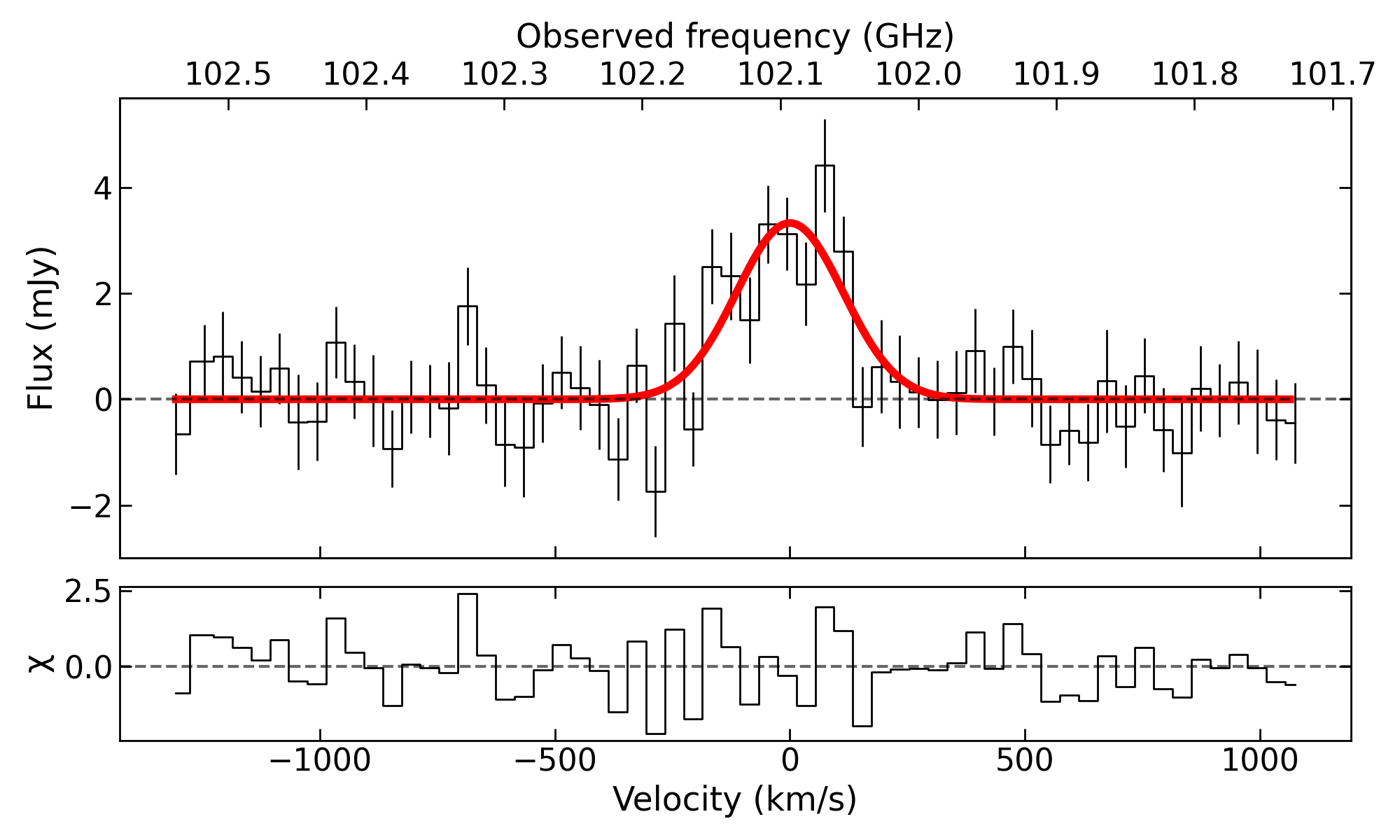}
	\caption{CO line velocity-integrated emission map (\textit{Left}) and spectrum (\textit{Right}) of cid\_108. Solid contour levels in the left panels start at 2$\sigma$ and increase linearly. Dashed contour indicate the [-3,-2]$\times\sigma$ level. The beam of each observation is shown by the grey ellipse at the bottom-left corner of each map. The grey shaded area in the spectrum of cid\_864 marks the $\pm\rm rms_{cube}$ range. CO maps and spectra of the rest of the AGN sample are available on Zenodo at the following link: \url{https://doi.org/10.5281/zenodo.13149280}. }
	\label{app:fig:ALMA1}
\end{figure*}

Ten observations target the \codueuno\ line, fifteen the \cotredue\ and three the \coquattrotre. 
We also present here the \cotredue\ NOEMA data of J1333+1649 (project code S21CG, PI Mainieri), a SUPER AGN (drawn from the WISSH sample) that was not previously included in \citetalias{circosta2021} because subsequently observed. Almost all the other AGN in \kashz\ and SUPER are equatorial or in the southern hemisphere to be observable by VLT. 
As a sanity check, we also queried the IRAM archive at the Centre de Donn\'ees astronomiques de Strasbourg for NOEMA observations of the rest of sample, however there are no public NOEMA observations covering the CO emission lines of our targets. 

\subsection{CO emission of \kashz\ AGN}
\label{kashz:sec:kashz_data_red}
We present here and in Appendix \ref{app:chap:kashz_alma} the data reduction and analysis of the ALMA archival observations analyzed in this work. We refer to the results of \citetalias{circosta2021} for the \kashz\ ALMA targets that are shared with SUPER and that were previously analyzed. We list all the \kashz\ ALMA targets, including those shared with SUPER, in Table \ref{kashz:tab:almatargets}, along with information on the parent X-ray deep field, AGN type, redshift, X-ray intrinsic photon index, X-ray absorption column density and absorption-corrected X-ray luminosity in the 2--10 keV rest-frame band with a flag indicating how we computed it. 

We retrieved the calibrated measurement sets using the dedicated service provided by the European ALMA Regional Center. We produced continuum maps and (continuum-subtracted) spectral cubes using the task {\tt tclean} in CASA 6.4, ran in ``mfs'' or ``velocity'' mode, respectively. The continuum was identified using all the spectral windows available in the ALMA observation, masking the channels with line emission, and then subtracted with the {\tt uvcontsub} CASA task. 
We imaged all fields using a natural weighting scheme of the visibilities and pixel size of $\simeq1/5$ of the beam FWHM, with the aim of maximizing the sensitivity of our data. 
We produced the final cubes by setting a channel width that allows to sample the line FWHM with at least 7 spectral resolution elements, chosen as the best trade off to sample the line profile while maintaining a good signal-to-noise ratio. CO fluxes were estimated by fitting the 1D spectrum, and validated by the comparison with the spatially-integrated line flux and two-dimensional fits of the line velocity-integrated maps (0th order moment). Non-detections were defined as sources with signal-to-noise ratio S/N$<$3 in the velocity-integrated maps. 
The steps and details of the ALMA data analysis are presented in Appendix \ref{app:chap:kashz_alma}, and the results in Table \ref{kashz:tab:almaresults}. 
We show the velocity-integrated maps and spectra of cid\_108 as example in Fig. \ref{app:fig:ALMA1}. Those of the rest of the sample are available on Zenodo at the following link: \url{https://doi.org/10.5281/zenodo.13149280}. 

Since our sample is heterogeneous in terms of targeted CO transition (\codueuno, \cotredue, \coquattrotre), we convert the measured CO fluxes to the \counozero\ transition assuming the line ratios measured for a sample of IR-selected AGN up to $z\simeq4$ ($r_{41}=0.37\pm0.11$, $r_{31}=0.59\pm0.18$, $r_{21}=0.68\pm0.17$, \citealp{kirkpatrick2019}), and derive the \counozero\ line luminosity as: 
\begin{equation}
	L'_{\rm CO}[{\rm K~km~s^{-1}~pc^2}]=3.25\times10^7S\Delta v\frac{D^2_{\rm L}}{(1+z)^3\nu_{\rm obs}^2}
	\label{kashz:eq:lco}
\end{equation}
where $S\Delta v$ is the line velocity-integrated flux in $\rm Jy\ km\ s^{-1}$, $D_{\rm L}$ is the luminosity distance in Mpc and $\nu_{\rm obs}$ is the observed centroid frequency of the line in GHz \citep[e.g.,][]{carilli_2013}. The uncertainty on \lpco is computed from error propagation of the uncertainties on the measured fluxes and the errors on the CO ladder. The latter are the larger source of uncertainty and thus dominate the error on \lpco. 
We note that the CO ladder used for AGN is consistent to that used for the non-active galaxies of the control sample, thus our analysis is set in a conservative framework.

Table \ref{kashz:tab:almaresults} summarizes all the values of interest derived from the ALMA data analysis of the \kashz\ ALMA targets presented in this work: CO properties and dust-continuum measurements, the chosen weighting scheme, the beam of the ALMA cubes and continuum maps, the channel width of the final cubes, the mean rms of the final cubes and velocity-integrated maps, and CO(1-0) luminosity. 
We apply the same CO ladder to convert the \cotredue\ luminosity of SUPER ALMA AGN measured by \citetalias{circosta2021} to \lpco, which we report in Table \ref{kashz:tab:super+kashz}.

	\section{Molecular gas properties of AGN at cosmic noon}
	\label{kashz:sec:results}
    We present in this Section our comparative, quantitative analysis of the cold molecular gas content of AGN host galaxies and non-active galaxies at $z=1-2.6$. The AGN sample selection is described in Sect. \ref{kashz:sec:samp_sel} and the control sample of non-active galaxies is built as described in Sect. \ref{kashz:sec:control} (see Fig. \ref{kashz:fig:MS}). 
    We address the effects of AGN feedback on the properties of host galaxies at $z=1-2.6$ by assessing whether they differ from the respective control sample in terms of: \textit{i)} CO vs. FIR luminosity, \textit{ii)} distribution of molecular gas fraction, for which we consider the observational proxy $f_{\rm gas}=$\lpco/\mstar, \textit{iii)} CO luminosity vs. stellar mass, and \textit{iv)} molecular gas fraction (i.e., \lpco/\mstar) vs. stellar mass. 
    Moreover, we also run the analysis removing the AGN at $z>1.8$ that fall below the MS and do not have a match in the control sample, and galaxies in the low redshift bin without a counterpart in the AGN sample (i.e., below the MS; see Sect. \ref{kashz:sec:control} and \ref{app:mstar_sfr_distros}). We anticipate that the results obtained excluding such AGN and non-active galaxies are consistent with those of the full samples. 

    The total AGN sample totals 46 AGN and presents a CO detection rate of $\simeq$50\%. Roughly half of the AGN have a constrained $\Lfir$ and about $\simeq$44\% have an upper limit for both quantities. 
    All but five AGN have constrained stellar mass, of which three (cid\_178, cid\_1605, cid\_467) are undetected in CO and thus do not allow to derive a limit on their gas fraction. 
    We summarize the properties of \kashz\ and SUPER ALMA AGN in Table \ref{kashz:tab:super+kashz}. The control sample totals 98 galaxies, with a CO detection rate of $\simeq$90\%. 
	
	We performed our analysis within a Bayesian framework as presented in \citetalias{circosta2021}, which allows us to take into account the upper (and lower) limits on both dependent and independent variables. There is only one main difference in this work with respect to \citetalias{circosta2021}: \kashz\ and literature AGN are heterogeneous in terms of targeted CO transition, while SUPER AGN were all observed in CO(3-2). We relied on the \counozero\ luminosity by assuming a CO SLED that we uniformly applied to the CO measurements of \kashz\ and SUPER AGN (see Sect. \ref{kashz:sec:kashz_data_red}) and a second one for the star-forming galaxies of the control sample (see Sect. \ref{kashz:sec:control}). Our analysis remains free of the additional uncertainties inherent to the $\alphaco$ conversion factor and the conversion of FIR luminosity into SFR. Figures of this Section also show the molecular mass axis, derived assuming $\alpha_{\rm CO}=3.6\Msun/(\rm K \kms pc^2)$, yet this serves for illustration purposes only, since we only consider \lpco\ in our quantitative analysis.

	\begin{figure}[!h]
		\centering
		\includegraphics[width=0.5\textwidth]{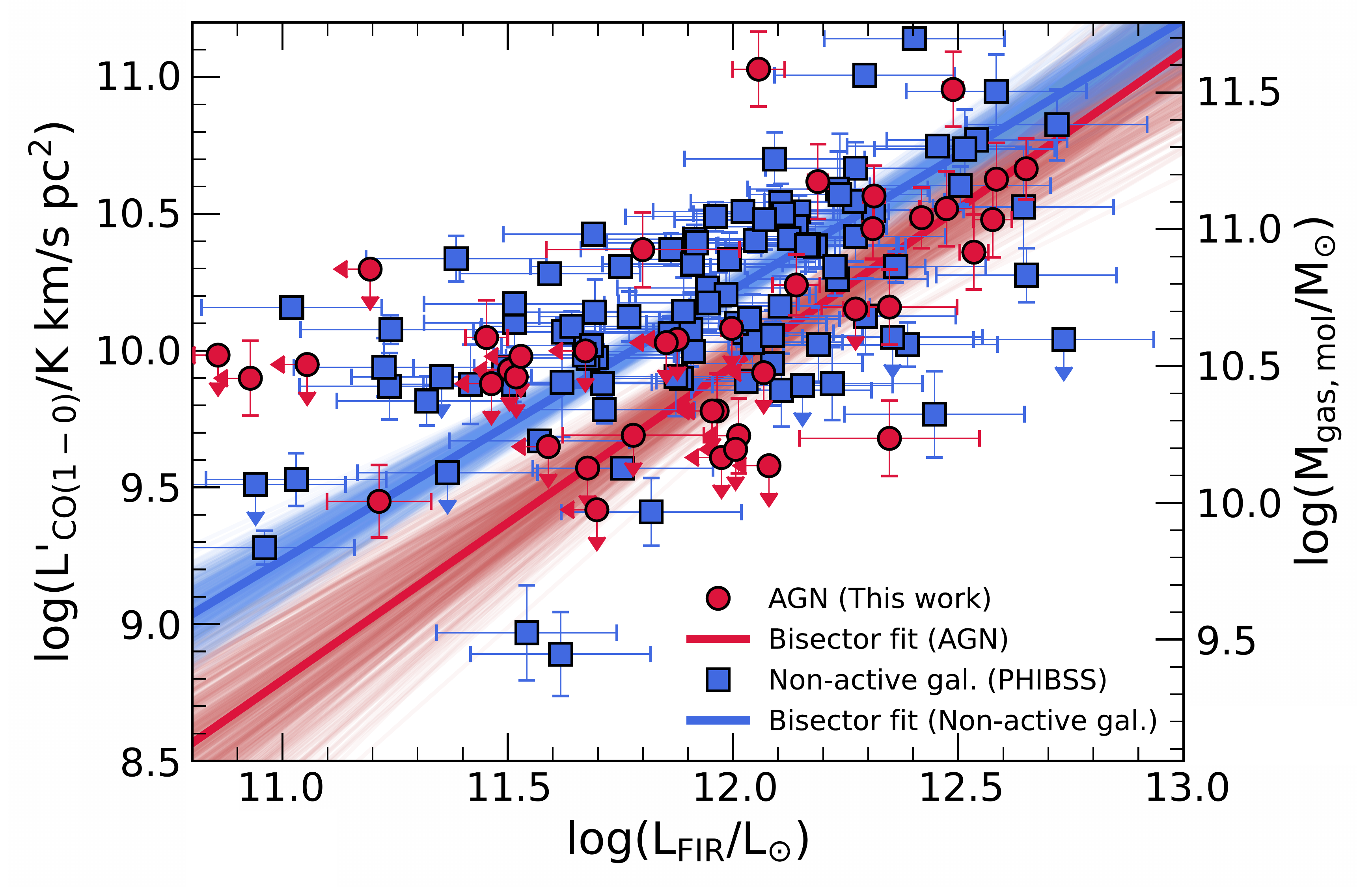}
		\caption{ CO luminosity vs. FIR luminosity bisector fits of AGN host galaxies (red circles) and non-active galaxies of their control sample (blue squares). Thick lines mark the bisector fits obtained by adopting a Bayesian framework (see main text). The dispersion of the fits is given by plotting 500 realizations of the bisector fits. 
			The vertical axis on the right is derived by assuming $\alpha_{\rm CO}=3.6\Msun/(\rm K \kms pc^2)$ and serves for illustration purposes only, since we only consider \lpco\ in our quantitative analysis (see Sect. \ref{kashz:sec:cofirlum}). }
		\label{kashz:fig:lco-fir}
	\end{figure}
	
	\subsection{CO vs. FIR luminosity}
	\label{kashz:sec:cofirlum}
	In this Section, we aim to quantify whether AGN host galaxies and non-active galaxies at cosmic noon follow a different distribution in the CO and FIR luminosity parameter space, i.e. an observational proxy of the integrated Schmidt-Kennicutt law. We apply the Bayesian method to produce bisector fits as developed by \citetalias{circosta2021}, which we briefly explain here. We fit a linear model to the data applying the ordinary least-squares (OLS) bisector fit method \citep{isobe1990}, that is, we take into account the uncertainties on \lpco\ and $\Lfir$ separately, so to consider the upper limits on both quantities, and then we derive the bisector of the two lines in a Bayesian framework, assuming uniform priors for free parameters. When building the likelihood function of constrained values, we assumed their uncertainties as Gaussian distributed. Upper limits were included in the likelihood function as error functions \citep[e.g., see][]{lamperti2019}, built integrating the Gaussian likelihood from minus infinity to the value of the upper limit, that is, 3$\sigma$ for both CO and FIR luminosity. We then sample the yielded likelihood function through the {\tt emcee} library \citep{foreman-mackey2013_emcee}, a python implementation of the invariant MCMC (Monte Carlo Markov Chain) ensemble sampler of \citet{goodman2010_mcmcm}. We derive the marginalized posterior distribution by sampling the posterior distribution in the parameter space, using the best fit obtained through the python module {\sf scipy.optimize} \citep{sciPy2020-NMeth} as initial guess. We include an additional intrinsic scatter to the relation as a third free parameter to account for the possibility of underestimated uncertainties, given the wide range spanned by our sources ($\sim$2.5~dex both in \lpco\ and in $\Lfir$). Best-fit parameters were then derived as the median of the sampled marginalized posterior distribution of the OLS best-fit parameters, assigning as uncertainties the 16th and 84th percentiles. The final values of slope ($a$) and intercept ($b$) for the bisector fit were then derived from the OLS best fit values following \citet{isobe1990}. Fits were performed by normalizing both x and y variables to the mean of the range covered by the data points so to reduce the correlation between the best fit parameters. 
	\renewcommand{\arraystretch}{1.15}
	\begin{table*}[!h]
		\centering
            \caption{Best fit parameters for AGN and non-active galaxies. }
		\begin{tabular}{lcccccccc}
                \hhline{~====}
			\multicolumn{1}{c}{}             & \multicolumn{2}{c}{AGN hosts} & \multicolumn{2}{c}{Non-active galaxies}                                                      \\
			\hline\hline
			\textit{Bisector fits}                              & $m$                           & $b$                          & $m$                 & $b$                        \\
			\hline
			$\log$(\lpco$/10^{10}~{\rm K~km~s^{-1}pc^2})=m\log(\Lfir/10^{12} \Lsun)+b$               
                                                                    & $ 1.16_{-0.12}^{+0.15} $           & $-0.06_{-1.12}^{+0.06} $        & $0.99 ^{+0.07}_{-0.06}$ & $0.22^{+0.02}_{-0.02}$    \\
			$\log$(\lpco$/10^{10}~{\rm K~km~s^{-1}pc^2})=m\log (M_{\rm \star}/10^{11}\Msun)+b$       
                                                                    & $0.97_{-0.54}^{+0.45} $       & $-0.11_{-0.20}^{+0.09} $       & $1.24 ^{+0.14}_{-0.13}$ & $0.35^{+0.04}_{-0.04}$     \\
			$\log [($\lpco$/\Mstar)/(10^{10}~{\rm K~km~s^{-1}pc^2}/10^{11}\Msun)]=m\log (M_{\rm \star}/10^{11}\Msun)+b$         
                                                                    & $-1.02_{-0.27}^{+0.29} $          & $-0.61_{-0.11}^{+0.10} $         & $-1.1^{+0.1}_{-0.1}$ & $-0.37^{+0.04}_{-0.04}$   \\
			\hline\hline
			\textit{Distribution}                     & \multicolumn{2}{c}{$\mu$}                         & \multicolumn{2}{c}{$\mu$}                       \\
			\hline
			$\log ($\lpco$/\Mstar)$                    &  \multicolumn{2}{c}{$-1.09^{+0.21}_{-0.19}$}      &   \multicolumn{2}{c}{$-0.62^{+0.07}_{-0.06}$}  \\

			\hline
		\end{tabular}
		\label{kashz:tab:bisecfits}
	\end{table*}
	\renewcommand{\arraystretch}{1}
	
\begin{figure}[!h]
		\centering
		\includegraphics[width=0.45\textwidth]{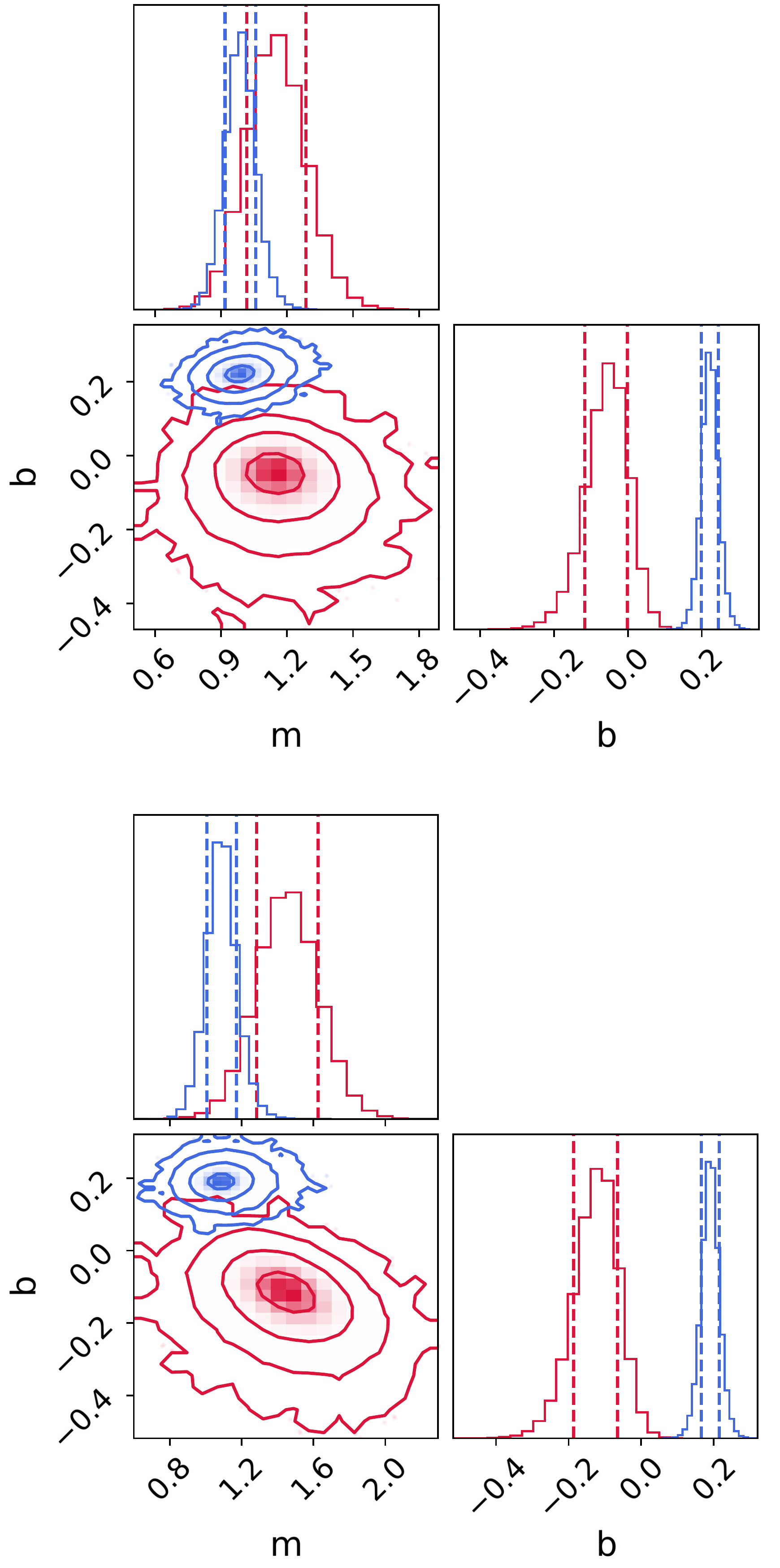}
		\caption{Corner plots of best fit parameters for the $\log$(\lpco$/10^{10}~{\rm K~km~s^{-1}pc^2})=m\log(\Lfir/10^{12} \Lsun)+b$ relation (see Sect. \ref{kashz:sec:cofirlum}) for the full sample (top) and when removing AGN without a match in the control sample (bottom). Results of AGN host galaxies are shown in red, those of non-active galaxies in blue. Contours show the 1 to 4$\sigma$ levels, inner to outer, and increase linearly. }
		\label{kashz:fig:cornercofir}
	\end{figure}

	Figure \ref{kashz:fig:lco-fir} shows the results in the integrated Schmidt-Kennicutt plane. 
	We report in Table \ref{kashz:tab:bisecfits} the best fit parameters of the relation $\log$(\lpco$/10^{10}~{\rm K~km~s^{-1}pc^2})=m\log(\Lfir/10^{12} \Lsun)+b$ for both AGN and control samples. 
	Our main aim here, however, resides in probing if AGN and control samples are indeed different and in quantifying the shift of the two distributions. We thus compare the corner plots of the best-fit parameters of AGN and non-active galaxies in Fig. \ref{kashz:fig:cornercofir}. We find that the relations for AGN and non-active galaxies are different at the 3$\sigma$ level both for the full AGN and non-active galaxy samples and when excluding galaxies without a good match (see Fig. \ref{kashz:fig:cornercofir}), that is, excluding AGN at $z=1.8-2.55$ and non-active galaxies at $z=1-1.8$ that are below the MS (see Sect. \ref{kashz:sec:control}). 

	\subsection{Molecular gas fraction distribution}
	\label{kashz:sec:distrofgas}
    We compared the molecular gas fraction distribution of AGN and control samples, measured as the observational proxy \lpco$/\Mstar$, to further test the difference between AGN hosts and non-active galaxies. Similarly to \citetalias{circosta2021}, we do not address the distribution of the ratio of \lpco\ and $\Lfir$ (observational proxy for the SF depletion time), since around half of \kashz\ and SUPER targets present an upper limit in at least one of the two quantities, and a quarter of them show an upper limit for both. 
	Our aim is to derive the total gas fraction distribution and mean value ($\mu_{f_{\rm gas}}$) for the AGN and non-active galaxy samples to quantify how they relate to one another. Once more, we carry out our analysis in a Bayesian framework, which allows us to obtain posteriors for the mean of the distribution, by folding in the information contained on upper limits. We exclude three \kashz\ and SUPER AGN (cid\_178, cid\_1605, cid\_467) because they have an upper limit for both \lpco and $\Mstar$.

    \begin{figure}[!h]
		\centering
		\includegraphics[width=0.45\textwidth]{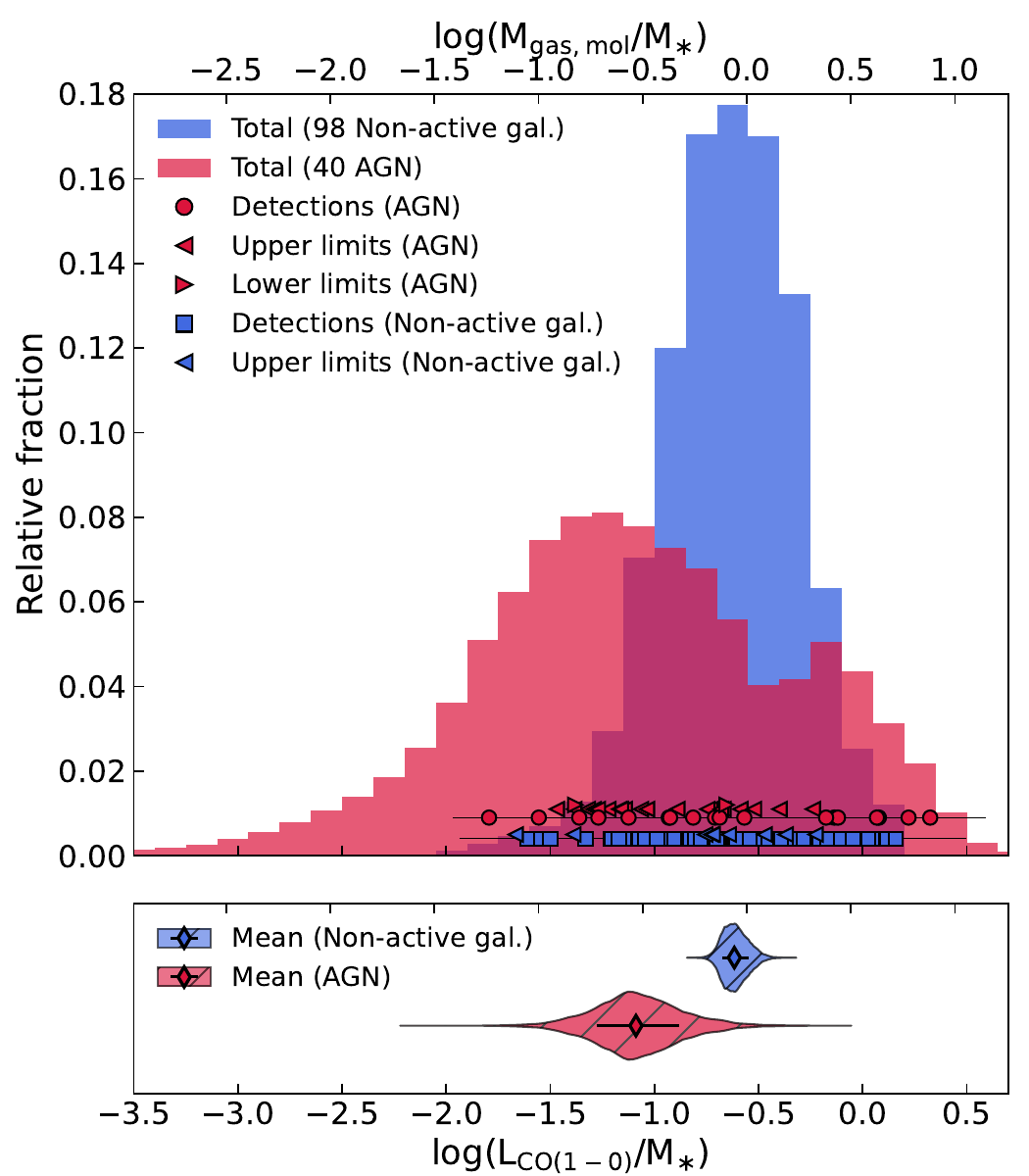}
		\caption{Distribution of \lpco$/\Mstar$ (observational proxy of the gas fraction $f_{\rm gas}$) for AGN host galaxies (red) and non-active galaxies of the control sample (blue). Upper panel: Filled histograms show the total distributions, obtained by joining the sampled posterior distribution of each target for both samples. Blue squares indicate the measure of each non-active galaxy (98 targets), red circles that of each AGN (40 targets). Blue and red triangles indicate upper limits if leftward and lower limits if rightward for the gas fraction of non-active galaxies and AGN, respectively. Lower panel: Violin plots show the sampled posterior distribution of the mean of the hierarchical Gaussian prior used in our Bayesian analysis. Diamonds show the measured mean value $\mu_{f_{\rm gas}}$ of AGN (red) and non-active galaxies (blue). The top axis is derived by assuming $\alpha_{\rm CO}=3.6\Msun/(\rm K \kms pc^2)$ and serves for illustration purposes only, since we only consider \lpco\ in our quantitative analysis (see Sect. \ref{kashz:sec:distrofgas}). }
		\label{kashz:fig:distrofgas}
	\end{figure}
 
	Following \citetalias{circosta2021} (and \citealp{mullaney2015,scholtz2018}), we assumed that the distribution of molecular gas fraction of AGN and non-active galaxies follows a Gaussian distribution, supported by the tests performed by \citetalias{circosta2021} to demonstrate the validity of such an assumption both on the xCOLD-GASS reference survey \citep{saintonge2017_xCOLDgass} and in PHIBSS galaxies at cosmic noon. We thus adopt their Bayesian hierarchical method: we assumed that the prior distribution, common to both samples, is Gaussian and defined by two hyper-parameters, that is, the mean $\mu_{f_{\rm gas}}$ and the standard deviation $\sigma_{f_{\rm gas}}$, for both of which we set uniform priors. Following the approach described in Sect. \ref{kashz:sec:cofirlum}, we build the likelihood assuming Gaussian-distributed uncertainties for constrained gas fractions and error functions for the upper/lower limits. The total posterior distributions of the gas fractions of AGN hosts and non-active galaxies (red and blue histograms, respectively; see Fig. \ref{kashz:fig:distrofgas}) were built by joining the sampled posterior distribution of each respective target, and, as such, cover the full range of $f_{\rm gas}$ spanned by each target with the advantage of carrying the information on how upper/lower limits are weighted by the prior Gaussian distribution. 
	The same Figure also shows the sampled posterior distribution of the mean of the hierarchical Gaussian prior that we obtained from our Bayesian analysis (red and blue violin plots, bottom panels). 
	As for the other quantities, we consider the 50th percentile as the best value and the 16th and 84th percentiles as uncertainties. We find that the mean $\log($\lpco$/\Mstar$) of both our AGN samples is lower than that of non-active galaxies (see diamond-shaped points in the bottom panel of Fig. \ref{kashz:fig:distrofgas}), as also found by \citetalias{circosta2021} for the sole SUPER sample. 
 
    The mean gas fractions of AGN sample and control sample are different at the 2$\sigma$ level, i.e., comparable to the 2.2$\sigma$ result of \citetalias{circosta2021} when considering only SUPER AGN. 
    However, the total $f_{\rm gas}$ distributions of AGN and non-active galaxies are significantly different according to the KS test (p-value < $10^{-7}$), and evidently skewed at low molecular gas fractions (see Fig. \ref{kashz:fig:distrofgas}, upper panel). 
 
    \subsection{Molecular gas mass and gas fraction vs. stellar mass}
    \label{kashz:sec:fgas-mstar}
    We compared the molecular gas mass and gas fraction of AGN and non-active galaxies as a function of stellar mass through bisector fits as described in Sect. \ref{kashz:sec:cofirlum}. Best fit parameters are summarized in Table \ref{kashz:tab:bisecfits}. As in the previous Section, we discard three \kashz\ and SUPER AGN (cid\_178, cid\_1605, cid\_467) for the $\log($\lpco$/\Mstar)$ vs. $\log\Mstar$ case since they have upper limits in both stellar mass and CO emission.

\begin{figure}[!h]
		\centering
		\includegraphics[width=0.5\textwidth]{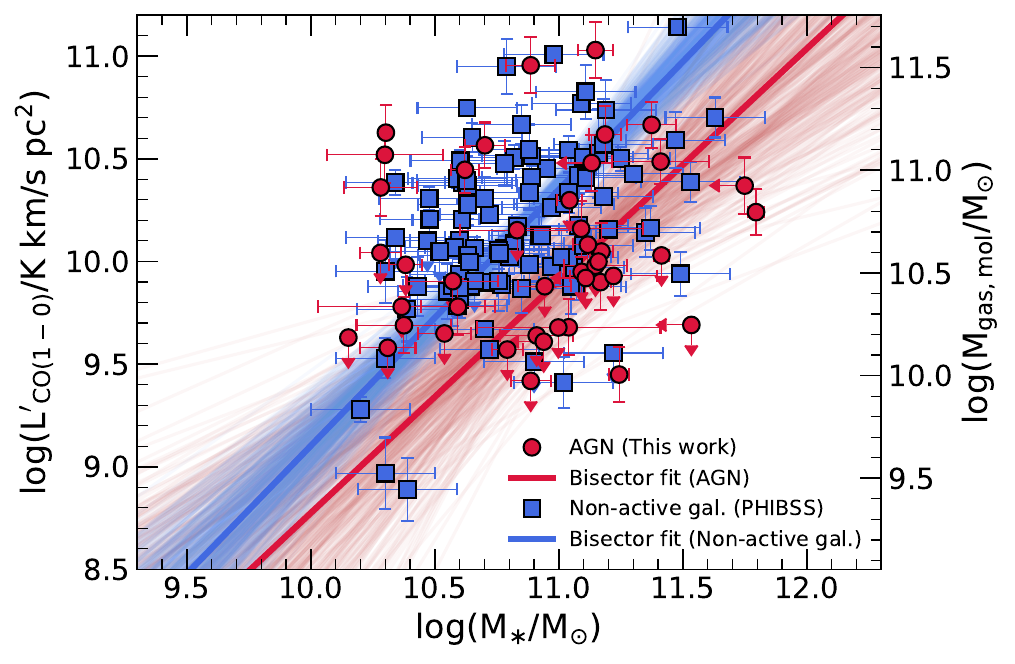}
		\caption{CO luminosity vs. stellar mass bisector fits of AGN host galaxies (red circles) and non-active galaxies of the control sample (blue squares). Thick lines mark the bisector fits obtained by adopting a Bayesian framework (see main text). The dispersion of the fits is given by plotting 500 realizations of the bisector fits. The vertical axis on the right is derived by assuming $\alpha_{\rm CO}=3.6\Msun/(\rm K \kms pc^2)$ and serves for illustration purposes only, since we only consider \lpco\ in our quantitative analysis (see Sect. \ref{kashz:sec:fgas-mstar}). }
		\label{kashz:fig:lco-mstar}
	\end{figure}

\begin{figure}[!h]
		\centering
		\includegraphics[width=0.5\textwidth]{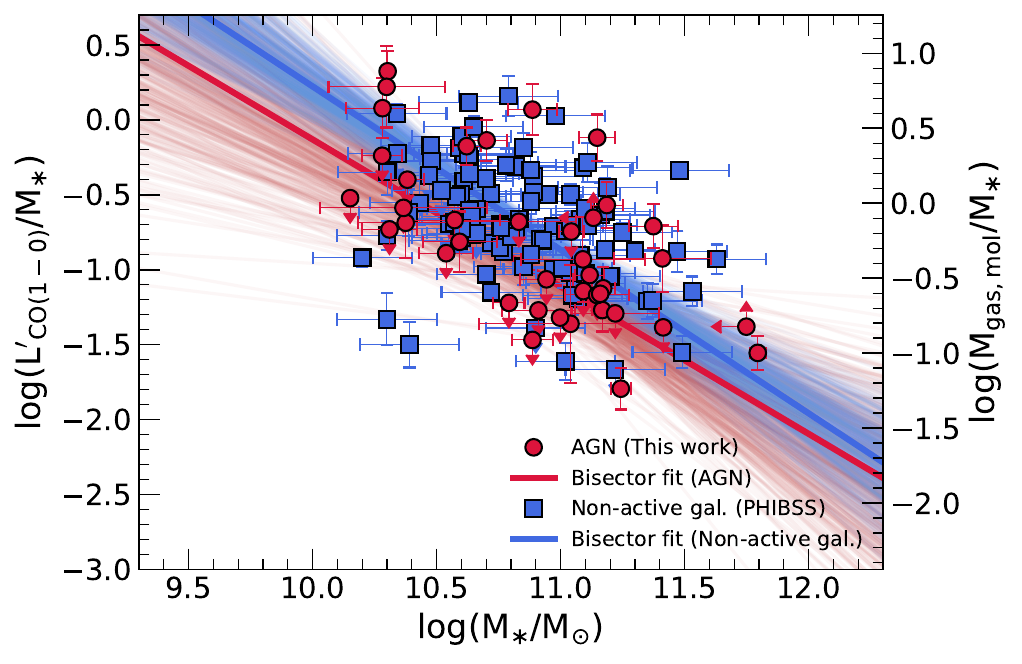}
		\caption{Gas fraction vs. stellar mass bisector fits of AGN host galaxies (red circles) and non-active galaxies of the control sample (blue squares). Thick lines mark the bisector fits obtained by adopting a Bayesian framework (see main text). The dispersion of the fits is given by plotting 500 realizations of the bisector fits. The vertical axis on the right is derived by assuming $\alpha_{\rm CO}=3.6\Msun/(\rm K \kms pc^2)$ and serves for illustration purposes only, since we only consider \lpco\ in our quantitative analysis (see Sect. \ref{kashz:sec:fgas-mstar}). }
		\label{kashz:fig:fgas-mstar}
	\end{figure}

    The bisector best fits of AGN hosts and non-active galaxies both in the log\lpco{} vs. log\mstar{} plane (see Fig. \ref{kashz:fig:lco-mstar}) and in the $\log($\lpco$/\Mstar)$ vs. $\log\Mstar$  (i.e., $f_{\rm gas}$ vs. $\log\Mstar$) plane (see Fig. \ref{kashz:fig:fgas-mstar}) are different at the $\lesssim2\sigma$ level. 
    This result holds in both parameter spaces also when excluding non-active galaxies and AGN hosts without a proper match in the other sample (see Sect. \ref{kashz:sec:control}), and also when the samples include only targets on the MS ($|\Delta\rm MS|\simeq0.3$dex; 14 AGN, of which 8 have upper limits on \lpco, and 50 galaxies). 
    Focusing on the \textit{y vs. x} Bayesian fits (i.e., considering uncertainties and upper limits on \lpco), the differences with non-active galaxies increase slightly ($\simeq2\sigma$), both for the full and the reduced samples (better match; only MS targets). 
    
    Thus, we find a $\lesssim$2$\sigma$ difference in molecular content and gas fraction between AGN and host galaxies for fixed stellar mass. 
    
    \subsection{Summary}    
    In this Section, we compared the properties of AGN hosts and non-active galaxies in the following parameter spaces, finding the listed differences: 
    \begin{itemize}
        \item $L'_{\rm CO(1-0)}$ vs. $L_{\rm FIR}$, where AGN and non-active galaxies are different at the 3$\sigma$ level (see Sect. \ref{kashz:sec:cofirlum} and Figs. \ref{kashz:fig:lco-fir}-\ref{kashz:fig:cornercofir});
        \item $\log($\lpco$/\Mstar)$ distribution (proxy of molecular gas fraction), where AGN and non-active galaxies show significantly different total distributions according to the KS test (p-value $<10^{-7}$), mean values consistent at the 2$\sigma$ level and consistent posterior distributions of the mean values (see violin plots in Fig. \ref{kashz:fig:distrofgas}), indicating that the significant difference in the total distributions is driven by the skewness to lower values of the AGN distribution (see Sect. \ref{kashz:sec:distrofgas} and upper panel of Fig. \ref{kashz:fig:distrofgas});
        \item $L'_{\rm CO(1-0)}$ vs. $\Mstar$ 
        and $\log($\lpco$/\Mstar)$ vs. $\Mstar$, in both of which AGN and non-active galaxies are different at the $\lesssim$2$\sigma$ (see Sect. \ref{kashz:sec:fgas-mstar} and Figs. \ref{kashz:fig:lco-mstar}-\ref{kashz:fig:fgas-mstar}).
    \end{itemize}
    Results obtained by excluding those AGN without a match in the control sample (i.e., galaxies well below the main sequence, $\Delta\rm MS<-0.5 dex$, at $z=1.8-2.6$) are consistent with the ones summarized here. Thus, CO depletion at fixed FIR luminosity and different molecular gas fraction distributions, with AGN skewed at lower values, are indicative of an intrinsic difference between the total molecular gas content of AGN hosts and non-active galaxies and are not induced by biases in the control sample matching. 
    Moreover, such a tail in the molecular gas fraction of the AGN sample without a correspondence in the non-active galaxies is also present when considering only SUPER AGN, as presented in \citetalias{circosta2021} (see their Fig. 4). 
    We note that AGN host galaxies present a trend of molecular gas fraction vs. stellar mass similar to that of non-active galaxies, with decreasing gas fraction for increasing stellar mass, that resembles the mean general trend of star-forming galaxies \citep[e.g., see][for a review]{tacconi2020,saintonge_catinella2022}. The reduced differences ($\simeq2\sigma$) in \lpco{} and $f_{\rm gas}$ between AGN and non-active galaxies for fixed stellar mass could be due to the large error bars on the best fit parameters, the large fraction of upper limits on the AGN side, and to the distribution of CO upper limits. In fact, there is no preferential locus for CO upper limits (see Figs. \ref{kashz:fig:lco-mstar} and \ref{kashz:fig:fgas-mstar}) for fixed stellar mass, while CO upper limits and CO detection are quite segregated, with CO upper limits (detections) mostly located at the lower (upper) end of the $\Lfir$ distribution (see Fig. \ref{kashz:fig:lco-fir}), contributing to increase the CO depletion effect. This is true also when restricting the AGN sample to host galaxies within 0.3dex distance from the MS. 

    Lastly, we note that the CO SLEDs assumed for AGN hosts and non-active galaxies are consistent with each other, thus the reduced CO luminosity of AGN hosts is not a by-product of the assumed CO SLED. This is indeed a conservative assumption, since, despite the large uncertainties, AGN are seen to show larger ${\rm CO}(J\rightarrow J-1)/\rm CO(1-0)$ ratios even at $J<5$ \citep[e.g.,][]{kirkpatrick2019}. 
    Moreover, AGN hosts and non-active galaxies span the same range of CO flux at fixed $J$ and similarly cover the same \lpco{} luminosity range, thus there is no bias between the two samples in this sense.

	\section{Comparison with cosmological simulations}
        \label{sec:ward22}
        \citet{ward2022} recently analyzed the output of three cosmological simulations (TNG, \citealp{springel2018,pillepich2018,naiman2018,nelson2018,marinacci2018}, data release: \citealp{nelson2019}; EAGLE, \citealp{crain2015,schaye2015}, data release: \citealp{mcalpine2016}; SIMBA, \citealp{dave2019}) with approaches similar to those taken by observers, to assess the predictions regarding the effects of AGN feedback on the properties of AGN host galaxies and how these relate with observational constraints and results. 
        \citeauthor{ward2022} interestingly found that simulations predict that powerful AGN tend to be located in gas rich galaxies and that the gas-depleted fraction of AGN hosts is lower than that of non-active galaxies. 

          \begin{figure}[!h]
		\centering		
            \includegraphics[width=0.45\textwidth]{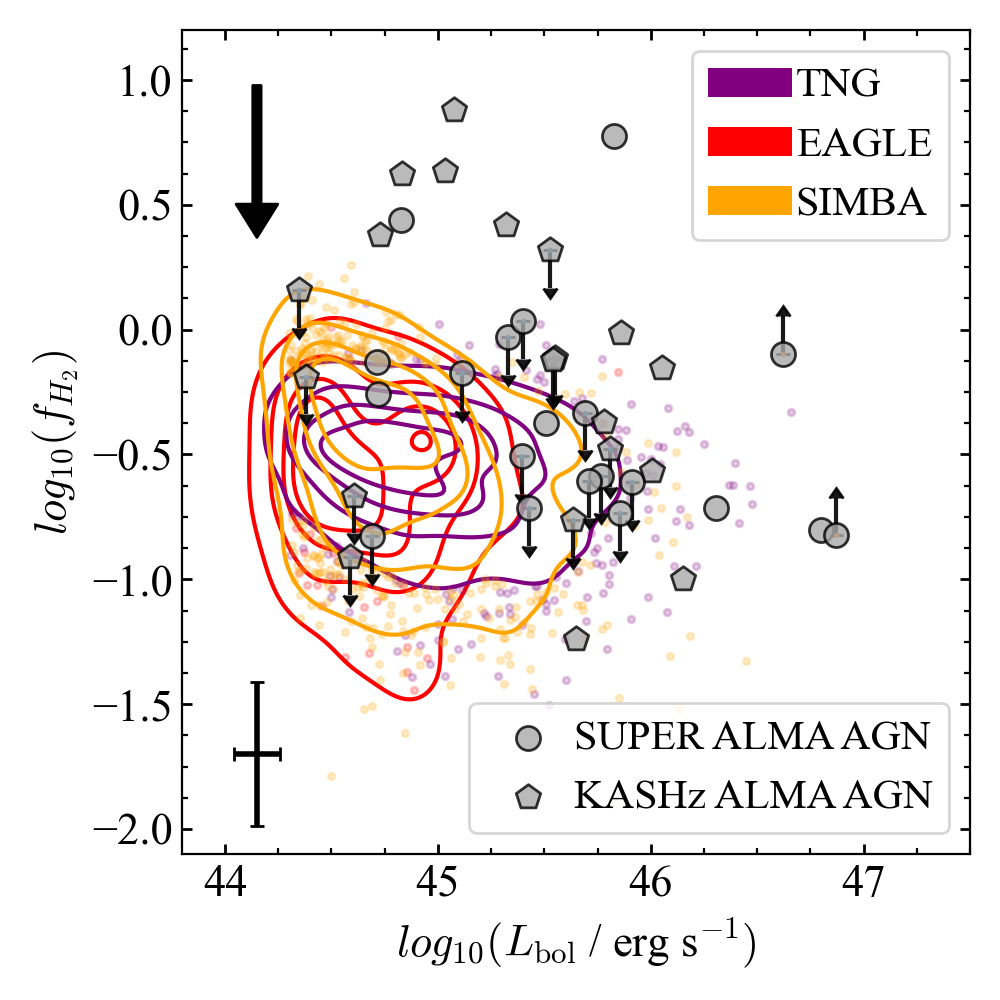}
		\caption{Distribution of gas fraction $f_{\rm H_2}$ vs. bolometric luminosity $\Lbol$. Galaxies in the simulations are selected to match the observed parameter range in stellar mass and SFR, and to host AGN with $\Lbol>10^{44.3}$ \lumcgs and $\lambda_{\rm Edd}>0.01$. Contours contain the 90\% of the selected systems and extreme outliers are shown as individual points. Color coding for simulations is as follows: EAGLE, TNG and SIMBA in purple, red, orange, respectively. Observational points are marked in grey as pentagons for \kashz\ AGN (presented in this work) and circles for SUPER AGN from \citetalias{circosta2021}, scaled by $\alphaco=3.6 \Msun\rm\ (K\ km\ s^{-1}\ pc^2)^{-1}$. The black arrow marks the downward shift for $\alphaco=0.8 \Msun\rm\ (K\ km\ s^{-1}\ pc^2)^{-1}$. The bottom left point shows the mean error bar of the observational sample. }
		\label{kashz:fig:fgaslbol_ward22}
	\end{figure}
 
          \begin{figure*}[!ht]
		\centering
		\includegraphics[width=\textwidth]{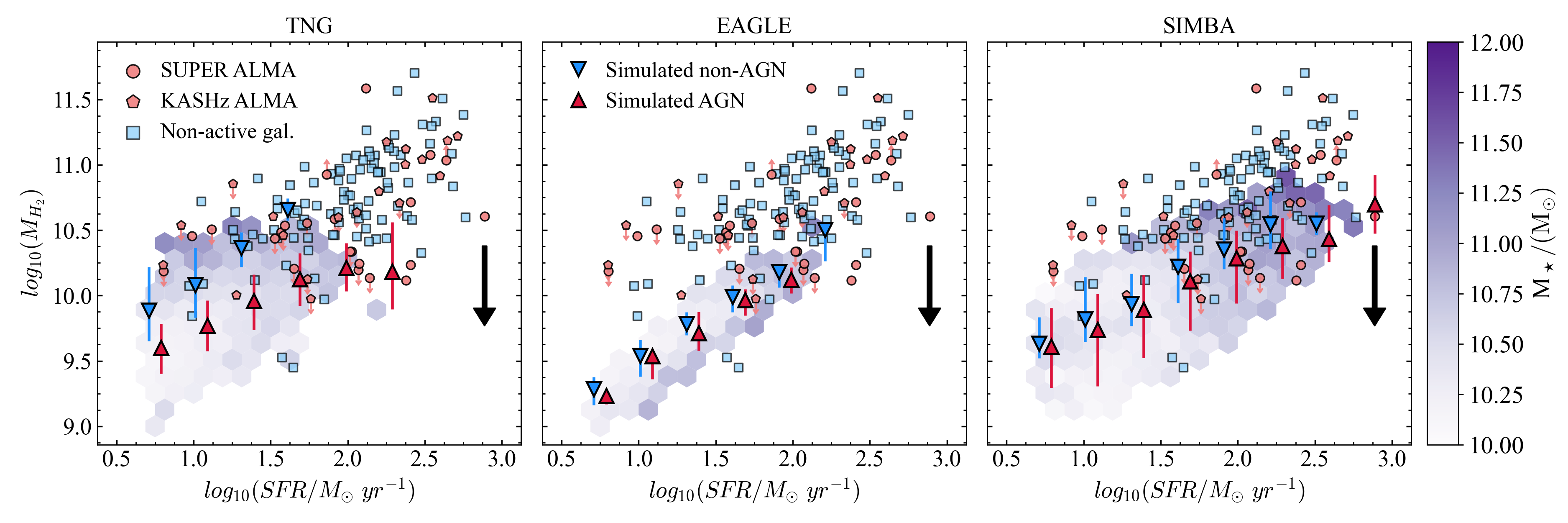}
		\caption{Comparison of molecular gas fraction against SFR of AGN (red) and non-active galaxies (blue) from observations and simulations. SUPER ALMA targets are shown as red circles, \kashz\ AGN as red pentagons, observed non-active galaxies as blue squares.
  Blue downward triangles and red upward triangles show the median and 16th,84th percentiles of non-active galaxies and AGN in the simulations, respectively, grouped in SFR bins of 0.3 dex. Hexagons show the distribution of simulated AGN and are color-coded based on the mean stellar mass of the host galaxies in each bin. 
  Observed molecular gas fractions are computed assuming $\alphaco=3.6 \Msun\rm\,(K\ km\ s^{-1}\ pc^2)^{-1}$. Galaxies in the simulations are filtered to match the observed properties of the observed galaxies. The black arrow marks the downward shift for $\alphaco=0.8 \Msun\rm\ (K\ km\ s^{-1}\ pc^2)^{-1}$.  }
		\label{kashz:fig:mh2sfr_ward22}
	\end{figure*}
 
         \begin{figure}[!h]
		\centering
		\includegraphics[width=0.45\textwidth]{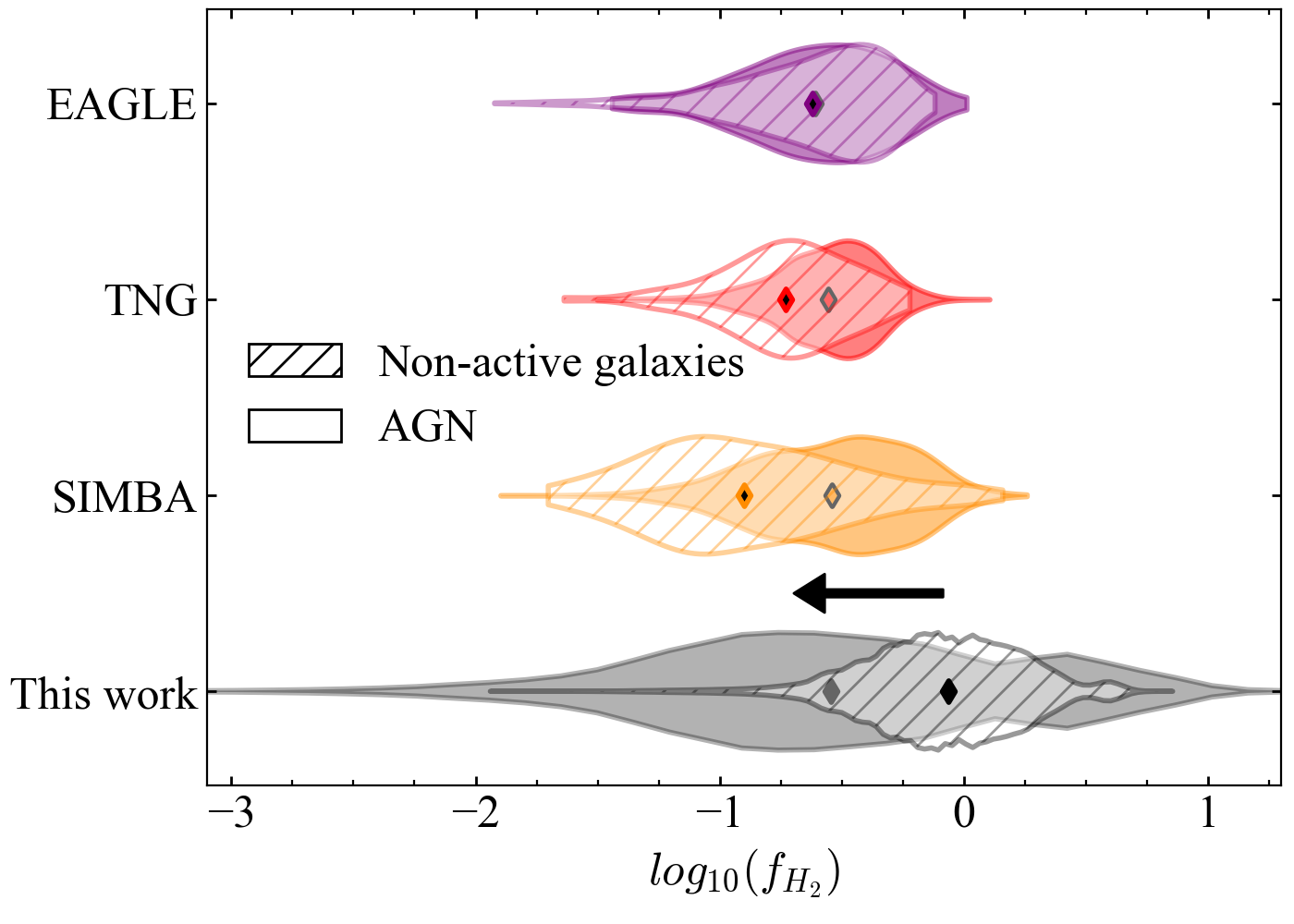}
		\caption{Gas fraction distributions of AGN host galaxies (filled violin plots) and non-active galaxies (hatched violin plots) from cosmological simulations (EAGLE, TNG and SIMBA in purple, red, orange, respectively) and observations presented in this work (grey), scaled for $\alphaco=3.6 \Msun\rm\ (K\ km\ s^{-1}\ pc^2)^{-1}$. Diamond-shaped points mark the mean value of each distribution. Gas fraction distributions from the observations correspond to the total gas fraction distributions built in Sect. \ref{kashz:sec:distrofgas}. Galaxies in the simulations are filtered as for Fig. \ref{kashz:fig:fgaslbol_ward22}. The black arrow marks the leftward shift of observed distributions (black) for $\alphaco=0.8 \Msun\rm\ (K\ km\ s^{-1}\ pc^2)^{-1}$.   } 
		\label{kashz:fig:fgas_ward22}
	\end{figure}

        The molecular gas phase cannot be directly traced in cosmological simulations because, due to computational limitations, the ISM is poorly resolved. However, the amount of molecular hydrogen can be measured through models that are calibrated in the post-processing phase \citep[e.g.,][using for instance the models by \citealp{gnedin2011}]{lagos2015,diemer2018,diemer2019,dave2019}. When tested against observational data, such models are in good agreement with the galaxy populations observed in the local Universe \citep{lagos2015,diemer2019}, while they slightly underestimate the amount of molecular gas observed at $z\simeq2$ (by a factor of 1.5 in EAGLE, \citealp{lagos2015}, and by a factor of 2--3 in TNG, \citealp{popping2019}).  

        Following \citet{ward2022}, we compare the predictions for the molecular gas fraction of galaxies at cosmic noon of TNG, EAGLE and SIMBA tuned to the observed range of \mstar, SFR and \lbol\ spanned by our sample. We thus filter the simulations' output at $z=2$ to select galaxies with stellar masses larger than $10^{10} \Msun$ and lower than $10^{11.7} \Msun$ and $\rm \log(SFR/\Msun~yr^{-1})$ within 0.7 and 2.8 (the lower and upper bounds of the observed sample; see Fig. \ref{kashz:fig:MS}). 
        We then identify AGN hosts in the simulations by combining bolometric luminosity and Eddington ratio ($\lambda_{\rm Edd}$) selection criteria. We use $\log\Lbol>44.3$ \lumcgs to span the same bolometric luminosity range as our observations, and we also consider $\lambda_{\rm Edd}>1\%$ to remove those AGN accreting in a radiatively inefficient fashion. We identify non-active galaxies in the simulations as those systems with $\lambda_{\rm Edd}<0.1\%$, yet we note that results are identical if relaxing the selection ($\lambda_{\rm Edd}<1\%$).
        Similarly to the analysis of Sect. \ref{kashz:sec:distrofgas}, we remove from the observed sample those three AGN (cid\_178, cid\_1605, cid\_467) for which both stellar mass and CO luminosity are upper limits. 
        	 
        Simulations only allow to compute the molecular gas fraction as $f_{\rm H_2}=M_{\rm H_2}/\Mstar$, thus we have to assume an $\alphaco$ value to convert the observed $f_{\rm gas}$ in $f_{\rm H_2}=\alphaco f_{\rm gas}$. 
        One of the backbones of our work is not to elect a value of $\alphaco$ since it is not trivial to discriminate whether each galaxy is more MS- ($\alphaco\simeq3.6 \Msun\rm\ (K\ km\ s^{-1}\ pc^2)^{-1}$; e.g., \citealp{daddi2010a,genzel2015}) or more starburst-like \citep[$\alphaco=0.8 \Msun\rm\ (K\ km\ s^{-1}\ pc^2)^{-1}$; e.g.,][]{downes_solomon_1998,solomonvandenbout2005,calistrorivera2018,amvrosiadis2023} or somewhere in the middle \citep[$\alphaco=1.8-2.5 \Msun\rm\,(K\ km\ s^{-1}\ pc^2)^{-1}$,][]{cicone2018_ngc6240,Herrero-Illana2019,MontoyaArroyave2023}. Yet, a good fraction of the AGN sample corresponds to MS host galaxies (see Fig. \ref{kashz:fig:MS}, thus we present the comparison between observed data and simulations assuming $\alphaco=3.6 \Msun\rm\ (K\ km\ s^{-1}\ pc^2)^{-1}$ and show the shift corresponding to the lowest $\alphaco$ value ($\alphaco=0.8 \Msun\rm\ (K\ km\ s^{-1}\ pc^2)^{-1}$) in the plots. 

        Figure \ref{kashz:fig:fgaslbol_ward22} shows the $f_{\rm H_2}$ vs. \lbol\ distribution of AGN in the simulations and in our observed sample. Density contours show the distribution of simulated sources in TNG (red), EAGLE (purple) and SIMBA (orange), while observed targets are shown as grey pentagons for ALMA \kashz\ AGN (presented in this work) and grey circles for SUPER AGN from \citetalias{circosta2021} and this work. 
        Figure \ref{kashz:fig:mh2sfr_ward22} shows the comparison of simulations and observations in the $f_{\rm H_2}$ vs. SFR plane. We show AGN and non-active galaxies of the simulations computing their median molecular gas fraction in SFR bins of 0.3dex width, and use the 18th, 64th percentiles as errorbars. We also show the un-binned distribution of AGN host galaxies as hexagons, color-coded based on the mean stellar mass of the galaxies in each bin. 
        Figure \ref{kashz:fig:fgas_ward22} compares the gas fraction distributions (violin plots) and mean values (diamonds) of AGN hosts and non-active galaxies from cosmological simulations and the total gas fraction distributions and mean values of observed AGN and non-active galaxies as derived in Sect. \ref{kashz:sec:distrofgas}.  
         
        While the \kashz\ survey in theory spans a very broad range of X-ray and bolometric luminosity ($\log (L_{\rm X}/$\lumcgs$) = 41-45.3 $; $\log (\Lbol/$\lumcgs$) \simeq 42-47$ using the bolometric correction of \citealp{duras_2020}), the ALMA \kashz\ AGN have bolometric luminosity $\log (\Lbol/$\lumcgs$)\geq44.3$, comparable to the lower bound of the observed AGN used in \citet{ward2022}. As a consequence, despite the better sampling of the intermediate-luminosity regime, the \lbol\ range spanned by observed and simulated AGN in Fig. \ref{kashz:fig:fgaslbol_ward22} hardly overlap, as seen by \citet{ward2022}. To bridge such a gap, both observations and simulations need to take a leap forward. On the one hand, we need to gather observational information of  AGN with lower bolometric luminosity, possibly through targeted programs. On the other hand, simulations still struggle in reproducing the high-$\Lbol$ population of AGN (Fig. \ref{kashz:fig:fgaslbol_ward22}), due to their small box size that prevents to reproduce rare and short lived objects like AGN with $\Lbol>10^{46}$ \lumcgs. We note that this mismatch in $\Lbol$ is still present even by relaxing the SFR and $\Mstar$ filtering, as seen in \citet{ward2022}. 
        
        Additionally, cosmological simulations suggest that AGN hosts should match or exceed non-active galaxies in gas fraction, which is the opposite of the observed trend (see Fig. \ref{kashz:fig:fgas_ward22}). We note that also this result still holds even by relaxing the mass and SFR filtering of the simulations' output, as can be seen in the left column panels of Fig. 7 in \citet{ward2022}. 
        EAGLE and Simba also suggest that AGN host galaxies show similar molecular gas mass for fixed SFR compared to non-active galaxies (see Fig. \ref{kashz:fig:mh2sfr_ward22}), which is at odds with our result in Sect. \ref{kashz:sec:cofirlum}. Yet, AGN hosts in TNG show molecular gas depletion for fixed SFR, even though with an opposite trend compared to the observations: the discrepancy increases for increasing SFR in the simulations, while in the observations it increases for decreasing SFR  (see Fig. \ref{kashz:fig:lco-fir}). Figure \ref{kashz:fig:mh2sfr_ward22} also shows that simulations do not fully reproduce the observed properties of AGN host galaxies. In fact, SIMBA is the only cosmological simulation that covers the full $\Mstar$ and SFR range, while TNG and EAGLE are basically missing AGN hosts with $\Mstar>10^{11}\Msun$ and $\rm\log (SFR/\Msun~yr^{-1})\gtrsim2$.
        Yet, a direct comparison to the simulations is difficult, as their limited box sizes and subgrid models for AGN feedback and molecular gas estimation reduce the overlap with our observations, especially at the high-mass, high-$\Lbol$ end.  
        
        Another pressing issue that arises from the comparisons in Figures \ref{kashz:fig:fgaslbol_ward22}--\ref{kashz:fig:fgas_ward22} is related to the choice of $\alphaco$. On the one hand, the gas fraction distributions of AGN hosts in the simulations are in fair agreement with the observed distributions of \kashz\ and SUPER AGN in the case of $\alphaco=3.6 \Msun\rm\ (K\ km\ s^{-1}\ pc^2)^{-1}$ (see Fig. \ref{kashz:fig:fgas_ward22}), in the sense that their mean values are consistent, but simulations seem to lack more gas-rich systems, as is evident also as a function of bolometric luminosity in Figs. \ref{kashz:fig:fgaslbol_ward22}. 
        On the other hand, simulations fail to reproduce the observed $f_{\rm H_2}$ distributions in the case of $\alphaco=0.8 \Msun\rm\ (K\ km\ s^{-1}\ pc^2)^{-1}$ (see Fig. \ref{kashz:fig:fgas_ward22}) but, at the same time, would show a better agreement with the observed $f_{\rm H_2}$ vs. \lbol\  distribution (see Fig. \ref{kashz:fig:fgaslbol_ward22}), at least in the range of \lbol, \mstar\ and SFR used to filter the simulations' output. 
        The clear impact of the adopted value of $\alphaco$ on the observed results in Figs. \ref{kashz:fig:fgas_ward22} and \ref{kashz:fig:fgaslbol_ward22} further stresses the importance of measuring tracers that can be used to derive or set a prior for $\alphaco$, for instance metallicity and distance from the MS \citep[e.g.,][]{elbaz2007,noeske2007,accurso2017}, two quantities that are not available for the full sample of AGN used in this work. 
	
	\section{Discussion: do AGN at cosmic noon gas-deplete their hosts?}
        \label{sec:discussion}
    Our results show that AGN at $z=1-2.6$, uniformly selected and analyzed, without prior knowledge on feedback tracers (e.g. ionised outflows or radio jets), and non-active galaxies, matched in \z, $\Mstar$, SFR,
    are statistically different in terms of $L'_{\rm CO(1-0)}$ vs. $L_{\rm FIR}$ (3$\sigma$ level; see Sect. \ref{kashz:sec:cofirlum}), and $\log($\lpco$/\Mstar)$ distribution (proxy of the molecular gas fraction distribution; p-value $< 10^{-7}$; see Sect. \ref{kashz:sec:distrofgas}).
    Yet, even though there are hints of gas depletion even at fixed stellar mass (see Fig. \ref{kashz:fig:lco-mstar}-\ref{kashz:fig:fgas-mstar}), AGN and non-active galaxies are statistically consistent in the $L'_{\rm CO(1-0)}$ vs. $\Mstar$, $f_{\rm gas}$ vs. $\Mstar$ and  $\log($\lpco$/\Mstar)$ vs. $\Mstar$ two-dimensional spaces (see Sect. \ref{kashz:sec:fgas-mstar}). 

    The control sample of non-active galaxies is built to match the properties shown by AGN host galaxies, thus in principle the sample of selected non-active galaxies is an analog of our AGN hosts with the only difference of not hosting an AGN. Moreover, we robustly controlled for consistency in SFR and $\Mstar$ between the two samples (see Sect. \ref{kashz:sec:control}) and that our results stay unvaried when considering a subsample of AGN and non-active galaxies. Additionally, we trace the SF level in our galaxies through $\Lfir$, i.e. corresponding to SF averaged over 100 Myr. Thus, we interpret our comparisons of AGN hosts and non-active galaxies in terms of differences (or absence of) in the molecular gas content due to the presence of the AGN. 
    
    In particular, we find that AGN hosts are, on average, more CO depleted for fixed FIR luminosity (see Fig. \ref{kashz:fig:lco-fir}). In other words, AGN hosts show a reduced amount of cold molecular gas compared to matched non-active galaxies, and such reduced CO luminosities may convert into reduced SF in AGN hosts in the future.  
    We stress that this difference holds also when we exclude AGN and non-active galaxies that are not well matched, thus does not depend on biases in building the control sample. Moreover, CO depletion is differential in $\Lfir$, meaning that it increases for decreasing $\Lfir$. 
    CO content and molecular gas fraction for fixed stellar mass of AGN and non-active galaxies  are different at the 2$\sigma$ level (see Figs. \ref{kashz:fig:lco-mstar}--\ref{kashz:fig:fgas-mstar}). Such a reduced difference for fixed stellar mass could be driven by the differential CO depletion level for fixed SFR (see Fig. \ref{kashz:fig:lco-fir}): at fixed stellar mass, we are sampling galaxies that span a broad range of SF levels, which differentially depend on stellar mass (e.g., see Fig. 8 of \citealp{saintonge_catinella2022}). In this sense, the mixing of host galaxies with different SF levels and the differential CO depletion observed for fixed $\Lfir$ (see Fig. \ref{kashz:fig:lco-fir}) likely dilute the differences between AGN and non-active galaxies for fixed stellar mass. Yet, we do not find more enhanced differences considering only targets on the MS, possibly due to the small sample size (14 AGN host galaxies on the MS). 
	Differences in the one-dimensional distributions of $f_{\rm gas}$ are more prominent because the contribution of CO detections and upper limits in the AGN sample is better decoupled: in fact, the mean value of the distribution is mainly driven by CO detections, while CO upper limits (and in this case, $f_{\rm gas}$ upper limits) shape the left wing of the distribution, producing the skewness at lower values that is missing in the distribution of non-active galaxies. 

    We confirm the hints observed by \citetalias{circosta2021} in SUPER ALMA AGN, using a uniformly selected and homogeneously analyzed AGN sample that is twice in size: molecular gas depletion is significant also when considering moderate-luminosity AGN ($44\lesssim\log(\Lbol/{\rm erg~s^{-1}})\lesssim47$). 
    This work thus supports the scenario for which AGN play a key role in shaping the life of galaxies, in particular by altering the molecular gas reservoir of their hosts, as was observed in higher-$\Lbol$ source ($46\lesssim\log(\Lbol/{\rm erg~s^{-1}})\lesssim48$; e.g., \citealp{harrison2012,brusa2018,vietri2018,perna2018,bischetti2021}). This is also coherent with the recent results in \citet{friascastillo2024} in moderate-luminosity, lensed AGN ($43\lesssim\log(\Lbol/{\rm erg~s^{-1}})\lesssim46$) at end of cosmic noon ($z=1.2-1.6$), which are seen as CO depleted for fixed FIR luminosity compared to star forming galaxies. 
    We also investigated the possibility of a dependency of the molecular gas fraction with properties of the AGN, like the bolometric luminosity, yet found no significant correlation, as already seen in previous studies (e.g., \citealp[]{shangguan2020a}; \citetalias{circosta2021}).  

    In this work, we assumed a CO SLED at $J<5$ for AGN hosts that is consistent with that of non-active galaxies. This is a conservative assumption: 
    works focused on determining the CO ladder of AGN found them to show higher $r_{J,1}$ values compared to non-active galaxies even at $J<5$, despite the large error bars \citep[e.g.,][]{kirkpatrick2019,boogaard2020_aspecs}. Moreover, the backbone of our analysis, and other previous studies, is to compare the gas content of AGN host galaxies and non-active galaxies in terms of \lpco, thus without assuming an $\alphaco$ value. In this way, the difference found between the two samples in molecular gas mass and fraction represents a lower limit of the real discrepancy: in fact, this would be unchanged if we were to assume the same $\alphaco$ value for both AGN hosts and non-active galaxies, while the difference would increase in the case of the reasonable assumption of AGN hosts showing lower $\alphaco$ values than non-active galaxies \citep[e.g., see][and references therein]{bolatto2013}. 
    Given all our conservative assumptions, our results indeed provide evidence for AGN reducing the molecular gas reservoir of their hosts and thus their evolution. 

    \citet{bischetti2021} recently investigated the effects of AGN feedback on the molecular gas reservoir of WISSH quasars. Among the various tests, the authors compared the gas fraction of WISSH AGN with that of star-forming galaxies after correcting for the evolution of the gas fractions with redshift and distance from the MS ($\rm\Delta MS=sSFR/sSFR_{MS}$). 
	In fact, analyses focused on assessing the molecular gas content of star-forming galaxies up to $z\simeq4$ \citep[e.g.,][]{genzel2015,tacconi2018} pointed out a dependency of the gas fraction on redshift and distance from the MS. 
    Our analysis already accounts for the redshift evolution of the MS when building the control sample by matching AGN and galaxies in two redshift bins ($z\simeq1-1.8$, $z\simeq2-2.6$), within which the MS evolution is not large \citep[$\simeq0.2$dex; e.g.,][]{whitaker2012}. 
    However, not correcting for the distance from the MS could still affect the comparison of non-active galaxies and AGN hosts. Unfortunately, such a correction is not viable for the AGN sample used in this work: about half of our targets have upper limits on $\Lfir$, making impossible to constrain how much they deviate from the MS. Nonetheless, we tentatively measure the $\rm\Delta MS=sSFR/sSFR_{\rm MS}$ parameter of the AGN sample considering $\Lfir$ at face value and compute the gas fractions  corrected for the distance from the MS ($f_{\rm gas}^{\rm corr}$) using Equation 15 of \citet{genzel2015} in which $\rm sSFR_{\rm MS}$ is parameterized as in \citet{whitaker2012}. We conservatively assume $\alphaco=3.6\Msun\rm\ (K\ km\ s^{-1}\ pc^2)^{-1}$ for both AGN and non-active galaxies.  Applying the analysis of Sect. \ref{kashz:sec:distrofgas}, we find that the $f_{\rm gas}^{\rm corr}$ total distributions of AGN and non-active galaxies are still significantly different according to the KS test (p value<0.001), with consistent mean values, both using the full sample and excluding those AGN and non-active galaxies without a good match (see Sect. \ref{kashz:sec:control}). Following the analysis of Sect. \ref{kashz:sec:fgas-mstar}, we also find that AGN and non-active galaxies are consistent with each other within 2$\sigma$ in the $f_{\rm gas}^{\rm corr}$ vs. $\Mstar$ parameter space. 

    Yet, there are still some effects that we are not fully considering, or cannot fully consider, when studying AGN feedback. 
    First, we are comparing two phenomena that occur on very different timescales \citep[e.g.,][]{harrison2017,harrison2018,harrison_ramosalmeida_2024}: the timescales of visible AGN activity do not necessarily mirror the AGN duty cycle, plus the timescales of feeding and feedback in general are uncertain. For instance, the AGN might have become less luminous or non-visible by the time the effects of its impact have become observable, or even the galaxy might have accreted more gas from its surroundings (e.g., from the CGM) reducing the overall depletion effect of AGN activity. Additionally, we have no means of understanding how long the currently observed AGN has been active or if it is the $n$-th active phase in a cycle of flickering accretion. In other words, we cannot know how much energy the AGN has input in the surrounding medium before being observed, and thus on the total energy input in the system on timescales that are relevant for SF activity. 
		
    Second, we are comparing phenomena and effects possibly happening on spatial scales that are very different: high-resolution studies in the local Universe unveiled AGN that create regions of CO depletion in their surroundings \citep[within the central few 100~pc to kpc; e.g.,][]{sabatini2018,rosario2019,fluetsch2019,feruglio2020,ellison2021}, where the bulk of the molecular gas is possibly in a warmer phase due to AGN heating \citep[e.g., through accretion radiation in the form of X-ray photons,][]{fabbiano2019}. Studies regarding SF activity in local AGN confirm such a scenario where AGN have a significant impact only in their surroundings, either reducing \citep[e.g.,][]{molina2023} or enhancing \citep[e.g.,][]{cresci2015a,garciaburillo2021_gatos,lammers2023} the central SF without affecting it on larger scales. 
    Yet, this work supports the scenario in which AGN at cosmic noon are indeed affecting the total molecular gas mass of galaxies (e.g., total molecular gas mass, future total SFR), and thus a significant portion of the full system.  
    An additional possibility is that AGN excite CO higher up in the rotational ladder without making the host molecular gas poor \citep[e.g., ][]{paraficz2018,fogasy2020,spingola2020}. However, such a scenario cannot be probed with the present sample since many AGN were targeted in only one CO transition.

    Third, the SUPER and \kashz\ surveys have selected AGN only based on their X-ray luminosity and redshift, with no priors regarding AGN-driven outflows and AGN feedback effects. 
    Here and in \citetalias{circosta2021}, we investigated AGN feedback by studying the overall properties of the AGN sample compared to those of non-active galaxies. Literature results finding AGN hosts to be significantly gas depleted, or with reduced depletion timescales, are usually associated with sources showing massive AGN-driven outflows on kpc-scales in at least one gas phase and that are often caught at the peak of their power \citep[e.g.,][but see also \citealp{scholtz2021,lamperti2021}]{carniani2017,brusa2018,loiacono2019}. One might then infer that gas depletion is significant and widespread only in those sources where feedback is ``caught in the act''. However, gas depletion in AGN host galaxies could be also due to earlier AGN event(s), so that the reduction in gas content and/or SF activity is a product of the integrated past activity of the AGN \citep[e.g.,][]{piotrowska2022}. A detailed comparison between [OIII] outflows and CO properties of \kashz\ and SUPER targets is out of the scope of the present study and will be the subject of a future work.
	
	Lastly, this work provides a uniformly-selected and -analyzed, sizable sample to compare observations and predictions of cosmological simulations. We compared the observed cosmic noon AGN and sample of non-active galaxies with the properties of AGN and galaxies from three cosmological simulations \citep[TNG, EAGLE and SIMBA, following][]{ward2022}, by filtering their output at $z=2$ to match the properties of the observed sample of AGN (see Sect. \ref{sec:ward22}).  
    One of the issues encountered in \citet{ward2022}, and also in this work (see Fig. \ref{kashz:fig:fgaslbol_ward22}), is that simulations and data hardly overlap, especially in terms of bolometric luminosity. 
    Our observed sample provides a better coverage of the $44<\log(\Lbol/{\rm erg~s^{-1}})<46$ than previously available. However, a robust sampling of observed AGN at $\log(\Lbol/{\rm erg~s^{-1}})<44$, where the bulk of simulated AGN lies, is still lacking. 
    Moreover, the three simulations predict that AGN hosts should match non-active galaxies in molecular gas mass for fixed SFR (see Fig. \ref{kashz:fig:mh2sfr_ward22}) and that they should even exceed non-active galaxies in gas fraction (see Fig. \ref{kashz:fig:fgas_ward22}). The only exception is TNG, which predicts AGN to be gas depleted for fixed SFR but with a trend (increasing gas depletion for increasing SFR) that is opposite to what seen in the observations (see Fig. \ref{kashz:fig:lco-fir}). Another issue encountered in these comparisons is that only SIMBA covers the full parameter range of observed AGN hosts, while TNG and EAGLE are missing AGN hosts with $\Mstar>10^{11}\Msun$ and $\log(\rm SFR/\Msun~yr^{-1}\gtrsim2$.
    Lastly, We also highlight another (well-known) issue: the selection of the ``right'' $\alphaco$ value. In fact, how well (or not well) the simulations agree with the observed properties strongly depends on the chosen value for $\alphaco$ (see Sect. \ref{sec:ward22} and Figs. \ref{kashz:fig:fgas_ward22}-\ref{kashz:fig:fgaslbol_ward22}). For the present sample, for instance, additional observations tailored at measuring the metallicity and better constrain the distance from the MS could allow to take advantage of the continuous $\alphaco(z,\Delta MS)$ function of \citet{accurso2017} and solve this issue.

\section{Summary and Conclusions}
\label{sec:summary}
This work addresses the effects of AGN activity on the molecular gas content of host galaxies with a carefully-selected, sizable sample of AGN at cosmic noon, to shed light on whether AGN have any impact on the fuel of future star formation. 

\begin{itemize}
  \item  We collected all the available ALMA observations of CO ($J<5$) emission lines of \kashz\ AGN at cosmic noon, a survey built to study AGN feedback in X-ray selected AGN as traced by the ionized gas component. We complement our \kashz\ ALMA sample with SUPER ALMA targets from \citetalias{circosta2021} (observed in \cotredue). Being heterogeneous in the CO transitions observed (\codueuno, \cotredue, \coquattrotre), we assumed a reasonable CO excitation ladder \citep[from][]{kirkpatrick2019} to convert our measurements to the \counozero. Other key quantities of the AGN+galaxy system were retrieved through SED fitting with the CIGALE code \citep{boquien2019_cigale,yang2020_xcigale}, collecting the most up-to-date multi-wavelength broad-band photometry as provided by the deep-field survey collaborations (CANDELS/GOODS-S, COSMOS, HELP collaboration). 

  \item The AGN sample thus totals 46 AGN at $z=1-2.6$, i.e. a sample size that is twice that of the SUPER AGN presented in \citetalias{circosta2021} and with a larger CO detection rate ($\simeq50$\% vs. 40\%). We built the control sample of non-active galaxies (98 galaxies), matched in \z, $\Mstar$, SFR to our AGN sample, by capitalizing on the PHIBSS project \citep[and references therein]{tacconi2018}, on the ASPECS survey \citep[and references therein]{boogaard2020_aspecs} and on the ALMA/NOEMA survey of sub-$mm$ galaxies in the COSMOS, UDS, and ECDFS fields by \citet{birkin2021}.

  \item We then quantitatively compared AGN and non-active galaxies in a Bayesian framework, as developed by \citetalias{circosta2021}, robustly considering upper and lower limits in any of the considered quantities. The samples were compared in terms of \textit{i)} $L'_{\rm CO(1-0)}$ vs. $L_{\rm FIR}$ (see Sect. \ref{kashz:sec:cofirlum}), \textit{ii)} $\log($\lpco$/\Mstar)$ distribution (proxy of the gas fraction distribution; see Sect. \ref{kashz:sec:distrofgas}), \textit{iii)} $L'_{\rm CO(1-0)}$ vs. $\Mstar$, and \textit{iv)} $\log($\lpco$/\Mstar)$ vs. $\Mstar$ (see Sect. \ref{kashz:sec:fgas-mstar}).
  Our results confirm the hints observed by \citetalias{circosta2021} on SUPER AGN: AGN hosts at cosmic noon, selected without priors regarding AGN feedback, are significantly gas depleted compared to non-active galaxies. 
  Differences in the $\log($\lpco$/\Mstar)$ vs. $\Mstar$ and $L'_{\rm CO(1-0)}$ vs. $\Mstar$ two-dimensional spaces are possibly washed out by the rather uniform distribution of CO upper limits with respect to stellar mass (see Figs. \ref{kashz:fig:lco-mstar}-\ref{kashz:fig:fgas-mstar}), and the differential CO depletion observed for fixed $\Lfir$ (see Fig. \ref{kashz:fig:lco-fir}).
  
  \item Building on the study of \citet{ward2022}, we also compared our observed samples of AGN and non-active galaxies with the output of three cosmological simulations (TNG, EAGLE, SIMBA), filtered to match the properties of the observed AGN, AGN hosts and non-active galaxies (see Sect. \ref{sec:ward22}). We find that SIMBA is the only simulation that can well reproduce the mass and SFR of host galaxies, TNG is the only simulation predicting gas depletion for fixed SFR in AGN hosts, but with a trend opposite to that of observations (see Fig. \ref{kashz:fig:mh2sfr_ward22}), and that all three equally fail at reproducing the observed distribution of molecular gas fractions (see Fig. \ref{kashz:fig:fgas_ward22}) and bolometric luminosity (see Fig. \ref{kashz:fig:fgaslbol_ward22}). 
\end{itemize}

The latest (and upcoming) facilities have opened (and will broaden) new windows to assess AGN feedback in the high-\z\ Universe, especially regarding high-spatial-resolution studies. ALMA has proven its outstanding capabilities over the past decade but high-resolution observations, matching the spatial resolution of NIR data, are still highly time-consuming, yet necessary to fully understand how AGN affect the molecular gas distribution within their hosts and bridge the gap between the local Universe and cosmic noon. 
Nonetheless, JWST/MIRI will allow us to probe molecular transitions previously inaccessible at cosmic noon, like the roto-vibrational $\rm H_2$ lines, which can shed new light on the effects of AGN activity on the molecular gas phase. 
Three SUPER targets (one of which is shared with \kashz), selected based on their prominent \oiii\ outflows, will be observed by JWST/MIRI/MRS (PI: V. Mainieri; wavelength coverage: $\lambda\simeq6.5-27.9~\mu$m) to map the warm-molecular gas phase, thus completing their multi-phase characterization. We will be able to determine the dynamics of the molecular gas and to derive the total (ionized+molecular) mass outflow rate and kinetic energy for these outflows, some of the most difficult wind parameters to measure with accuracy, thus providing a key constrain for current models of AGN feedback \citep[e.g.,][]{costa2020}. 
    
\begin{acknowledgements}
EB thanks R. Decarli, L. Boogaard, G. Sabatini and L. Barchiesi for useful discussion. We acknowledge financial support from INAF under the Large Grant 2022 ``The metal circle: a new sharp view of the baryon cycle up to Cosmic Dawn with the latest generation IFU facilities'', from ASI under grants ASI-INAF I/037/12/0 and n. 2017-14-H.O, and from the grant PRIN MIUR 2017PH3WAT (``Black hole winds and the baryon life cycle of galaxies''). MD and EB gratefully acknowledge INAF funding through the ``Ricerca Fondamentale 2022'' program (mini-grant 1.05.12.04.04).

SRW acknowledges funding from the Deutsche Forschungsgemeinschaft (DFG, German Research Foundation) under Germany's Excellence Strategy: EXC-2094-390783311. 

AF acknowledges the support from project "VLT-MOONS" CRAM 1.05.03.07. AP acknowledges support from Fondazione Cariplo grant no. 2020-0902. 

AM and MG acknowledge support from grant PRIN-MUR 2020ACSP5K\_002 financed by European Union – Next Generation EU.

IL acknowledges support from PID2022-140483NB-C22 funded by AEI 10.13039/501100011033 and BDC 20221289 funded by MCIU by the Recovery, Transformation and Resilience Plan from the Spanish State, and by NextGenerationEU from the European Union through the Recovery and Resilience Facility. 

MP acknowledges support from the research project PID2021-127718NB-I00 of the Spanish Ministry of Science and Innovation/State Agency of Research (MCIN/AEI/10.13039/501100011033). 

DMA thanks the Science Technology Facilities Council (STFC) for support from the Durham consolidated grant (ST/T000244/1). 

CC has received funding from the European Union’s Horizon 2020 research and innovation programme under grant agreement No 951815 (AtLAST). 

CH acknowledges funding from a United Kingdom Research and Innovation grant (code: MR/V022830/1). 

IEL received funding from the European Union’s Horizon 2020 research and innovation program under Marie Skłodowska-Curie grant agreement No. 860744 "Big Data Applications for Black Hole Evolution Studies" (BiD4BESt; \url{https://www.bid4best.org/}).

This paper makes use of ALMA data as listed in Table \ref{app:tab:almaobs}. ALMA is a partnership of the ESO
(representing its member states), NSF (USA), and NINS
(Japan), together with the NRC (Canada), NSC and ASIAA
(Taiwan), and KASI (Republic of Korea), in cooperation with
the Republic of Chile. The Joint ALMA Observatory is
operated by the ESO, AUI/NRAO, and NAOJ. 

This paper is based on observations carried out with the IRAM Interferometer NOEMA. IRAM is supported by INSU/CNRS (France), MPG (Germany) and IGN (Spain).

The project leading to this publication has received support from
ORP, that is funded by the European Union’s Horizon 2020 research
and innovation program under grant agreement No 101004719 [ORP]. This work made use of computing resources and technical support from INAF-OAS Bologna and the Italian node of the European ALMA Regional Center, hosted by INAF-Istituto di Radioastronomia. EB thanks M. Dadina for granting the use of sbuccia whenever needed. 

This research made use of the {\tt spectral-cube} (\url{https://github.com/radio-astro-tools/spectral-cube}), {\tt numpy} \citep{numpy_harris2020array}, {\tt astropy} \citep{astropycoll2013,astropy_2018} and {\tt scipy} \citep{sciPy2020-NMeth} Python packages. 
This research made use of data from HerMES project (\url{http://hermes.sussex.ac.uk/}). HerMES is a Herschel Key Programme utilising Guaranteed Time from the SPIRE instrument team, ESAC scientists and a mission scientist.
The HerMES data was accessed through the Herschel Database in Marseille (HeDaM - \url{http://hedam.lam.fr}) operated by CeSAM and hosted by the Laboratoire d'Astrophysique de Marseille. 

\end{acknowledgements}

\bibliographystyle{aa}
\bibliography{mybib.bib}

\newpage

\appendix
\section{Derivation of X-ray luminosity of \kashz\ AGN}
\label{app:xraylum}
Based on our spectral fits of NIR IFS data (which will be presented in Scholtz et al., in prep), we could provide a new estimate of the redshift of \kashz\ AGN, with which we confirmed or updated that of the X-ray catalogs.
We confirm and provide a more precise redshift estimate for $\simeq$80\% of the \kashz\ sample, compared to the values present in the parent X-ray survey catalogs. For $\simeq10$\% of the sources, we measured spectroscopic redshifts that differ more than $|\Delta z|=0.1$ with respect to the ones in the parent catalogs. The remaining $\simeq10$\% are undetected in our IFS data, thus we either keep the X-ray redshift or update it with spectroscopic estimates as collected from the literature. 
We retrieve the X-ray properties of our sources from the parent X-ray catalogues and spectral fits performed by the survey collaborations for the \xuds\ \citep{kocevski2018}, SXDS \citep{akiyama2015}, COSMOS \citep{lanzuisi2013,lanzuisi2015,marchesi2016} and CDFS fields \citep{luo2017,liu2017}, updating them to the new spectroscopic redshift in case this differs by more than $|\Delta z|=0.1$. For SSA22 AGN, we collect the sources' spectra from the \chandra\ Source Catalog v.2.0 and fit them ourselves to retrieve intrinsic photon index, column density and absorption corrected X-ray luminosity, since the survey catalog only reports the X-ray flux. The COSMOS collaboration did not estimate the intrinsic parameters of AGN lacking a robust redshift estimate. Thus, for those COSMOS AGN with an unreliable redshift for which we could provide a spectroscopic redshift, we collect the X-ray spectra directly from the COSMOS-Legacy collaboration and retrieve the intrinsic X-ray properties through spectral fitting. Regarding the X-ray properties of those AGN in common with the SUPER sample, we refer to results as obtained from X-ray spectral fitting by \citetalias{circosta2018} and \citetalias{circosta2021}. 

\section{Comparison of \mstar\ and SFR distributions}
\label{app:mstar_sfr_distros}
\begin{figure}[!th]
	\centering
	\includegraphics[width=0.45\textwidth]{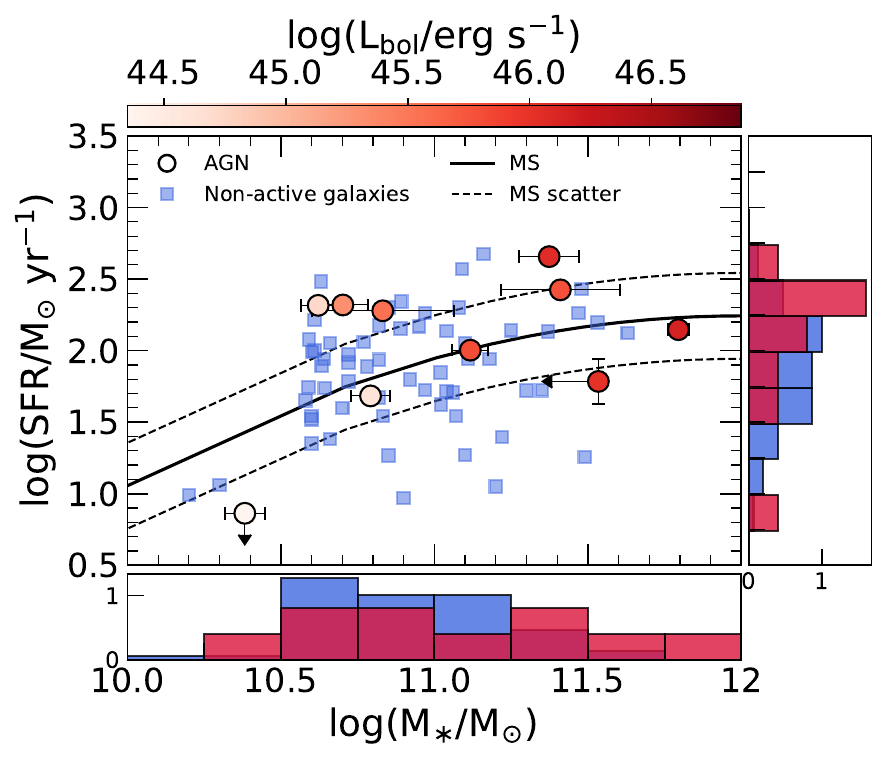}\\
	\includegraphics[width=0.45\textwidth]{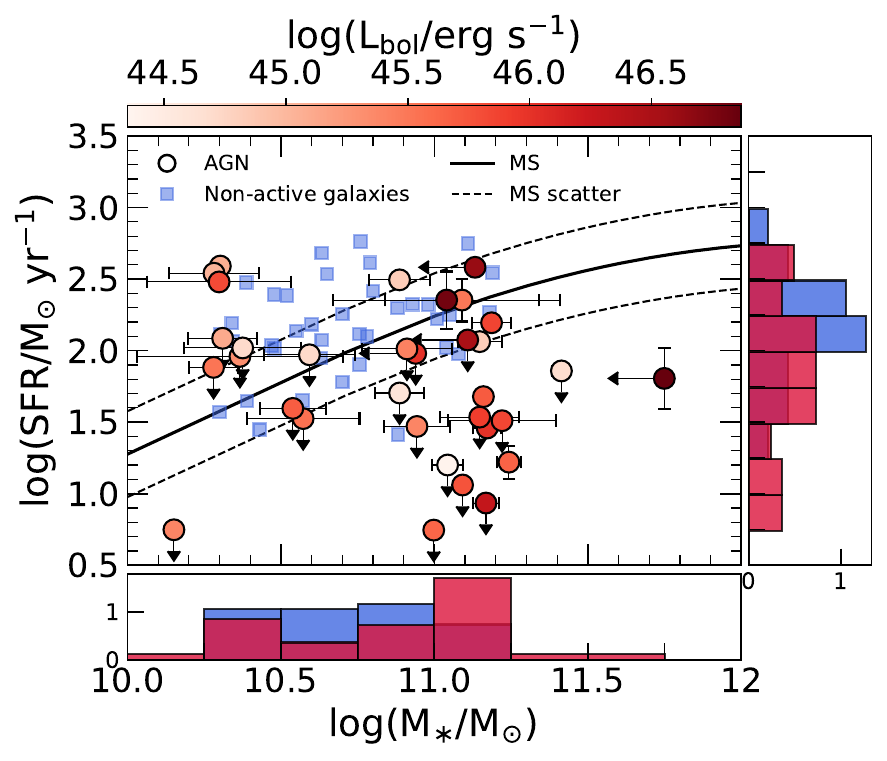}
	\caption{Comparison of SFR and stellar mass of AGN host galaxies (red circles) and non-active galaxies of the control sample (blue squares) for low-\z\ bin ($z=1-1.8$; Top) and high-\z\ bin ($z=1.8-2.55$; Bottom). Red circles are color-coded based on their AGN bolometric luminosity, as retrieved from SED fitting from this work or from the literature. The solid line marks the MS at the mean redshift of each bin from \citet{schreiber2015} and dashed lines show its scatter (equal to 0.3 dex). The distribution of SFR (right) and $\Mstar$ (bottom) for AGN (red) and non-active galaxies (blue) in the side panels are intended for illustration purposes only, since the upper limits are considered at face value. A robust comparison of the distribution of SFR and \mstar applying the hierarchical method described in Sect. \ref{kashz:sec:distrofgas} is presented in Fig. \ref{kashz:fig:sfr_mstar_distros}.  }
	\label{kashz:fig:MS_zbins}
\end{figure}
\begin{figure}[!h]
	\centering
	\includegraphics[width=0.45\textwidth,height=0.5\textheight]{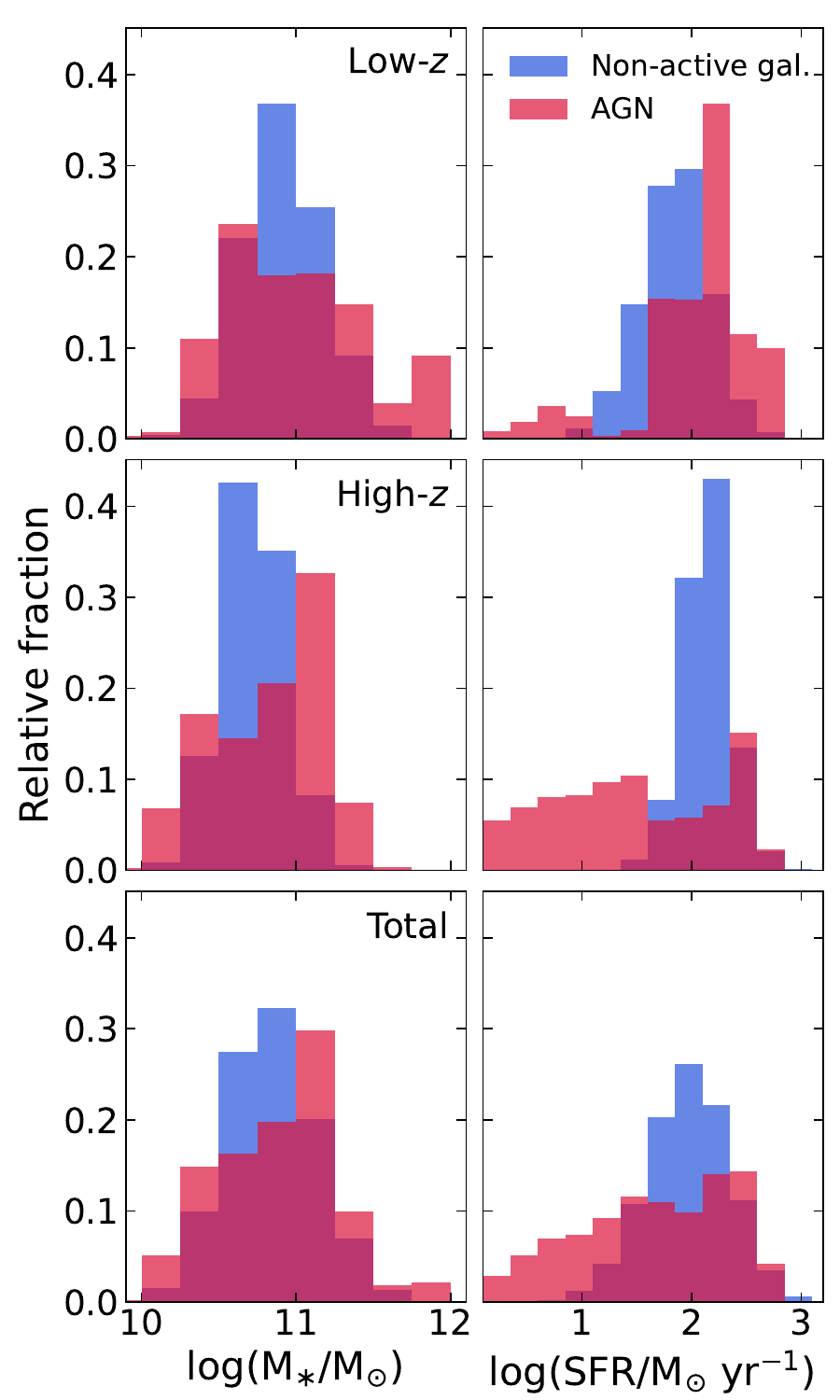}
	\caption{Total posterior distributions of $\Mstar$ (left) and SFR (right) of AGN hosts (AGN sample 1, red) and non-active galaxies (blue). The comparison is carried out for low-\z\ bin, high-\z\ bin and full redshift range, from top to bottom.  }
	\label{kashz:fig:sfr_mstar_distros}
\end{figure}

 
We compare the total posterior distributions of \mstar\ and SFR of AGN hosts and non-active galaxies of the full samples and devided in the two redshift bins used to match the two samples. The \mstar\ and SFR distributions of AGN hosts and non-active galaxies were built applying the Bayesian hierarchical method used for the molecular gas fractions, as described in Sect. \ref{kashz:sec:distrofgas}. The KS test returns a p-value larger than 5\% for all cases but the high-\z\ bin of SFR. Figure \ref{kashz:fig:sfr_mstar_distros} shows the total posterior distributions of both samples and all the considered redshift ranges. 

\section{ALMA data analysis of \kashz\ AGN}
\label{app:chap:kashz_alma}
We present in this section the ALMA data analysis and some notes about single sources. Table \ref{kashz:tab:almatargets} summarizes the info of all \kashz\ ALMA targets. 
The \kashz\ molecular sample includes also: 
\begin{itemize} 
	\item Four\footnote{\kashz\ ALMA targets that are also in SUPER: cdfs\_419, cdfs\_427, cdfs\_522, cdfs\_614.} AGN that are shared with SUPER that were not presented in \citetalias{circosta2021} because data became available later on; 
    \item The ALMA observations of  cdfs\_278, cdfs\_419 and cdfs\_614 that were recently presented in \citep{yang2022_candels25998};
	\item The ALMA program 2018.1.00251.S (PI: M. Brusa) covering \codueuno\ (and CO(5-4)) of lid\_1639 (XID5395, \citealp{brusa2016}), the analysis of which will be presented in Ricci et al. (2024, in prep.);
	\item A few AGN that are already well known for their gas kinematics and AGN feedback properties: lid\_1565 \citep[XID2028,][]{cresci2015a,brusa2018,scholtz2020,cresci2023,veilleux2023}, cdfs\_587 \citep[GMASS0953,][]{popping2017,talia2018,loiacono2019}, cdfs\_794  \citep[ALESS067.1,][]{chen2017_aless0671,calistrorivera2018}.
\end{itemize}

Our ALMA cube analysis was performed as follows: \textit{i)} we identify the source emission by finding the peak emission in the vicinity of the coordinates of the target in the line velocity-integrated map produced with the channels at $\pm1000 \rm ~ km/s$ around the expected line frequency peak based on the source redshift, with a visual assessment that the identified peak is not associated to random noise; \textit{ii)} we extract the source spectrum from the peak spaxel or, in case of partially-resolved emission, from the aperture that maximizes the signal-to-noise ratio on the moment 0 map and than divided by the beam size in pixels; \textit{iii)} we fit the line with one Gaussian to measure its centroid and FWHM; \textit{iv)} we re-center the cube to the observed frequency and produce the line velocity-integrated (0th order) moment maps on a spectral range corresponding to $\pm3\times$ the line width $\sigma$; \textit{v)} we extract the spectrum from the re-centered cube, re-selecting the spectrum extraction area (or peak pixel) on the new 0th-moment map, and we rebin it to a channel width that samples the FWHM of the line at least 7 times, chosen as best trade-off to also maintain a good signal-to-noise ratio; \textit{vi)} lastly, we fit the line with a single- and a double-Gaussian function, storing the respective parameters and relative errors. Regarding step \textit{v)}, the spectrum is extracted from the 3$\sigma$ mask of the target in the moment 0 map in case this is larger than the corresponding beam, otherwise the spectrum is extracted from the peak pixel of the line velocity-integrated map produced in step \textit{iv)}. We derive the uncertainty of the integrated CO flux from the root-mean square noise (rms) of the line-free range of the spectrum ($\rm rms_{\rm spec}$): $\Delta I_{\rm CO}= \rm rms_{\rm spec} \times \sqrt{FWHM\times \Delta v}$, where $\Delta v$ is the channel width in $\kms$. 
We produced the final cubes by setting a channel width that allows us to sample the line FWHM at least 7 times in the {\tt tclean}, chosen as the best trade-off to sample the line profile while maintaining a good signal-to-noise ratio. We consider a source as detected if its emission peak in the velocity-integrated map is at least 3 times the rms of the map.
\begin{table}[!h]
	\centering
	\caption{Summary of the ALMA observations reduced and analyzed in this work. }
	\label{app:tab:almaobs}
 \resizebox{0.48\textwidth}{!}{
	\begin{tabular}{lccl}
		\hline\hline
		\multicolumn{1}{l}{ID} &
		\multicolumn{1}{c}{Band} &
		\multicolumn{1}{c}{Project code} &
		\multicolumn{1}{l}{MOUS UID} \\
		\hline
		cdfs\_258 & 3 & 2019.1.00678.S & uid://A001/X14d8/X3eb\\
		cdfs\_313 & 3 & 2018.1.00164.S & uid://A001/X133d/X7a8\\
		cdfs\_419 & 3 & 2019.1.00678.S & uid://A001/X14d8/X3ef\\
		cdfs\_427 & 3 & 2013.1.00092.S & uid://A001/X13f/X16\\
		cdfs\_458 & 3 & 2013.1.00092.S & uid://A001/X13f/X16\\
		cdfs\_522 & 4 & 2016.1.00990.S & uid://A001/X87c/X9e8\\
		cdfs\_587 & 3 & 2015.1.00228.S & uid://A001/X2f6/X45e\\
		cdfs\_614 & 3 & 2019.1.00678.S & uid://A001/X14d8/X3eb\\
		cdfs\_794 & 3 & 2013.1.00470.S & uid://A001/X12a/X3c\\
		cid\_108 & 3 & 2016.1.00171.S & uid://A001/X879/X23f\\
		cid\_1286 & 3 & 2016.1.01001.S & uid://A001/X8c4/Xf\\
		cid\_178 & 3 & 2016.1.00171.S & uid://A001/X879/X23b\\
		cid\_499 & 3 & 2016.1.00171.S & uid://A001/X879/X233\\
		cid\_72 & 3 & 2016.1.01001.S & uid://A001/X8c4/X1b\\
		cid\_86 & 4 & 2016.1.00726.S & uid://A001/X885/X249\\
		cid\_864 & 3 & 2016.1.00171.S & uid://A001/X879/X237\\
		lid\_1565 & 3 & 2016.1.00171.S & uid://A001/X879/X237\\
		lid\_1639 & 3 & 2018.1.00251.S & uid://A001/X133d/X316\\
		xuds\_358 & 3 & 2019.1.00337.S & uid://A001/X14d7/X168\\
		xuds\_477 & 3 & 2019.1.00337.S & uid://A001/X14d7/X168\\
		xuds\_481 & 3 & 2019.1.00337.S & uid://A001/X14d7/X168\\
		\hline
	\end{tabular}
 }
\end{table}

Almost all of our CO lines are well fitted by a single Gaussian component based on the F-test, a commonly used statistical test to compare the goodness of two fits through their $\chi^2$ and degrees of freedom. 
For the purposes of our study and having checked that the single- and double-Gaussian models return consistent values, we refer to the results obtained with the single-Gaussian modeling as a measure of the total CO flux of our host galaxies. We report the velocity integrated map (derived in the $\pm3\times$FWHM, with FWHM$=$300 $\kms$) and spectrum of cid\_108 in Fig. \ref{app:fig:ALMA1} as example of CO detections. Maps and spectra of the rest of the sample are available on Zenodo  at the following link: \url{https://doi.org/10.5281/zenodo.13149280}. 

As a sanity check of the robustness of our ALMA analysis, we compared the CO flux derived from the spectral fits with the flux obtained integrating over the spectrum extraction region in the moment 0 maps (either the peak pixel value or the flux in the 3$\sigma$ mask in case this is smaller or larger than the corresponding beam, respectively) and as derived from the 2D fits of the moment 0 maps with the CASA task {\tt imfit}. The CO flux measured from the spectral fits is well consistent with both estimates from the moment 0 maps for all targets, with the exception of the CO emission of cdfs\_794. The value measured with {\tt imfit} is consistent with that obtained by \citet{calistrorivera2018}, where the spectrum was extracted using a circular region of 1\arcsec\ radius, which is underestimated by the flux obtained from the spectrum extracted from the 3$\sigma$ mask. We thus re-extract such spectrum using the same approach as in \citet{calistrorivera2018} and find a \cotredue\ flux that is consistent with the value integrated from the 1\arcsec\ circle, the 2D fit result of {\tt imfit} and that of \citeauthor{calistrorivera2018}. The CO properties of cdfs\_794 presented in this paper are thus those corresponding to the 1\arcsec-circle spectrum extraction region. 

We define non-detections the sources with S/N$\lesssim3$ in the velocity-integrated maps (11 CO lines of 10 \kashz\ AGN). For the non-detections, we assume a FWHM$=300\kms$ \citep[e.g.,][]{damato2020,pensabene2021}, consistent with the median FWHM of the detected targets, and estimate the flux upper limits as $I_{\rm CO}^{\rm ul}=3~{\rm rms_p}~\Delta v\sqrt{n}$, where rms$\rm_p$ is the rms of the peak channel in the cube with channel width of $\Delta v=40\ \kms$ and $n$ is the number of channels sampling the FWHM (i.e., $n=7.5$ for $\Delta v=40\ \kms$ and FWHM of 300 $\kms$). 
Velocity integrated maps derived in the $\pm3$~FWHM (with FWHM$=$300 $\kms$) and spectra of non-detections are available on Zenodo  at the following link: \url{https://doi.org/10.5281/zenodo.13149280}. 

We assessed the continuum emission of our targets with the {\tt tclean} task of CASA 6.4 used in ``mfs'' mode
on line-free channels of all the available spectral windows. Two AGN are detected in continuum (see Table \ref{kashz:tab:almaresults}). We measure the continuum flux from the peak pixel in the continuum map or from the 3$\sigma$ mask if this is more extended than one beam. Regarding continuum non-detections, we report in Table \ref{kashz:tab:almaresults} the 3$\sigma$ upper limit. The photometric points obtained from ALMA Band 3 and 4 continuum are included in the SED fitting of the KASHz molecular sample (see Sect. \ref{app:sec:sedfit}).


\section{SED fitting of \kashz\ and SUPER AGN}
\label{app:sec:sedfit}
We measure the stellar mass $\Mstar$, the FIR luminosity $\Lfir$ of the host galaxies and the AGN bolometric luminosity $\Lbol$ of the \kashz\ ALMA targets by performing SED fitting with CIGALE v.2022.1 \citep{boquien2019_cigale,yang2020_xcigale}, a publicly available, python-based SED-fitting code. We also update the SED fitting of the SUPER AGN presented in \citetalias{circosta2021} that are drawn from the CDFS and COSMOS fields, to take advantage of the new photometry releases from the parent-survey collaborations (see Sect. \ref{sec:sed_2}). 

\subsection{CIGALE SED fitting code}
\label{sec:sed_1}
The CIGALE code adopts a multicomponent fitting approach to separate the AGN contribution from the galaxy emission. It employs model templates that encompass nebular emission, dust-attenuated stellar emission, dust emission driven by star formation, and AGN emission spanning from X-rays to radio wavelengths, including both the primary accretion disk emission and the emission reprocessed by the dusty torus surrounding the central engine.
This is an ``energy-balance'' code, meaning that it takes into account the energy balance between the UV-optical attenuation and the FIR re-emission by dust. The code provides the values of the parameters of interest (as selected by the user) in two ways. The first corresponds to the solution returning the lowest $\chi^2$, and the second to the output of a Bayesian-like analysis, for which the returned value of each parameter corresponds to the mean value of the probability distribution function (PDF) built accounting for all the different models that are used in the fitting process. In this work, we refer to the Bayesian output as the best-fit SED. 
One of the major updates in the code was the inclusion of the X-ray module\footnote{X-CIGALE is merged in CIGALE from CIGALE v.2022.0.}, which allows to set a prior on the $\alpha_{\rm OX}$\footnote{The parameter $\alpha_{\rm OX}=0.3838\times\log(L_{\rm 2keV}/L_{2500\AA}$) corresponds to the slope of a nominal power-law connecting the rest-frame UV and X-ray emission of AGN \citep{tananbaum1979}.} of the AGN by feeding the code with the intrinsic (i.e., absorption corrected) X-ray fluxes.
Since the AGN X-ray emission significantly surpasses that of the host galaxy, such a prior assumption allows to delimit the range of AGN UV luminosity for a chosen scatter with respect to the $\alpha_{\rm OX}$-$L_{2500\AA}$ relation (e.g., \citealp{lusso2016,martocchia2017}), and thus to better decouple the emission of galaxy and AGN in this range. 
We use the same setup of models presented in \citetalias{circosta2018} to reduce the possible sources of systematics between the \kashz\ and SUPER AGN: stellar population models with solar metallicity \citep{bruzual_charlot2003}, a delayed SF history, the modified \citet{calzetti2000} attenuation law and a \citet{chabrier2003} initial mass function, plus the nebular emission following \citet{inoue2011,nagao2011}. We model the contributions from the cold dust associated with star formation and AGN emission with the libraries of \citet{dale2014} and \citet{fritz2006}, with updates from \citet{feltre2012}, respectively.
Thus, the main difference between our SED-fitting setup and that of \citetalias{circosta2021} resides in the inclusion of the X-ray module. 

\subsection{Updates on multi-band photometry}
\label{sec:sed_2}
We collect the Near-UV to FIR photometry referring to the latest Near-UV-to-FIR multi-wavelength catalogs provided by the collaborations of the deep surveys our targets are extracted from: 
\begin{itemize}
	\item \cosmos: We retrieve the Near-UV/optical to MIR photometry from \cosmos2020 \citep[catalog version: classic][]{waever2022_cosmos2020}; we complement the multiwavelength info of our \cosmos\ targets with the ``super-deblended'' FIR to (sub-)$mm$ photometric catalog of \citet{jin2018_cosmos_superdebl}. Two sources of the \kashz\ molecular sample (lid\_1565, cid\_72) are not included in the superdeblended catalog, thus, following \citetalias{circosta2018}, we collect their 24--500 $\mu$m photometry from the previous PACS Evolutionary Probe (PEP) and \textit{Herschel} Multi-tiered Extragalactic Survey (HerMES) DR4 catalogs by \citet{lutz2011} and \citet{hurley2017}, respectively, using a match radius of 2\arcsec. Moreover, cid\_72 has only one match in the COSMOS2020 catalog at a distance of 6\arcsec\ from the coordinates listed in \citet{marchesi2016}, while the best match with the COSMOS2015 source list \citep{laigle2016} falls at less than 0.5\arcsec. We thus collect the Near-UV/optical to MIR photometry of cid\_72 from COSMOS2015; 
	\item \cdfs: we collect the Near-UV to MIR photometry from the ASTRODEEP-GS43 catalog of the CANDELS/GOODS-S field \citep[Cosmic Assembly Near-infrared Deep Extragalactic Legacy Survey/Great Observatories Origins Deep Survey-South,][]{merlin2021_cdfs}. FIR photometry is retrieved from the \textit{Herschel} Extragalactic Legacy Project (HELP) collaboration \citep{shirley2019,shirley2021}. , As done for \cosmos\ targets, we complement the FIR photometry of \cdfs\ AGN that are not included in the HELP catalog (cdfs\_313, cdfs\_522) 
    with previous results from PEP and HerMES\footnote{The HerMES data was accessed through the Herschel Database in Marseille (HeDaM - \url{http://hedam.lam.fr}) operated by CeSAM and hosted by the Laboratoire d'Astrophysique de Marseille.} \citep{oliver2012_hermesFIR,hurley2017}; 
	\item \xuds: we collect the Near-UV to FIR photometry from the ``best photometry'' catalog compiled by the HELP collaboration \citep[N-UV to MIR source catalogs used by the HELP collaboration:][]{almaini2007_uds,furusawa2008_uds,tudorica2017_megacamUDS}. 
\end{itemize}

We include in the photometry of our sources also the ALMA continuum photometric points (or upper limits), from Band 3 to Band 7, when available, either drawn from this work (see Table \ref{kashz:tab:almaresults}) or from \citet{scholtz2018,lamperti2021} and \citetalias{circosta2021}. 
Given the significant update in terms of SED sampling, flux depth and photometry extraction of the available photometry for COSMOS and CDFS targets, we also refit the SED of the SUPER targets in \citetalias{circosta2021} drawn from these two surveys (see Table \ref{kashz:tab:super+kashz}) for consistency. 
We flag as upper limits all the HELP FIR fluxes below the 2$\sigma$ level and feed to CIGALE the corresponding 3$\sigma$ level value as upper limit. We consider those above the 2$\sigma$ limit as reliable based on the implementation of the flux extraction procedure by HELP collaboration, which uses the optical/NIR position of a target as prior to deblend its flux in the FIR bands. The ``Super-deblended'' catalog already flags the upper limits, thus we follow their classification, with the only exception of cid\_1205: the source is placed in a crowded field, thus its FIR photometry is well detected but highly contaminated by a nearby source. We flag all the Superdeblended filters of cid\_1205 as upper limits. 
For the X-ray prior to work, one has to feed CIGALE with the absorption-corrected (i.e., intrinsic) X-ray fluxes. We compute the X-ray photometric points in the 0.5--2 keV and 2--10 keV observed energy range from the rest-frame, absorption-corrected X-ray luminosity using the intrinsic photon index, either obtained from the direct fit of the X-ray spectra from the literature \citep[for CDFS and COSMOS targets,][\citetalias{circosta2018,circosta2021}]{lanzuisi2013,lanzuisi2015,liu2017} or as derived by the X-ray deep survey collaborations in the survey catalogs \citep{luo2017,akiyama2015,kocevski2018,marchesi2016}, as flagged in Table \ref{kashz:tab:almatargets}. Since the survey catalogs do not provide an uncertainty for the intrinsic X-ray luminosity, we assigned a uniform $\pm$30\% uncertainty for each X-ray photometric point in our sample. The mean (and median) uncertainty of the observed fluxes in the COSMOS-Legacy field is $\simeq20\%$ \citep{civano2016}, thus here we include a mean 10\% to account for additional uncertainties due to the absorption correction. 
The photometric bands used for each of the X-ray deep fields are listed in Table \ref{app:tab:phot_summary}. 
We summarize in Table \ref{app:tab:sed_input_parameters} all the model parameters that were set to a value different than the default of X-CIGALE in our SED fitting runs. We note that the chosen X-CIGALE setup tests $\sim290$M SED models for each source. 

\begin{figure}[!h]
\centering
\adjincludegraphics[clip,trim={0 0 {0.1\width} 0},width=0.44\textwidth]{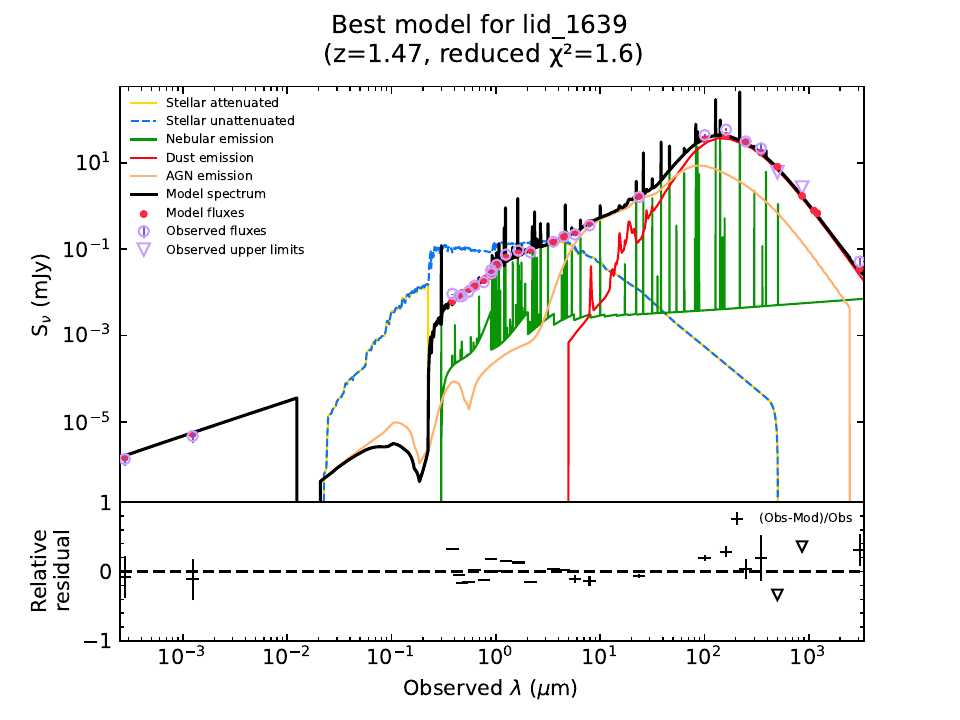}
\caption{Observed-frame, best-fit SED models of lid\_1639. Purple empty circles mark observed flux densities, purple empty triangles the observational upper limits and red filled circles indicate the best-fit model prediction. Yellow (blue dashed) lines are for the stellar (un)attenuated emisson; green lines for the nebular emission; red lines for dust emission; orange lines for the AGN component and black lines for the total, best-fit model. The SEDs of the rest of the AGN sample are available on Zenodo at the following link: \url{https://doi.org/10.5281/zenodo.13149280}. }
\label{app:fig:sed_example}
\end{figure}
\subsection{Highlights and caveats on the results}
\label{sec:sed_3}
The results obtained with the described photometry and CIGALE setup provide adequate fits of the SEDs of our targets, based on the reduced $\chi^2$ value and the overall agreement between the best fit model and the photometry throughout the full X-ray-to-FIR band (see Table \ref{kashz:tab:super+kashz} and Fig. \ref{app:fig:sed_example}\footnote{Fig. \ref{app:fig:sed_example} shows the SED of lid\_1639 as example. The SEDs of the rest of the AGN sample are available on Zenodo at the following link: \url{https://doi.org/10.5281/zenodo.13149280}.}), with the exception of cdfs\_313. 
Moreover, \citet{mountrichas2021} and \citet{buat2021} show that if there is a large discrepancy between the mean of the posterior distribution and the best-fit value of a certain parameter of CIGALE, then the result for that parameter is not reliable. We checked for such a consistency in $\Mstar$, SFR and $L_{\rm AGN}$ (i.e., $\Lbol$). All of our results match the acceptance range defined by the authors (that is, the ratio of the Bayesian and the best fit value falls in the 0.5--5 range). 
We managed to obtain a good fit of the SED of cdfs\_313 by removing the X-ray part of the SED: it is possible that the X-ray flux, and thus the link set by CIGALE with the UV emission of this target, was not representative of the AGN accretion, for instance due to the low number of counts in the CDFS spectrum ($25\pm5$ cts, \citealp{luo2017}) or to faster source variability in the X-rays compared to the rest of the SED. 

We compared the results for SUPER AGN to those of \citetalias{circosta2018} and \citetalias{circosta2021}. For most of the targets, the main parameters of interest for our analysis, which are $\Mstar$, $L_{\rm FIR}$ and $L_{\rm AGN}$, are consistent with those of \citetalias{circosta2018} and \citetalias{circosta2021} within the typical uncertainties derived by \citet{pacifici2023} (0.1dex in stellar mass and 0.3dex in SFR, hence $L_{\rm FIR}$ for our purposes). Few targets present a larger difference with respect to the SED fitting of \citetalias{circosta2018} and \citetalias{circosta2021} but this can be explained as due to the updated SED fitting code (e.g., the inclusion of the X-ray module) and the update in the photometry, especially in the FIR band. 

We checked the agreement between the best fit SED of lid\_1639 and the ALMA photometry in Band 6 (Ricci et al., 2024 in prep). The redward side of the FIR bump is well consistent with the flux in ALMA Band 6, however, given the good sampling of the FIR band for this source, we decided to exclude ALMA Band 6 from its filter set because of the high resolution of the observation  ($\sim0.6$\arcsec), which might underestimate the target flux by missing the emission at larger scales. A different case is instead cid\_1205: since we have considered all of its FIR data as upper limits, we decided to include the ALMA Band 7 flux of \citet{lamperti2021} despite the even higher resolution of the corresponding observation ($\sim0.2$\arcsec). We consider this as the best choice to not overestimate the FIR bump of this system, also supported by the fact that \citeauthor{lamperti2021} conclude that their ALMA Band 6 fluxes mostly sample emission due to dust heated by SF. 

Lastly, for some of the very near-UV-bright Broad Line AGN in our sample, we can only constrain an upper limit to the stellar mass. Since the optical/NIR galaxy emission is outshined by the AGN, there is a strong degeneracy between the AGN component and the stellar component which can unfortunately result in stellar masses constrained only as upper limits, which we estimate at 3$\sigma$. 

\onecolumn

\section{Tables}
\begin{table*}[!h]
	\caption{Summary of \kashz\ ALMA targets}
	\label{kashz:tab:almatargets}
	\centering
			\begin{tabular}{lcccccccccc}
				\hline\hline
				ID        & X-ray field   & RA    & DEC       & Type & \z\        & $\Gamma$  &  $ \log N_{\rm H}$  & $\log L_{2-10}$  &  flag\_Lx  & CO transition\\
                                &               & (deg)  &  (deg)  &        &           &           &  ($\rm cm^{-2}$)  & (\lumcgs)     &           &           \\
                     (1)           &   (2)            & (3)  &  (4)  &   (5)     &   (6)        &   (7)        &  (8)  & (9)     &  (10)         &    (11)       \\
                     \hline    
				cdfs\_258   & CDFS          & 53.060 & -27.852 & 2      & 1.542       & 1.8    &   24.2  & 43.5  &  5       & CO(2-1)   \\    
				cdfs\_313   & CDFS          & 53.072 & -27.834 & 0      & 1.612$^c$   & 3.0    &   $<$20    & 41.6  &  1        & CO(2-1)  \\    
				cdfs\_419$^a$   & CDFS          & 53.097 & -27.715 & 2      & 2.142       & 1.8    &   24.0  & 43.6  &  5       & CO(3-2)   \\    
				cdfs\_427$^a$   & CDFS          & 53.100 & -27.715 & 2      & 2.302       & 1.8    &   22.5  & 43.3  &  5       & CO(3-2) \\ 
				cdfs\_458   & CDFS          & 53.107 & -27.718 & 2      & 2.297       & 1.77   &   23.1  & 44.2  &  5       & CO(3-2)   \\    
				cdfs\_522$^a$   & CDFS          & 53.118 & -27.782 & 2      & 2.308       & 1.92   &   $<$20   & 42.1   &  1       & CO(4-3)   \\    
				cdfs\_587   & CDFS          & 53.131 & -27.773 & 2      & 2.225       & 1.8    &   23.2 & 44.8  &  1        & CO(3-2) \\
				cdfs\_614$^a$   & CDFS          & 53.137 & -27.700 & 2      & 2.453       & 1.8    &   24.1  & 43.9  &  5       & CO(3-2) \\
				cdfs\_794   & CDFS          & 53.179 & -27.920 & 2      & 2.122       & 2.32   &   $<$20   & 42.6  &  1        & CO(3-2)  \\    
		          cid\_72     & COSMOS-L       & 150.09 & 2.39907 & 1      & 2.472       & 1.8    &   $<$20   & 44.6  &  1        & CO(3-2)  \\    
				cid\_86     & COSMOS-L       & 150.11 & 2.29590 & 2      & 2.099       & 1.8    &   $<$20   & 44.6  &  2        & CO(4-3)  \\    	
                    cid\_108    & COSMOS-L       & 150.05 & 2.47742 & 1      & 1.258       & 1.8    &   22.5 & 44.0  &  1        & CO(2-1)  \\    
				cid\_178    & COSMOS-L       & 149.58 & 2.05112 & 1      & 1.351       & 1.8    &   22.7 & 44.0  &  1        & CO(2-1)  \\    
				cid\_346$^b$    & COSMOS-L   & 149.93 & 2.11873 & 1      & 2.218       & 1.74   &   23.0 & 44.5  &  4        & \cotredue\ \\    
				cid\_357$^b$    & COSMOS-L   & 149.99 & 2.13197 & 1      & 2.132       & 1.88   &   $<$20   & 44.4  &  4        & \cotredue\ \\    
				cid\_451$^b$    & COSMOS-L   & 150.00 & 2.25862 & 2      & 2.442       & 1.8    &   23.9 & 45.2  &  4        & \cotredue\ \\    
				cid\_499    & COSMOS-L       & 149.91 & 2.32746 & 1      & 1.455       & 1.8    &   $<$20   & 44.7  &  1        & CO(2-1)  \\    
				cid\_864    & COSMOS-L       & 149.88 & 2.31819 & 2      & 1.617       & 1.8    &   $<$20   & 43.4  &  1        & CO(2-1)  \\    
				cid\_970$^b$    & COSMOS-L   & 150.23 & 2.36176 & 1      & 2.506       & 1.74   &   $<$20   & 44.7  &  4        & \cotredue\ \\    
				cid\_971$^b$    & COSMOS-L   & 150.24 & 2.33262 & 2      & 2.469       & 1.8    &   $<$20   & 43.9  &  4        & \cotredue\ \\    
                    cid\_1205$^b$   & COSMOS-L   & 150.01 & 2.33296 & 2      & 2.258       & 1.8    &   23.5  & 44.2  &  4        & \cotredue\ \\    
				cid\_1215$^b$   & COSMOS-L   & 150.06 & 2.32905 & 1      & 2.447       & 1.8    &   22.9 & 44.3  &  4        & \cotredue\ \\    
				cid\_1286   & COSMOS-L       & 150.14 & 2.26507 & 2      & 2.199       & 1.8    &   23.0 & 43.7  &  1        & CO(3-2)  \\    
				lid\_1565   & COSMOS-L       & 150.54 & 1.61846 & 1      & 1.594       & 1.8    &   $<$20   & 44.9  &  1       & CO(2-1)  \\    
				lid\_1639   & COSMOS-L       & 150.74 & 2.17052 & 1      & 1.472$^d$   & 1.8    &   23.0 & 44.5  &  1       & CO(2-1)  \\    
				xuds\_358   & \xuds\           & 34.322 & -5.2300 & 2      & 2.182       & 1.7    &   20.1  & 42.9  &  2      & CO(3-2)   \\    
				xuds\_477   & \xuds\           & 34.510 & -5.0091 & 0      & 1.086       & 1.7    &   19.9  & 42.6  &  2      & CO(2-1)   \\    
				xuds\_481   & \xuds\           & 34.657 & -4.9806 & 1      & 1.407       & 1.7    &   22.2 & 44.1  &  1       & CO(2-1)  \\           
				\hline
			\end{tabular} 
		\tablefoot{$^a$ Target shared with the SUPER survey, ALMA data analysis is presented in this work. $^b$ Target shared with the SUPER survey, ALMA data analysis is presented in \citetalias{circosta2021}. $^c$ Redshift from \citet{luo2017}. $^d$ Redshift from \citet{marchesi2016}. The Type column reports the AGN classification based on the \kashz\ spectral fits: Type 1, Broad Line AGN; Type 2, Narrow Line AGN; Type 0: undetected in NIR spectra. The flag\_Lx column expresses how we derived the X-ray properties of the targets: 
		flag\_Lx=1, values as in the parent survey catalogs; 
		flag\_Lx=2, redshift of the source is updated based on \kashz\ NIR spectral fitting, $N_{\rm H}$ and photon index of parent survey catalogs are corrected according to the new spectroscopic redshift estimate; 
		flag\_Lx=3, values are retrieved from spectral fit, this work;
		flag\_Lx=4, values are retrieved from spectral fit, \citetalias{circosta2021};
		flag\_Lx=5, values are retrieved from spectral fit, \citet{liu2017};
		flag\_Lx=6, values are retrieved from spectral fit \citet{liu2017} and corrected according to the new spectroscopic redshift estimated from \kashz\ NIR spectra; 
		flag\_Lx=7, values are retrieved from spectral fit, \citet{lanzuisi2013,lanzuisi2015} and corrected according to the new spectroscopic redshift estimated from \kashz\ NIR spectra. }
\end{table*}

\renewcommand{\arraystretch}{1.1}
\begin{table*}[!p]
	\caption{Summary of the properties of the AGN sample. }
	\label{kashz:tab:super+kashz}
	\centering
		\resizebox{\textwidth}{!}{
			\begin{tabular}{lcccccccc} 
				\hline\hline
				ID             &  \z\      &  $\log(\Lbol/\Lumcgs)$                & $\log(\Lfir/\Lumcgs)$                & $\log(\Mstar/\Msun)$             &  $\log($\lpco /K km s$^{-1}$ pc$^2$)       & SFR ($\Msun\rm~yr^{-1}$)       & sample  \\
                      (1)       & (2)        &  (3)                          &  (4)                            &   (5)                   &(6)                        & (7)               & (8)           \\
				\hline
                    J1333+1649$^*$       & 2.089 &   $ 47.91 \pm 0.02 $   &   $        -        $   &   $        -      $  &   $ 10.07 \pm  0.14  $   &   $        -         $  &  2  \\
                    X\_N\_102\_35$^*$ & 2.19  &   $ 46.82 \pm 0.02 $   &   $        -        $   &   $        -        $   &   $ <9.94            $   &   $        -         $  &  2  \\
                    X\_N\_104\_25$^*$ & 2.241 &   $ 45.97 \pm 0.4  $   &   $        -        $   &   $        -        $   &   $ <9.97            $   &   $        -         $  &  2  \\
                    X\_N\_128\_48$^*$ & 2.323 &   $ 45.81 \pm 0.4  $   &   $        -        $   &   $        -        $   &   $ <10.01           $   &   $        -         $  &  2  \\
                    X\_N\_44\_64 & 2.252 &   $ 45.51 \pm 0.07 $   &   $ 45.93  \pm 0.15  $   &   $ 11.09  \pm 0.25  $   &   $ 10.16   \pm 0.14 $      &   $ 226     \pm 78   $  &  2  \\
                    X\_N\_53\_3$^*$ & 2.434 &   $ 46.21 \pm 0.03 $   &   $ 46.41  \pm 0.11  $   &   $        -        $   &   $ 10.05   \pm 0.14 $    &   $ 681     \pm 173  $  &  2  \\
                    X\_N\_6\_27$^*$ & 2.263 &   $ 45.85 \pm 0.05 $   &   $ <45.9            $   &   $        -        $   &   $ 9.56   \pm  0.14 $    &   $<210              $  &  2  \\
                    X\_N\_81\_44 & 2.311 &   $ 46.8  \pm 0.03 $   &   $ 45.93  \pm 0.2   $   &   $ 11.04  \pm 0.37  $   &   $ 9.68    \pm 0.14 $      &   $ 226     \pm 104  $  &  2  \\
                    cdfs\_258   & 1.54  &   $ 44.61 \pm 0.35 $   &   $ 45.26  \pm 0.02  $   &   $ 10.80  \pm 0.06  $   &   $ <10.01           $       &   $ 48     \pm  2    $  &  1  \\
                    cdfs\_313   & 1.611 &   $ 45.32 \pm 0.11 $   &   $ 45.90  \pm 0.02  $   &   $ 10.70  \pm 0.08  $   &   $ 10.57   \pm 0.11 $       &   $ 208     \pm 11   $  &  1  \\
                    cdfs\_419   & 2.142 &   $ 45.63 \pm 0.13 $   &   $ <44.32           $   &   $ 11.00 \pm  0.02  $   &   $ <10.01           $       &   $<6                $  &  3  \\
                    cdfs\_427   & 2.302 &   $ 44.59 \pm 0.19 $   &   $ <45.28           $   &   $ 10.89 \pm  0.08  $   &   $ <9.91            $       &   $<51               $  &  3  \\
                    cdfs\_458   & 2.297 &   $ 46.00 \pm 0.05 $   &   $ 45.04  \pm 0.05  $   &   $ 11.17  \pm 0.05  $   &   $ 10.05   \pm 0.14 $       &   $ 29      \pm 3    $  &  1  \\
                    cdfs\_522   & 2.309 &   $ 45.08 \pm 0.02 $   &   $ 46.17  \pm 0.02  $   &   $ 10.30  \pm 0.02  $   &   $ 10.63   \pm 0.13 $       &   $ 390     \pm 19   $  &  3  \\
                    cdfs\_587   & 2.225 &   $ 45.86 \pm 0.11 $   &   $ 45.77  \pm 0.02  $   &   $ 11.19  \pm 0.06  $   &   $ 10.62   \pm 0.14 $       &   $ 156     \pm 8    $  &  1  \\
                    cdfs\_614   & 2.453 &   $ 45.03 \pm 0.02 $   &   $ 46.12  \pm 0.03  $   &   $ 10.28  \pm 0.15  $   &   $ 10.36   \pm 0.14 $       &   $ 347     \pm 25   $  &  3  \\
                    cdfs\_794   & 2.122 &   $ 44.83 \pm 0.02 $   &   $ 46.07  \pm 0.02  $   &   $ 10.89  \pm 0.10  $   &   $ 10.96   \pm 0.14 $       &   $ 312     \pm 16   $  &  1  \\
                    cid\_108    & 1.258 &   $ 45.78 \pm 0.02 $   &   $ 46.00  \pm 0.02  $   &   $ 11.41  \pm 0.19  $   &   $ 10.49   \pm 0.11 $       &   $ 266     \pm 14   $  &  1  \\
                    cid\_1205   & 2.255 &   $ 45.85 \pm 0.16 $   &   $ <45.09           $   &   $ 11.22 \pm  0.18  $   &   $ <9.93            $       &   $<32               $  &  3  \\
                    cid\_1215   & 2.45  &   $ 45.83 \pm 0.05 $   &   $ 46.06  \pm 0.06  $   &   $ 10.30  \pm 0.23  $   &   $ 10.52   \pm 0.14 $       &   $ 303     \pm 41   $  &  3  \\
                    cid\_1253   & 2.147 &   $ 44.83 \pm 0.13 $   &   $ 45.64  \pm 0.06  $   &   $ 11.15  \pm 0.07  $   &   $ 11.03   \pm 0.14 $       &   $ 116     \pm 15   $  &  2  \\
                    cid\_1286   & 2.199 &   $ 45.55 \pm 0.05 $   &   $ <45.10           $   &   $ 10.57 \pm  0.18  $   &   $ <10.42           $       &   $<33               $  &  1  \\
                    cid\_1605   & 2.121 &   $ 46.01 \pm 0.10 $   &   $ <45.56           $   &   $ <10.94           $   &   $ <9.61            $       &   $<96               $  &  2  \\
                    cid\_166    & 2.448 &   $ 46.87 \pm 0.02 $   &   $ 45.38  \pm 0.21  $   &   $ <11.75           $   &   $ 10.37  \pm  0.14 $       &   $ 64     \pm  32   $  &  2  \\
                    cid\_178    & 1.356 &   $ 46.01 \pm 0.08 $   &   $ 45.36  \pm 0.16  $   &   $ <11.53           $   &   $ <10.02           $       &   $ 61    \pm   22   $  &  1  \\
                    cid\_247    & 2.412 &   $ 45.43 \pm 0.10 $   &   $ <45.59           $   &   $ 10.91 \pm  0.15  $   &   $ <9.64            $       &   $<103              $  &  2  \\
                    cid\_2682   & 2.435 &   $ 45.76 \pm 0.10 $   &   $ <44.64           $   &   $ 11.09 \pm  0.03  $   &   $ <9.95            $       &   $<11               $  &  2  \\
                    cid\_337    & 2.226 &   $ 45.39 \pm 0.12 $   &   $ <45.05           $   &   $ 10.94 \pm  0.11  $   &   $ <9.88            $       &   $<30               $  &  2  \\
                    cid\_346    & 2.219 &   $ 46.62 \pm 0.02 $   &   $ 46.16  \pm 0.04  $   &   $ <11.13           $   &   $ 10.48  \pm  0.14 $       &   $ 382    \pm  37   $  &  3  \\
                    cid\_357    & 2.136 &   $ 45.11 \pm 0.17 $   &   $ <45.66           $   &   $ 10.31 \pm  0.11  $   &   $ <9.58            $       &   $<122              $  &  3  \\
                    cid\_38     & 2.192 &   $ 45.91 \pm 0.05 $   &   $ <45.11           $   &   $ 11.15 \pm  0.13  $   &   $ <9.98            $       &   $<34               $  &  2  \\
                    cid\_451    & 2.45  &   $ 46.30 \pm 0.07 $   &   $ <44.51           $   &   $ 11.17 \pm  0.04  $   &   $ 9.9    \pm  0.14 $       &   $<9                $  &  3  \\
                    cid\_467    & 2.288 &   $ 46.51 \pm 0.07 $   &   $ <45.65           $   &   $ <11.11           $   &   $ <9.9             $       &   $<119              $  &  2  \\
                    cid\_499    & 1.457 &   $ 45.87 \pm 0.04 $   &   $ 45.58  \pm 0.02  $   &   $ 11.12  \pm 0.06  $   &   $ <10.46           $       &   $ 101    \pm  5    $  &  1  \\
                    cid\_72     & 2.473 &   $ 45.52 \pm 0.04 $   &   $ <45.46           $   &   $ 10.28 \pm  0.08  $   &   $ <10.48           $       &   $<76               $  &  1  \\
                    cid\_852    & 2.232 &   $ 45.71 \pm 0.06 $   &   $ <45.26           $   &   $ 11.16 \pm  0.02  $   &   $ <10.0            $       &   $<48               $  &  2  \\
                    cid\_86     & 2.097 &   $ 45.65 \pm 0.05 $   &   $ 44.80  \pm 0.12  $   &   $ 11.24  \pm 0.04  $   &   $ 9.45    \pm 0.13 $       &   $ 17      \pm 4    $  &  1  \\
                    cid\_864    & 1.617 &   $ 44.73 \pm 0.13 $   &   $ 45.89  \pm 0.02  $   &   $ 10.62  \pm 0.06  $   &   $ 10.45   \pm 0.11 $       &   $ 207     \pm 10   $  &  1  \\
                    cid\_970    & 2.501 &   $ 45.69 \pm 0.03 $   &   $ <45.17           $   &   $ 10.54 \pm  0.11  $   &   $ <9.65            $       &   $<39               $  &  3  \\
                    cid\_971    & 2.473 &   $ 44.72 \pm 0.23 $   &   $ <45.55           $   &   $ 10.60 \pm  0.15  $   &   $ 9.78   \pm  0.14 $       &   $<94               $  &  3  \\
                    lid\_1289   & 2.408 &   $ 45.33 \pm 0.06 $   &   $ <45.54           $   &   $ 10.37 \pm  0.34  $   &   $ <9.78            $       &   $<91               $  &  2  \\
                    lid\_1565   & 1.593 &   $ 46.15 \pm 0.05 $   &   $ 45.72  \pm 0.05  $   &   $ 11.80  \pm 0.03  $   &   $ 10.24   \pm 0.11 $       &   $ 140     \pm 17   $  &  1  \\
                    lid\_1639   & 1.472 &   $ 46.05 \pm 0.13 $   &   $ 46.23  \pm 0.02  $   &   $ 11.37  \pm 0.10  $   &   $ 10.67   \pm 0.11 $       &   $ 454     \pm 23   $  &  1  \\
                    lid\_1852   & 2.444 &   $ 45.40 \pm 0.08 $   &   $ <44.32           $   &   $ 10.15 \pm  0.02  $   &   $ <9.63            $       &   $<6                $  &  2  \\
                    lid\_206    & 2.33  &   $ 44.71 \pm 0.09 $   &   $ <45.60           $   &   $ 10.38 \pm  0.19  $   &   $ 9.69   \pm  0.14 $       &   $<104              $  &  2  \\
                    lid\_3456   & 2.146 &   $ 44.69 \pm 0.08 $   &   $ <45.43           $   &   $ 11.41 \pm  0.03  $   &   $ <10.03           $       &   $<72               $  &  2  \\
                    xuds\_358   & 2.182 &   $ 44.38 \pm 0.09 $   &   $ <44.78           $   &   $ 11.04 \pm  0.05  $   &   $ <10.69           $       &   $<16               $  &  1  \\
                    xuds\_477   & 1.087 &   $ 44.35 \pm 0.02 $   &   $ <44.44           $   &   $ 10.38 \pm  0.07  $   &   $ <10.32           $       &   $<7                $  &  1  \\
                    xuds\_481   & 1.406 &   $ 45.54 \pm 0.05 $   &   $ 45.86  \pm 0.03  $   &   $ 10.83  \pm 0.23  $   &   $ <10.67           $       &   $ 190    \pm  12   $  &  1  \\
				\hline	
			\end{tabular}
	}
	
	\tablefoot{Columns: (1) Source ID; (2) redshift; (3) bolometric luminosity; (4) FIR luminosity (8-1000 $\mu$m, star-formation only); (5) stellar mass; (6) CO luminosity; (7) SFR derived from $\Lfir$; (8) sample flag. 
 Bolometric luminosity, FIR luminosity and stellar mass are derived from SED fitting (see Sect. \ref{app:sec:sedfit}). SFRs are obtained from the FIR luminosity applying the \citet{kennicutt1998} relation corrected for a \citet{chabrier2003} IMF (i.e., reduced by 0.23 dex). CO luminosity \lpco\ is derived from Eq. \ref{kashz:eq:lco}. Errors are given at 1$\sigma$. The sample flag indicates the parent sample of the targets: 1 for \kashz\ only, 2 for SUPER only, 3 for targets shared by the two surveys. The CO luminosity of SUPER targets (flag=2 and 3) is taken from \citetalias{circosta2021}. The CO luminosity of SUPER targets shared with \kashz\ (flag=3) and drawn from CDFS are from this work. The SED fitting parameters of SUPER targets (flag=2 and 3) drawn from CDFS and COSMOS-Legacy are derived in this work (Sect. \ref{app:sec:sedfit}), those of the other SUPER AGN are taken from \citetalias{circosta2018} and \citetalias{circosta2021}. $^*$: Bright Broad Line AGN that are not considered in the analysis of Sect. \ref{kashz:sec:results} because missing one or more key parameters.}
	
\end{table*}
\renewcommand{\arraystretch}{1}

\begin{sidewaystable*}[p]
	\caption{Summary of results from the ALMA data analysis of \kashz\ AGN analyzed in this work}
	\label{kashz:tab:almaresults}
	\centering
		\resizebox{\textheight}{!}{
			\begin{tabular}{lccccccccccccc}
				\hline\hline   
                                name      & line        &\z\        & beam$_{\rm cube}$ &z$_{\rm CO}$  &$\Delta v$ & rms$_{\rm cube}$ & rms$_{\rm mom0}$ & FWHM      & $I_{\rm line}$      & $\log L'_{\rm CO(1-0)}$           & beam$_{\rm cont}$     & rms$_{\rm cont}$      & $S_{\rm cont}$   \\
								&            &           &  (arcsec$^2$)   &              &($\kms$)  & (Jy/beam)        & (Jy/beam $\kms$) & ($\kms$)         & (Jy $\kms$)          & ($\rm K \kms pc^2$)        & (arcsec$^2$)    & ($\mu$Jy)   & ($\mu$Jy)  \\
		                     (1)       & (2)        &  (3)      &  (4)              &   (5)       &(6)      & (7)               & (8)           & (9)           & (10)              & (11)                 & (12)               & (13)  & (14) \\
				\hline 
				J1333+1649 & CO(3-2)    & 2.089     & 2.79$\times$1.58 &    2.102     &  54     &      0.267       &     0.060      & $281\pm  57 $ & $0.281\pm 0.046$  & $  10.07 \pm 0.14 $ &           --         & --  &    $    -    $    \\
				cdfs\_258  & \codueuno\ & 1.540     & 2.38$\times$2.10 &     1.540    &   50    &      0.214       &      0.095     & $300       $  & $<0.213         $ & $  < 10.00        $ & 2.24 $\times$ 1.97   &  21  & $<67         $  \\
				cdfs\_313  & \codueuno\ & 1.611     & 4.22$\times$2.73  &    1.615     &   40    &      0.208       &     0.053      & $324\pm  26 $ & $0.714\pm 0.027$ & $  10.57 \pm 0.11 $ & 3.98 $\times$ 2.57   &  10  & $<36         $  \\
				cdfs\_419  & CO(3-2)    & 2.142     & 2.17$\times$1.82 &    2.142     &   40    &      0.330       &     0.092      & $300       $  & $<0.234         $ & $  < 10.01        $ & 2.11 $\times$ 1.80   &  27  & $<98         $  \\
				cdfs\_427  & CO(3-2)    & 2.302     & 0.81$\times$0.60 &    2.302     &   40    &      0.160       &    0.047       & $300       $  & $<0.162         $ & $  < 9.91         $ & 0.81 $\times$  0.60  &  10  & $<29         $ \\
				cdfs\_458  & CO(3-2)    & 2.301     & 0.81$\times$0.60 &     2.300    &   70    &      0.125       &     0.047      & $533\pm  99 $ & $0.224\pm 0.026$  & $  10.05 \pm 0.14 $ & 0.81 $\times$  0.60  &  10  & $<29        $  \\
				cdfs\_522  & CO(4-3)    & 2.308     & 0.20$\times$0.16 &    2.309     &   50    &      0.125       &     0.029      & $397\pm  25 $ & $0.943\pm 0.033$  & $  10.63 \pm 0.13 $ & 0.19 $\times$ 0.16   &  10  &  $222\pm 17$  \\
				cdfs\_587  & CO(3-2)    & 2.225     & 3.38$\times$2.14 &    2.225     &   100   &      0.166       &     0.081      & $834\pm  80 $ & $0.883\pm 0.050$  & $  10.62 \pm 0.14 $ & 2.96 $\times$ 1.98   &  18  & $<39         $ \\
				cdfs\_614  & CO(3-2)    & 2.453     & 2.17$\times$1.89 &    2.452     &   60    &      0.198       &     0.066      & $502\pm  88 $ & $0.412\pm 0.036$  & $  10.35 \pm 0.14 $ & 2.30 $\times$ 2.01   &  26  & $<81         $  \\
				cdfs\_794  & CO(3-2)    & 2.122     & 0.58$\times$0.50 &    2.121     &   220   &      0.203       &     0.110      & $695\pm  79$ & $2.086\pm 0.323$  & $  10.96 \pm 0.14 $ & 0.58 $\times$ 0.55   &  20  & $<59         $  \\
				cid\_108  & \codueuno\  & 1.258     & 1.78$\times$1.43 &    1.258     &   40    &      0.773       &     0.129      & $267\pm  39 $ & $0.947\pm 0.070$  & $  10.49 \pm 0.11 $ & 1.95 $\times$ 1.53   &  33  & $<155         $ \\
				cid\_1286  & CO(3-2)    & 2.199     & 0.37$\times$0.24 &    2.199     &   40    &      0.530       &    0.150       & $300       $  & $<0.567         $ & $  < 10.42        $ & 0.40 $\times$ 0.26   &  21  & $<69         $  \\
				cid\_178  & \codueuno\  & 1.356     & 2.43$\times$1.71 &    1.356     &   40    &      0.400       &  0.122         & $300       $  & $<0.283         $ & $  < 10.02        $ & 2.18 $\times$ 1.55   &  25  & $<54         $  \\
				cid\_499  & \codueuno\  & 1.457     & 1.59$\times$1.34 &    1.457     &   40    &      0.860       &  0.207         & $300       $  & $<0.675         $ & $  < 10.46        $ & 1.49 $\times$ 1.27   &  38  & $<116         $ \\
				cid\_72   & CO(3-2)     & 2.471     & 0.40$\times$0.26 &    2.471     &   40    &      0.595       &   0.154        & $300       $  & $<0.54          $ & $  < 10.48        $ & 0.42 $\times$ 0.28   &  21  & $<59         $ \\
				cid\_86   & CO(4-3)     & 2.096     & 0.44$\times$0.38 &    2.097     &31.5$^a$ &      0.078       &     0.017      & $235\pm  37 $ & $0.074\pm 0.006$  & $  9.45 \pm 0.13  $ & 0.42 $\times$ 0.36   &  74  & $<23         $  \\
				cid\_864  & \codueuno\  & 1.617     & 1.60$\times$1.40 &    1.616     &   30    &      0.785       &     0.125      & $242\pm  51 $ & $0.539\pm 0.069$  & $  10.45 \pm 0.11 $ & 1.51 $\times$ 1.33   &  43  & $<123         $ \\
				lid\_1565  & \codueuno\ & 1.593     & 1.61$\times$1.39 &    1.594     &   30    &      0.840       &     0.100      & $154\pm  46 $ & $0.344\pm 0.068$  & $  10.24 \pm 0.11 $ & 1.52 $\times$ 1.32   &  40  & $<107         $ \\
				lid\_1639  & \codueuno\ & 1.472     & 1.98$\times$1.23 &    1.471     &   60    &      0.186       &      0.053     & $603\pm  49 $ & $1.065\pm 0.064$  & $  10.67 \pm 0.11 $ & 1.87 $\times$ 1.15   &  11  &  $52\pm 12$   \\
				xuds\_358  & CO(3-2)    & 2.183     & 1.40$\times$0.89 &    2.183     &   40    &      1.200       &   0.337        & $300       $  & $<1.08          $ & $  < 10.69        $ & 1.48 $\times$ 0.95   &  53  & $<161         $ \\
				xuds\_477  & \codueuno\ & 1.087     & 1.34$\times$0.88 &    1.087     &   40    &      1.080       &   0.410        & $300       $  & $<0.855         $ & $  < 10.32        $ & 1.46 $\times$ 0.95   &  50  & $<125         $ \\
				xuds\_481  & \codueuno\ & 1.407     & 1.50$\times$1.01 &    1.407     &   40    &      1.330       &   0.500        & $300       $  & $<1.17          $ & $  < 10.67        $ & 1.43 $\times$ 0.95   &  48  & $<147         $ \\ \hline

   \end{tabular}
	}
		\tablefoot{Channel width ($\Delta v$) is given in $\kms$, rms$_{\rm cube}$ in units of Jy/beam and rms$_{\rm mom0}$ in units of Jy/beam $\kms$, FWHM in units of $\kms$, CO flux in units of Jy $\kms$, continuum flux in units of $\mu$Jy. Line ratios $r_{j1}$ were retrieved from of \citet{kirkpatrick2019}, along with their uncertainties: $r_{41}=0.37\pm0.11$, $r_{31}=0.59\pm0.18$, $r_{21}=0.68\pm0.17$. $^a$ Native channel-width of the ALMA observation. }

\end{sidewaystable*}

\begin{table*}[h]
\caption{Summary of the photometric data used for the SED-fitting modeling described in Appendix \ref{app:sec:sedfit}.} 
\label{app:tab:phot_summary} 
\centering
\resizebox{\textwidth}{!}{
	\begin{tabular}{ccccc}
		\hline\hline
		Field & $\lambda$ range & Reference & Telescope/Instrument & Bands \\ 
		\hline
		CDF-S & UV to NIR & \citet{merlin2021_cdfs} & CTIO-Blanco/Mosaic-Iwe & \textit{U} \\ 
		&               &                       & VLT/VIMOS & \textit{U, B, R}\\ 
		&               &                       & HST/ACS & F435W, F606W, F775W, F814W, F850LP \\
		&               &                       & HST/WFC3 & F098M, F105W, F125W, F160W  \\
		&               &                       & VLT/HAWK-we & \textit{K$_{S}$} \\
		& $3-500 $    $\mu$m          &  \citet{shirley2021}  & \textit{Spitzer}/IRAC & 3.6, 4.5, 5.8, 8.0 $\mu$m \\
		& & & \textit{Spitzer}/MIPS & 24 $\mu$m \\
		&                &                      & \textit{Herschel}/PACS & 70, 100, 160 $\mu$m \\
		&  &                    & \textit{Herschel}/SPIRE & 250, 350, 500 $\mu$m \\
		& $>1000$ $\mu$m & \citet{scholtz2018}, & ALMA & Band 7 (800$-$1100 $\mu$m) \\
		&				 & \citet{gonzalezlopez2020_aspecs1mm},           &       & Band 6 (1100–-1400 $\mu$m) \\
		&				 &  \citet{aravena2020}                     &       & Band 5 (1400–-1800 $\mu$m) \\
		&				 &  and this work                    &       & Band 4 (1800–-2400 $\mu$m) \\
		&				 &                       &      & Band 3 (2600$-$3600 $\mu$m) \\
		\hline
		COSMOS & UV to MIR & \citet{waever2022_cosmos2020}        &  CFHT/MegaCam & \textit{u, u$^{\ast}$} \\
		&               &                       & HST/ACS &  F814W \\
		&                &                             & Subaru/HSC & \textit{g, r, i, z, y} \\
		&                &                            & Subaru/Suprime-Cam & \textit{B, g$^{+}$, V, r$^{+}$, i$^{+}$, z$^{+}$, z$^{++}$} \\
		
		&                &                             & VISTA/VIRCAM & \textit{Y, J, H, K$_{s}$} \\
		&         &    & \textit{Spitzer}/IRAC & 3.6, 4.5, 5.8, 8.0 $\mu$m \\
		& $24-500$ $\mu$m $^a$ &     \citet{jin2018_cosmos_superdebl}   & \textit{Spitzer}/MIPS & 24 $\mu$m \\
		&                  &                             & \textit{Herschel}/PACS & 70, 100, 160 $\mu$m \\
		&  &                            & \textit{Herschel}/SPIRE & 250, 350, 500 $\mu$m \\
		& $>1000$ $\mu$m & \citet{scholtz2018}, & ALMA & Band 7 (800$-$1100 $\mu$m) \\
		&				 & \citet{circosta2021}           &       & Band 6 (1100–-1400 $\mu$m) \\
		&				 & and this work                     &       & Band 4 (1800–-2400 $\mu$m) \\
		&				 &                       &      & Band 3 (2600$-$3600 $\mu$m) \\
		\hline
		X-UDS         &  UV to FIR & \citet{shirley2021} & CFHT/MegaCam & \textit{u$^{\ast}$} \\
		& & & Subaru/Suprime & \textit{r} \\
		&                &               & Subaru/HSC     & \textit{g, i, z, y} \\
		&                &               & UKIDSS & \textit{J, H, K} \\
		&                &               & VISTA/VIRCAM & \textit{Ks} \\
		&          &    & \textit{Spitzer}/IRAC & 3.6, 4.5, 5.8, 8.0 $\mu$m \\
		&  &                              & \textit{Spitzer}/MIPS & 24 $\mu$m \\
		&                  &                             & \textit{Herschel}/PACS & 70, 100, 160 $\mu$m \\
		&  &                             & \textit{Herschel}/SPIRE & 250, 350, 500 $\mu$m \\
		& $>1000$ $\mu$m & \citet{scholtz2018}, & ALMA & Band 7 (800$-$1100 $\mu$m) \\
		&				 & and this work           &       & Band 6 (1100–-1400 $\mu$m) \\
		&				 &                      &       & Band 5 (1400–-1800 $\mu$m) \\
		&				 &                      &       & Band 4 (1800–-2400 $\mu$m) \\
		&				 &                       &      & Band 3 (2600$-$3600 $\mu$m) \\

		\hline
	\end{tabular} 
}

\tablefoot{$^a$ The $24-500$ $\mu$m photometry of lid\_1565 and cid\_72 is collected from \citet{lutz2011} and \citet{hurley2017}. }

\end{table*}

\renewcommand{\arraystretch}{1.1}
\begin{table*}[h]
\caption{ Input parameter values used for the SED-fitting procedure in Appendix \ref{app:sec:sedfit}.}
\label{app:tab:sed_input_parameters}
\centering
\resizebox{1\textwidth}{!}{
	\begin{tabular}{cccc}
		\hline\hline
		Template & Parameter & Value and range& Description \\
		\hline
		{\it Stellar emission} & IMF & \citet{chabrier2003}& \\
		& Z & 0.02 & Metallicity \\
		& Separation age & 10 Myr & Separation age between the young \\
		& & & and the old stellar populations \\
		Delayed SFH & Age &  0.10, 0.25, 0.5, 1.0, 1.5, 2.0, 2.5 Gyr & Age of the oldest SSP \\
		& $\tau$ & 0.10, 0.25, 0.5, 1.0, 3.0, 5.0, 10.0 Gyr & e-folding time of the SFH \\
		\hline 
		{\it Attenuation law} & $E(B-V)$ & 0.05, 0.1, 0.3, 0.5, 0.7, 0.9, 1.1, 1.3 & Attenuation of the \\
            &          &   & young stellar population \\
		  Modified Calzetti & Reduction factor & 0.93 & Differential reddening applied to \\
		&					&		& the old stellar population \\
		& $\delta$ & -0.6, -0.4, -0.2, 0.0 & Slope of the power law multiplying \\
		&			&		& the Calzetti attenuation law \\
		\hline
		{\it Dust emission} & $\alpha_{\textnormal{SF}}$ & 0.5, 1.0, 1.5, 2.0, 2.5, 3.0 & Slope of the power law combining \\
		&		&		& the contribution of different dust templates \\
		\hline
		{\it AGN emission} & $R_{\textnormal{max}}/R_{\textnormal{min}}$ & 60 & Ratio of the outer and inner radii \\
		& $\tau_{9.7}$ & 0.6, 3.0, 6.0 & Optical depth at 9.7 $\mu$m \\
		& $\beta$ & 0.00, -0.5, -1.0 & Slope of the radial coordinate \\
		& $\gamma$ & 0.0, 6.0 & Exponent of the angular coordinate \\
		& $\Theta$ & 100 degrees & Opening angle of the torus \\
		& $\psi$ & 0, 10, 20, 30, 40, 50, 60, 70, 80, 90 degrees & Inclination of the observer's line of sight\\
		& $f_{\textnormal{AGN}}$ & 0.05, 0.1, 0.15, 0.2, 0.25, 0.3, 0.35, 0.4, 0.45, & AGN fraction \\
		&	& 0.5, 0.55, 0.6, 0.65, 0.7, 0.75, 0.8, 0.85, 0.9 & \\
		\hline
		\textit{AGN intrinsic} & $\Gamma$ & 1.8$^a,b$, 1.9$^a,b$, 2.4$^a,b$, 3$^a,b$ & Intrinsic X-ray photon index \\
		\textit{ X-ray emission}   &  &  &  \\
		\hline
		{\it Nebular emission} & $U$ & $10^{-2}$ & Ionization parameter \\
		& $f_{\textnormal{esc}}$ & 0\% & Fraction of Lyman continuum \\
		&           &     & photons escaping the galaxy \\
		& $f_{\textnormal{dust}}$ & 10\% & Fraction of Lyman continuum \\
		&			&  & photons absorbed by dust \\
		\hline
\end{tabular}}
	\tablefoot{We divided the sample in different runs of CIGALE based on the photon index value obtained from the X-ray catalogs or X-ray fits to reduce the computational time. $^b$ Input values used only for cdfs\_313 and 
    cdfs\_794, which are AGN Type 1 with very steep intrinsic photon index \citep{luo2017}. }

\end{table*}
\renewcommand{\arraystretch}{1}

\end{document}